\documentstyle[aps,prc,preprint]{revtex}
\tightenlines
\newcommand{\bff}[1]{{\mbox{\boldmath $#1$}}}

\begin{document}

\draft

\title{\Large  Cranked Relativistic Hartree-Bogoliubov Theory: \\
probing the gateway to superheavy nuclei}

\author{A.\ V.\ Afanasjev$^{(1,2,3)}$, T.\ L.\ Khoo$^{(1)}$,
S.\ Frauendorf$^{(2,4)}$, G.\ A.\ Lalazissis$^{(5)}$, I.\ Ahmad$^{(1)}$}
\address{$^1$Physics Division, Argonne National Laboratory,
Argonne, IL 60439, USA}

\address{$^{2}$Department of Physics, University of Notre Dame,
Notre Dame, Indiana 46556, USA}

\address{$^{3}$Laboratory of Radiation Physics, Institute of Solid State
Physics, University of Latvia, \\ LV 2169 Salaspils, Miera str. 31, Latvia}

\address{$^{4}$IKH, Research Center Rossendorf, Dresden, Germany}

\address{$^5$Department of Theoretical Physics, Aristotle University 
of Thessaloniki, GR-54124, Thessaloniki, Greece}

\date{\today}

\maketitle

\begin{abstract}
The cranked relativistic Hartree+Bogoliubov theory has been applied for 
a systematic study of the nuclei around $^{254}$No, the heaviest elements 
for which detailed spectroscopic data are available. The deformation, 
rotational response, pairing correlations, quasi-particle and other 
properties of these nuclei have been studied with different parametrizations
for the effective mean-field Lagrangian. Pairing correlations are taken into
account by a finite range two-body force of Gogny type.
While the deformation properties are well reproduced, 
the calculations reveal some deficiencies of the effective forces both in the 
particle-hole and particle-particle channels. 
For the first time, the quasi-particle spectra of odd deformed nuclei 
have been calculated in a fully self-consistent way within the framework of 
the relativistic mean field (RMF) theory. The energies of the spherical 
subshells, from which active deformed states of these nuclei emerge, are 
described with an accuracy better than 0.5 MeV for most of the subshells 
with the NL1 and NL3 parametrizations. However, for a few subshells the 
discrepancies reach 0.7-1.0 MeV. In very heavy systems, where the level 
density is high, this level of accuracy is not sufficient for reliable 
predictions of the location of relatively small deformed shell gaps. The 
calculated moments of inertia reveal only small sensitivity to the RMF 
parametrization and, thus, to differences in the single-particle structure. 
However, in contrast to lighter systems, it is necessary to decrease 
the strength of the D1S Gogny force in the pairing channel in order 
to reproduce the moments of inertia.  
\end{abstract}

\pacs{PACS: 21.60.Cs, 21.60Jz, 27.90.+b, 21.10.Pc}

\input epsf

%%%%%%%%%%%%%%%%%%%%%%%%%%%%%%%%%%%%%%%%%%%%%%%%%%%%%%%%%%
\section{Introduction}
%%%%%%%%%%%%%%%%%%%%%%%%%%%%%%%%%%%%%%%%%%%%%%%%%%%%%%%%%%
   
 The possible existence of shell-stabilized superheavy nuclei, 
predicted with realistic nuclear potentials  
\cite{SGK.66,C.67,M.67} and the macroscopic-microscopic (MM) 
method \cite{Nilsson68,Nilsson69,MG.69}, has been a driving 
force behind experimental and theoretical efforts to investigate 
the superheavy nuclei. These investigations pose a number of 
experimental and theoretical challenges. The recent discovery of 
elements with Z=112 \cite{Z112}, Z=114 \cite{Z114a} and Z=116 
\cite{Z116} (for review of the present experimental situation see 
Refs.\ \cite{HM.00,A.00,Og01}) clearly shows great progress on 
the experimental side, but also indicates difficulties 
in the investigation of nuclei with low production cross-sections 
and analyses based only on 1 or 2 events.

 The theoretical challenges are also considerable since 
different theoretical methods predict different spherical shell 
closures.  Modern calculations based on the MM method with the 
Woods-Saxon \cite{Nilsson68,PS.91,CDHMN.96}, Nilsson \cite{Nilsson69}, and 
folded Yukawa \cite{MN.94} potentials indicate $Z=114$ and $N=184$ 
as the spherical shell closures. It is necessary to say, however, 
that some earlier calculations indicated $Z=126$ as a possible magic 
number (see Ref.\ \cite{HM.00} for a review).  There are differences 
in the predictions of self-consistent calculations, which depend both 
on the approach and on the effective force. Self-consistent calculations 
based on the 
Hartree-Fock method with Skyrme forces (SHF) predict spherical shell 
closures at $Z=126$ and $N=184$ for most of the forces 
\cite{CDHMN.96,RBBSRMG.97,BRRMG.99}. However, some forces indicate 
$Z=114$ (SkI4) and $Z=120$ (SkI3) as proton shell closures, while 
some predict no doubly magic superheavy nuclei at all. On the other 
hand, the relativistic mean field theory (RMF) prefers $Z=120$ and 
$N=172$ as spherical shell closures \cite{RBBSRMG.97,BRRMG.99}. 
However, $Z=114$  and $N=184$ also appear as the shell closures in 
some RMF calculations \cite{LSRG.96,NL-RA1}. For a detailed comparison of the predictions 
of the different Skyrme and RMF calculations, see Refs.\ 
\cite{BRRMG.99,RBM.01}. Hartree-Fock-Bogoliubov (HFB) calculations 
with the Gogny force give $Z=120, 126$ and $N=172, 184$ as spherical 
shell closures \cite{BBDGD.01}. 

 Clearly, the accuracy of predictions of spherical shell closures 
depends sensitively on the accuracy of describing the single-particle 
energies, which becomes especially important for superheavy nuclei, 
where the level density is very high. Variations in single-particle 
energy of $1-1.5$ MeV yield spherical shell gaps at different particle 
numbers, which restricts the reliability in extrapolating to the unknown 
region.

 Usually, the MM method describes the single-particle 
energies rather well. This is due to the fact that the experimental data on 
single-particle states are used directly in the parametrization of 
the single-particle potential. Moreover, different parametrizations 
of the single-particle potential are used in different mass regions. 
However, the extrapolation of the single-particle potential may be 
much less reliable since it is not determined self-consistently. 
For example, microscopic models predict that the appearance of the shell 
closures in superheavy nuclei is influenced by a central depression of 
the nuclear density distribution \cite{BRRMG.99,Detal99}. This effect 
cannot be treated in self-consistent way in current MM models.

 Although the nucleonic potential is defined in SHF and  RMF approaches 
in a fully self-consistent way, this does not guarantee that 
single-particle degrees of freedom are accurately described, as 
indicated by the large variety of the parametrizations (more than 60 
for SHF and about 20 for RMF \cite{RBM.01}). In addition, the parameters 
have been fitted in almost all cases to the bulk properties of  spherical 
nuclei. Single-particle information on spin-orbit splittings is used only 
in the fits of the parameters of the Skyrme and Gogny forces. 
The spin-orbit interaction is a 
relativistic effect, which arises naturally in the RMF theory. Thus, 
available RMF fits were obtained without the use of any single-particle 
information.
For heavy nuclei, the calculated RMF single-particle states were 
directly compared with experimental data only in spherical nuclei (see, 
for example, Refs.\ \cite{R.96,RBRMG.98} and references quoted therein).   
These comparisons, however, do not reveal the accuracy of the description 
of the single-particle states because the particle-vibration coupling, which 
can affect considerably the energies of single-particle states in spherical 
nuclei \cite{QF.78,RS.80,MBBD.85}, has been neglected.

 Compared with the MM method, self-consistent calculations have been 
confronted with experiment to a lesser degree and for a smaller 
number of physical observables (mainly binding energies and quantities
related to their derivatives). For many parametrizations, even the 
reliability of describing 
conventional nuclei is poorly known. In such a situation, it is 
important to perform a comprehensive study of the heaviest nuclei for 
which detailed spectroscopic information is available. {\it The 
results of such 
a study will allow us to better judge the reliability of predictions for 
superheavy nuclei.} The experimental data on deformed nuclei around $^{254}$No 
provides sufficient information for such a test. The purpose of this work is 
to compare the predictions of RMF theory with these data.

  RMF calculations have been compared with experiment in this mass region
in only Ref.\ \cite{BRBRMG.98}. However, the comparison was restricted to 
binding energies and quantities related to their derivatives. In addition, 
the pair correlations were treated in the BCS approximation, with no 
particle number projection. 

 In the present manuscript, the cranked relativistic Hartree-Bogoliubov 
(CRHB) theory \cite{A190,CRHB}, with approximate particle number projection 
by means of the Lipkin-Nogami method (CRHB+LN theory),  is employed 
for a detailed investigation of a wide set of experimental observables.
The use of the Lipkin-Nogami method has the clear advantage of avoiding
the collapse of pairing correlations at large shell gaps. We 
address for the first time the question of the blocking procedure in 
odd mass nuclei in the framework of the RMF theory, with effects of 
time-reversal symmetry breaking taken into account in a fully 
self-consistent way.  The calculated binding energies, deformations, 
moments of inertia and quasi-particle states are compared with 
experiment.

 The paper is organized in the following way. In Sect.\ \ref{theory},
a brief description of the CRHB+LN theory and of some specific features 
of the present calculations are given. In order to outline the general 
features of the evolution of physical observables as a function of proton 
and neutron number, systematic calculations with different RMF parametrizations 
are performed along the $Z=100$ (Fm) isotope and 
$N=152$ isotone chains. In addition, calculations are carried out 
for Cm, Cf and No nuclei, for which experimental data are available. 
The rotational properties are studied in detail in Sect.\ 
\ref{Rotation}, the deformations are discussed in Sect.\ \ref{Deformation} 
and the shell structure in Sect.\ \ref{shell-structure}. The quasiparticle 
spectra of selected nuclei  are compared with experiment in 
Sect.\ \ref{qp-energies}. Finally, Sect.\ \ref{Conclusion} summarizes 
our main conclusions.

%%%%%%%%%%%%%%%%%%%%%%%%%%%%%%%%%%%%%%%%
\section{Theoretical formalism}
\label{theory}
%%%%%%%%%%%%%%%%%%%%%%%%%%%%%%%%%%%%%%%

 In relativistic mean field (RMF) theory \cite{R.96,SW.86,Reinh89}
the nucleus is described as a system of point-like nucleons (Dirac 
spinors) coupled to mesons and photons. The nucleons interact 
by exchanging  scalar $\sigma$-mesons, vector $\omega$-, $\rho$- 
mesons and photons. The  isoscalar-scalar $\sigma $-mesons
generate strong intermediate range attraction between the nucleons. 
For the vector particles we have to distinguish the time-like 
 and the space-like components. In the case of photons,
they correspond to the Coulomb field and the magnetic field when 
currents are present. For the isoscalar-vector $\omega$-meson, 
the time-like component provides a very strong repulsion at short 
distances for all combinations of nucleons, $pp$, $nn$ and $pn$. 
For the isovector-vector $\rho$-meson, the time-like components 
give rise to a short range repulsion for like nucleons ($pp$ and $nn$) 
and a short range attraction for unlike nucleons ($np$). They also have 
a strong influence on the symmetry energy. The space-like components 
of the $\omega$- and $\rho$-mesons lead to an interaction between 
currents, which is attractive in the case of the $\omega$-meson for 
all combinations ($pp$, $nn$ and $pn$) and in the case of
the $\rho$-meson attractive for $pp$ and $nn$, but repulsive for 
$pn$. Within the mean field theory, these currents only occur in cases of 
time-reversal breaking mean fields, which applies to rotating or odd-mass 
nuclei.

 The cranked relativistic Hartree-Bogoliubov (CRHB) theory \cite{A190,CRHB} 
extends  the RMF theory to rotating nuclei and includes pairing correlations. 
If an approximate particle number projection is performed by means of the 
Lipkin-Nogami (LN) method \cite{L.60,N.64,NZ.64,PNL.73}, the abbreviation 
CRHB+LN  will be used. Since the theory is described in detail in 
Ref.\ \cite{CRHB}, only the features important for the present discussion 
will be outlined below.

%%%%%%%%%%%%%%%%%%%%%%%%%%%%%%%%%%%%%%%%%%%%%%%%%%%%%%%%%%%
\subsection{The CRHB+LN equations}
\label{CRHB-eq}
%%%%%%%%%%%%%%%%%%%%%%%%%%%%%%%%%%%%%%%%%%%%%%%%%%%%%%%%%%%
  
 The CRHB+LN equations for the fermions in the rotating frame are
given by \cite{CRHB}
\begin{eqnarray}
\pmatrix{
  \hat{h}_D' -\lambda'-\Omega_x \hat{J_x} & \hat{\Delta}        \cr
 -\hat{\Delta}^*& -\hat{h}_D'^{\,*} +\lambda'+\Omega_x \hat{J_x}^*   \cr}
  \pmatrix{ U(\bff r) \cr V(\bff r)}_k
  = E_k' \pmatrix{ U(\bff r) \cr V(\bff r)}_k ,
\end{eqnarray}
where
\begin{eqnarray}
  \hat{h}_D'&= & \hat{h}_D  + 4 \lambda_2 \rho - 2\lambda_2 Tr(\rho)
\,,  \\
\lambda' &=&  \lambda_1+ 2 \lambda_2 \,, \\
\label{equasi}
E_k' &=&  E_k - \lambda_2\,.
\end{eqnarray}
Here, $\hat{h}_D$ is the Dirac Hamiltonian for the nucleon with mass $m$;
$\lambda_1$ is defined from the average particle number constraints for 
protons and neutrons; $\rho_{\tau} =V^{*}_{\tau} V^{T}_{\tau}$ is the 
density matrix;  $U_k (\bff r)$ and $V_k (\bff r)$ are quasiparticle 
Dirac spinors; $E_k$ denotes the quasiparticle energies; and $\hat{J_x}$
is the angular momentum component.
The LN method corresponds to a restricted variation of $\lambda_2 \langle 
(\Delta N)^2 \rangle$ (see Ref.\ \cite{CRHB} for definitions of $\lambda_1$ 
and $\lambda_2$), where $\lambda_2$ is calculated self-consistently in each 
step of the iteration. The form of the CRHB+LN equations given above 
corresponds to the shift of the LN modification into the particle-hole 
channel. 

The Dirac Hamiltonian $\hat{h}_D$ contains an attractive scalar potential $S(\bff r)$
\begin{eqnarray}
S(\bff r)=g_\sigma\sigma(\bff r),
\label{Spot}
\end{eqnarray}
a repulsive vector potential $V_0(\bff r)$
\begin{eqnarray}
V_0(\bff r)~=~g_\omega\omega_0(\bff r)+g_\rho\tau_3\rho_0(\bff r)
+e \frac{1-\tau_3} {2} A_0(\bff r),
\label{Vpot}
\end{eqnarray}
and a magnetic potential $\bff V(\bff r)$
\begin{eqnarray}
\bff V(\bff r)~=~g_\omega\bff\omega(\bff r)
+g_\rho\tau_3\bff\rho(\bff r)+
e\frac {1-\tau_3} {2} \bff A(\bff r).
\label{Vmag}
\end{eqnarray}
The last term breaks time-reversal symmetry and induces currents.
In rotating nuclei, the time-reversal symmetry is broken by the 
Coriolis field. Without rotation, it is broken when the time-reversal orbitals 
are not occupied pairwise. In the Dirac equation, the space-like
components of the vector mesons $\bff\omega(\bff r)$ and 
$\bff\rho(\bff r)$ have the same structure as the space-like
component $\bff A(\bff r)$ generated by the photons. 
Since $\bff A(\bff r)$ is the vector potential of the magnetic
field, by analogy the effect due to presence of the vector
field $\bff V(\bff r)$ is called {\it nuclear magnetism} \cite{KR.89}.
It has considerable influence on the magnetic moments \cite{HR.88} 
and the moments of inertia \cite{KR.93,AKR.96,NM}. In the present calculations 
the spatial components of the vector  mesons are properly 
taken into account in a fully self-consistent way. The detailed description 
of the mesonic degrees of freedom in the CRHB+LN theory is presented 
in Ref.\ \cite{CRHB}.
              
 The pair field $\hat{\Delta}$ is given by
\begin{eqnarray}
\hat{\Delta} \equiv \Delta_{ab}~=~\frac{1}{2}\sum_{cd} V^{pp}_{abcd}
\kappa_{cd}
\label{gap}
\end{eqnarray}
where the indices $a,b,\dots$ denote quantum numbers which specify
the single-particle states with the space coordinates $\bff r$, as 
well as the Dirac and isospin indices $s$ and
$\tau$. It contains the pairing tensor  $\kappa$
\begin{eqnarray}
\kappa = V^{*}U^{T}
\label{kappa}
\end{eqnarray}
and the matrix elements $V^{pp}_{abcd}$ of the effective interaction in 
the particle-particle ($pp$) channel, for which the phenomenological 
non-relativistic Gogny-type finite range interaction 
\begin{eqnarray}
V^{pp}(1,2) = f \sum_{i=1,2} e^{-[({\bff r}_1-{\bff r} _2)/\mu_i]^2}
\times (W_i+B_i P^{\sigma}- H_i P^{\tau} - M_i P^{\sigma} P^{\tau})
\label{Vpp}
\end{eqnarray}
is used. The clear advantage of such a force is that it provides 
an automatic cutoff of high-momentum components. The motivation for 
such an approach to the description of pairing is given in 
Ref.\ \cite{CRHB,GELR.96}. In Eq.\ (\ref{Vpp}), $\mu_i$, $W_i$, $B_i$, $H_i$ 
and $M_i$ $(i=1,2)$ are the parameters of the force and $P^{\sigma}$ 
and $P^{\tau}$ are the exchange operators for the spin and isospin 
variables, respectively. Note  that a scaling  factor $f$ is introduced 
in Eq.\ (\ref{Vpp}). In our previous studies, the original (scaling factor 
$f=1.0$) parameter set D1S \cite{D1S,D1S-a} provided a good description of 
the moments of inertia in the $A\sim 75$ \cite{AF}, $A\sim 160-170$ 
\cite{J1Rare} and $A\sim 190$ \cite{A190,CRHB} mass regions. 
As discussed in Sect.\ \ref{Rot-pair}, it produces  pairing 
correlations in the $A\sim 250$ mass region that are 
too strong, and, thus, it has to be 
attenuated ($f<1.0$).
 
  As a  measure for the size of the pairing correlations in 
Hartree-(Fock)-Bogoliubov calculations, we use the pairing energy 
\begin{eqnarray}
E_{pairing}~=~-\frac{1}{2}\mbox{Tr} (\Delta\kappa).
\label{Epair}
\end{eqnarray}
%

%%%%%%%%%%%%%%%%%%%%%%%%%%%%%%%%%%%%%%%%%%%%%%%%%%%%%%%%%%
\subsection{The RMF parametrizations}
\label{RMFforces}
%%%%%%%%%%%%%%%%%%%%%%%%%%%%%%%%%%%%%%%%%%%%%%%%%%%%%%%%%%
 
 In the present study, the NL1 \cite{NL1}, NL-Z \cite{NLZ}, NL3 \cite{NL3}, 
NLSH \cite{NLSH} and NL-RA1 \cite{NL-RA1} parametrizations will be compared
in order to see how well observables, such as the moments of inertia, the
deformations, the quasiparticle energies, the separation energy and the
quantity $\delta_{2n}(Z,N)$ related to its derivative, agree with each 
other and with experiment. 

 These  sets differ in the experimental input used in the fitting
procedure. The binding energies of a number of spherical nuclei were 
included in the fit of all those sets, but the selection of nuclei
was different. NL1 and NL-Z employ the data mainly from the valley
of beta-stability, while additional information on neutron-rich 
nuclei has been used in the fit of the NL3 set. Moreover, 
there is a difference in the selection of additional observables 
used in the fit. Charge diffraction radii and surface thicknesses 
were included in the fit of NL1 and NL-Z sets \cite{NL1,NLZ}. The 
NL-Z set is a re-fit of NL1 where the correction for spurious 
center-of-mass motion is calculated from an actual many-body wave 
function \cite{NLZ}. On the contrary, NL3 and NLSH employ data 
on charge and neutron radii \cite{NL3,NLSH}. This (together with 
the fact that in the NL3 set more experimental data on neutron rich 
nuclei were used in fitting procedure) provides better a description 
of isospin, surface and symmetry properties of finite nuclei in the
NL3 and NLSH sets. Unfortunately, Ref.\ \cite{NL-RA1} does not
state which data the NL-RA1 set is fitted to.

 The sets NL1, NL3 and NLSH have been used extensively in RMF studies and 
tested on a wide range of physical observables related, for example, to the 
ground state properties, rotational properties, properties of giant resonances 
etc.; see Ref.\ \cite{R.96} for review. The sets NL-Z and NL-RA1 have been 
tested only for observables related to ground state properties. The set NL-Z 
is a re-fit of NL1 with a correction for spurious center-of-mass motion 
\cite{NLZ} given by 
\begin{eqnarray}
E_{cm}=-\langle \hat{\bff P}^2_{cm}\rangle/2mA , 
\label{ecm-micr}
\end{eqnarray}
where $\hat {\bff P}_{cm}$ is the total momentum operator in the center-of-mass
frame, $m$ is the nucleon mass and $A$ the mass number. This term is added after 
the variation is performed to circumvent two-body terms in the mean field equations. 
Thus,  the use of other prescriptions instead of Eq.\ (\ref{ecm-micr}) for the 
treatment of the center-of-mass motion with NL-Z  will affect only the binding 
energies and the quantities related to their derivatives. In all our 
calculations (including those with NL-Z), the correction for the spurious 
center-of-mass motion is approximated by its value in a nonrelativistic 
harmonic oscillator potential
\begin{eqnarray}
E_{cm}=-\frac{3}{4} 41 A^{-1/3} \quad {\rm MeV} . 
\label{ecm-ho}
\end{eqnarray}
This is consistent with the NL1,
NL3, NLSH and NL-RA1 parametrizations. As illustrated in Ref.\ 
\cite{BRRM.00}, Eq.\ (\ref{ecm-ho}) is a very good approximation to Eq.\ 
(\ref{ecm-micr}) in the $A\sim 250$ mass region: the difference between two 
prescriptions does not exceed 0.3 MeV, which is only $\approx$0.017\% correction 
to the typical binding energy and changes smoothly with the mass number $A$. 
Based on the results given in Fig.\ 2 of Ref.\ \cite{BRRM.00} one can 
estimate that in this mass region the use of Eq.\ (\ref{ecm-ho}) instead 
of Eq.\ (\ref{ecm-micr}) will affect two-particle separation energies 
$S_{2n}(Z,N)$ and $\delta_{2n}(Z,N)$ by at most 0.030 MeV.  
This justifies the use of Eq.\ (\ref{ecm-ho}) for NL-Z.

 The parametrization NL-RA1 has been introduced recently in Ref.\ \cite{NL-RA1}. 
A number of conclusions of this article has been strongly questioned in Ref.\ 
\cite{Bend01}, in part due to the use of an unrealistically strong pairing 
interaction. However, if a more realistic pairing is employed, this 
parametrization 
provides a rather good description of the binding energies (see Fig.\ 2 in 
Ref.\ \cite{Bend01}).

%%%%%%%%%%%%%%%%%%%%%%%%%%%%%%%%%%%%%%%%%%
\subsection{Details of the calculations}
%%%%%%%%%%%%%%%%%%%%%%%%%%%%%%%%%%%%%%%%%%

 The CRHB(+LN) equations are solved in the basis of an anisotropic
three-dimensional harmonic oscillator in Cartesian coordinates with
the deformation parameters $\beta_0=0.3$, $\gamma=0^{\circ}$ and 
oscillator frequency $\hbar \omega_0=41$A$^{-1/3}$ MeV.
All fermionic and bosonic states belonging to the shells up to $N_F=14$ 
and $N_B=16$ are taken into account in the diagonalization of the Dirac 
equation and the matrix inversion of the Klein-Gordon equations, respectively. 
The detailed investigation of $^{246,248,250}$Fm indicates that this truncation 
scheme provides reasonable numerical accuracy. The values of the kinematic
moment of inertia $J^{(1)}$, charge quadrupole moment $Q_0$, mass hexadecapole 
moment $Q_{40}$, binding energies, separation energies $S_{2n}(Z,N)$, and  
$\delta_{2n}(Z,N)$ obtained with truncation of the basis at $N_F=14$ and 
$N_B=16$ differ from the values obtained with $N_F=18$ and $N_B=18$ by less 
than 0.75\%, 0.9\%, 3.4\%, 0.1\%, 40 keV, and 40 keV, respectively. 
The convergence in energy of our calculations is similar to
that reported in non-relativistic calculations of Ref.\ 
\cite{WERP.02} based on the Gogny force.

%%%%%%%%%%%%%%%%%%%%%%%%%%%%%%%%%%%%%%%%%%%%%%%%%%%%%%%%%
\section{Rotational response}
\label{Rotation}
%%%%%%%%%%%%%%%%%%%%%%%%%%%%%%%%%%%%%%%%%%%%%%%%%%%%%%%%%

%%%%%%%%%%%%%%%%%%%%%%%%%%%%%%%%%%%%%%%%%%%%%%%%%%%%%%%%%
\subsection{The $^{254}$No ground band}
\label{rot254No}
%%%%%%%%%%%%%%%%%%%%%%%%%%%%%%%%%%%%%%%%%%%%%%%%%%%%%%%%%

\begin{figure}
\epsfxsize 13.5cm
\epsfbox{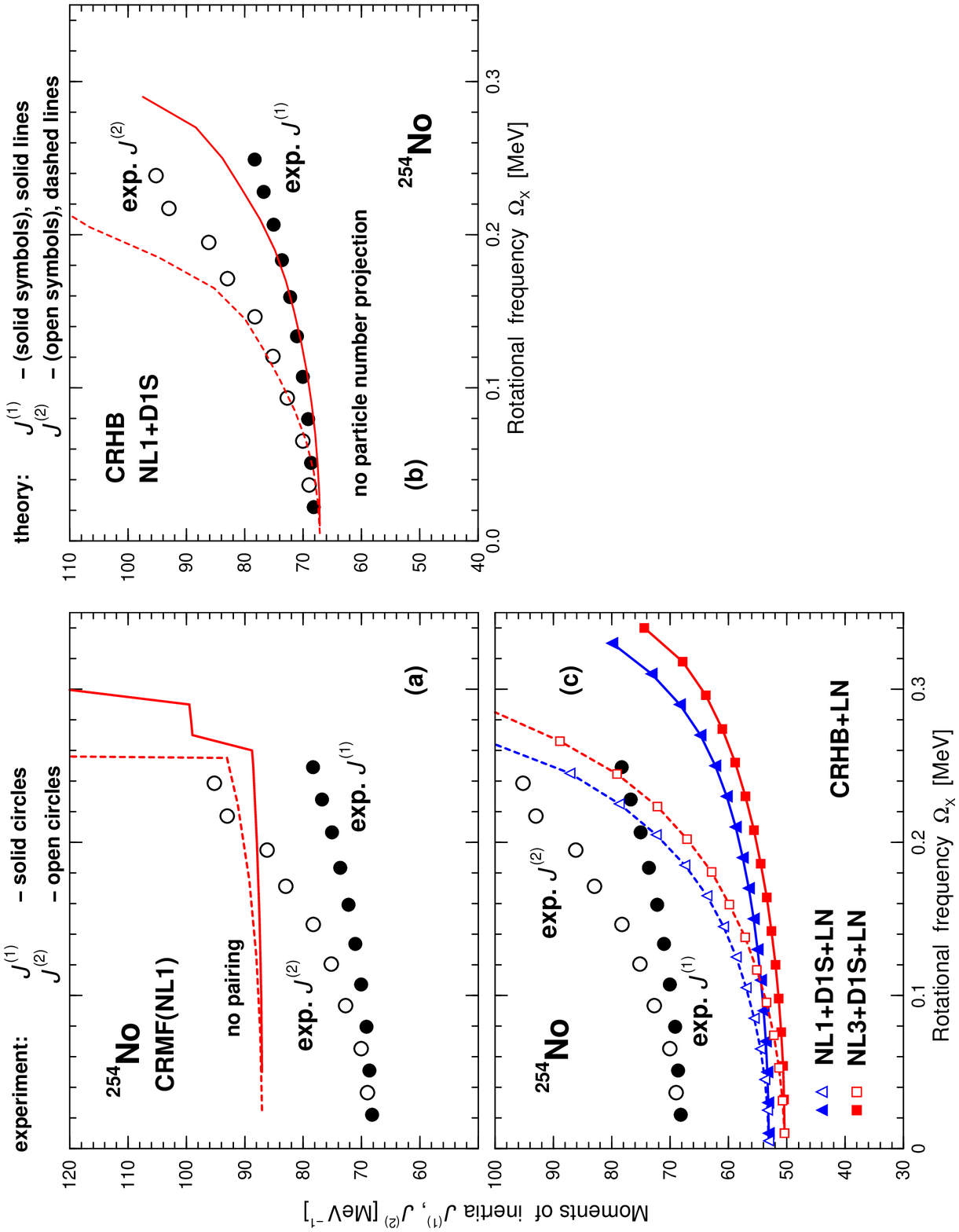}  
\vspace{0.5cm}
\caption{Experimental and calculated dynamic ($J^{(2)}$)
and kinematic ($J^{(1)}$) moments of inertia of the lowest 
band in $^{254}$No. Panel (a) shows the CRMF results without
pairing. The CRHB and CRHB+LN results are displayed in panels (b) 
and (c), respectively. The results of the CRHB(+LN) calculations 
are shown only up to the rotational frequency where a paired 
band crossing takes place. Note different scales of the ordinate 
on different panels.}
\label{no254-j2j1-comp}
\end{figure}

 The observed moments of inertia of the ground band in $^{254}$No 
\cite{No254-exp1,No254-exp2,No254-exp3} are compared with the 
calculated values in Fig.\ \ref{no254-j2j1-comp}. The 
CRMF calculations without pairing, based on the NL1 parameterization,
marked as CRMF(NL1) in  Fig.\ \ref{no254-j2j1-comp}a, provide an 
almost constant kinematic moment of inertia $J^{(1)}\approx 87$ 
MeV$^{-1}$ up to $\Omega_x \sim 0.26$ MeV and a dynamic moment of 
inertia which slightly increases with rotational frequency. A band 
crossing with another configuration takes place at $\Omega_x \sim 
0.26$ MeV. These calculations provide a reference point for how much 
the moments of inertia decrease due to pairing. It is interesting 
to note that the moments of inertia in the calculations without 
pairing are only one half of the rigid body value. This unexpected
result will be discussed in detail in forthcoming article \cite{AF}.

 The CRHB calculations without particle number projection (scaling factor 
$f=1$), marked as CRHB (NL1+D1S) in Fig.\ \ref{no254-j2j1-comp}b, agree 
very well with experiment up to $\Omega_x=0.18$ MeV. At higher frequency,  
experiment and theory diverge. With approximate particle number projection 
using the LN method, marked as NL1+D1S+LN  in  Fig.\ \ref{no254-j2j1-comp}c, 
the theory underestimates the experimental kinematic and dynamic moments of 
inertia by $\sim 25\%$. This result is in contrast with the good agreement 
obtained by the same method for superdeformed bands with $A\sim 190$ 
\cite{A190,CRHB} and for nuclei in the rare-earth \cite{J1Rare} and $A\sim 75$ 
\cite{AF} regions. 

 Different parametrizations of the RMF Lagrangian give quite similar results 
for the moments of inertia if $f=1$. For example, the results of the CRHB+LN 
calculations based on the NL3 parametrization \cite{NL3} (marked as NL3+D1S+LN 
in Fig.\ \ref{no254-j2j1-comp}c) provide moments of inertia
which are only slightly lower (by $\approx 3$ MeV$^{-1}$) than 
the ones obtained with the NL1 parametrization. The CRHB+LN calculations 
with the NL-Z \cite{NLZ}, NLSH \cite{NLSH} and NL-RA1 \cite{NL-RA1} 
parametrizations of the RMF Lagrangian give results which are quite 
similar to those obtained with NL3 
%%%%%%%%%%%%%%%%%%%%%%%%%%%%%%%%%
\footnote{The CRMF  calculations (without pairing) for the ground state band 
in $^{254}$No also show a weak dependence of the moments of inertia on the RMF 
parametrization, and, thus on details of the single-particle structure. A similar 
situation has been encountered earlier in the $A\sim 60$ 
\cite{A60} and $A\sim 150$ \cite{A150,ALR.98,Afan-unpub} mass regions of 
superdeformation.}
%%%%%%%%%%%%%%%%%%%%%%%%%%%%%%%%%%%%%%%%%%%%%%%%%
and, thus, they are not shown in Fig.\ \ref{no254-j2j1-comp}. 
The moments of inertia are
very similar despite the differences in the single-particle spectra 
near the Fermi level (see Figs.\ \ref{proton-no254-sp} 
and \ref{neutron-no254-sp}). Hence, the most likely 
reason for discrepancies between experiment and 
the CRHB+LN calculations lies in an inadequate parametrization of the Gogny 
force (D1S) in the particle-particle channel for this mass region, which
gives too strong pair correlations. This is not quite unexpected since no 
experimental data above $^{208}$Pb have been used when the D1S set was 
fitted. The study of other heavy nuclei around $^{254}$No also shows that the 
kinematic moments of inertia obtained in the CRHB+LN calculations with the 
original D1S force (scaling factor $f=1.0$) are systematically lower than 
experimental ones (by $\sim 20\%$ in even-even $^{236-244}$Pu nuclei).
Different parametrizations of the RMF Lagrangian give similar results and 
the deformations of these nuclei are well described in the calculations (see 
Sect.\ \ref{Deformation}). It is unlikely that other available parametrizations 
for the Gogny force such as D1 \cite{D1} and D1P \cite{D1P} will improve the 
situation, since they produce even stronger pairing than the D1S set 
in superdeformed bands of the $A\sim 190$ mass region \cite{A190}.

 The results of the calculations for the $^{254}$No rotational band obtained 
in the non-relativistic cranked HFB approach based on the Gogny force with 
D1S set of parameters  \cite{ER.00} seem to support this interpretation.
These calculations, which are performed without particle number projection, 
also come very close to the data. One presumes that the inclusion of 
particle number projection by means of the LN method will lower the 
calculated kinematic moment of inertia, as has been seen in the rare-earth 
region \cite{J1Rare}, leading to a similar situation as described above.

%%%%%%%%%%%%%%%%%%%%%%%%%%%%%%%%%%%%%%%%%%%%%%%%%%%%%%%
\subsection{Selection of the pairing strength}
\label{Rot-pair}
%%%%%%%%%%%%%%%%%%%%%%%%%%%%%%%%%%%%%%%%%%%%%%%%%%%%%%%

Quantitative information on the strength of the pair correlations can be 
extracted from the odd-even mass differences, excitation energies of high-K 
isomers or the moments of inertia. We use the moments of inertia for an 
adjustment of the strength of the Gogny force because they  are not too 
sensitive to the details of the single-particle spectrum (see above).

 Our CRHB+LN calculations indicate that in the $A\sim 250$ mass region 
the strength of pairing correlations should be reduced in order to 
reproduce the observed moments of inertia. The scaling factor $f$ 
of the Gogny D1S force (see Eq.\ (\ref{Vpp})) has been chosen to reproduce 
the experimental kinematic moment of inertia of $^{254}$No at rotational 
frequency $\Omega_x=0.15$ MeV. The values found for the various parametrizations 
of the RMF Lagrangian are given in Table \ref{Table-scaling}. These scaling factors, 
which are nearly the same,  are used in all subsequent calculations, unless 
otherwise  specified. The scaled CRHB+LN calculations reproduce the amplitude 
and the $\Omega_x$-dependence of the dynamic and the kinematic moments of inertia 
in $^{254}$No (see Fig.\ \ref{no5254-nl1nl3}b,d) rather well. With NL3, experiment 
and theory agree very well, while with NL1 some discrepancy develops above 
$\Omega_x=0.2$ MeV. Our choice of the scaling factor $f$ leads also to a reasonable 
description of the odd-even mass differences in the CRHB+LN calculations (see 
columns 5 and 6 in Table \ref{Table-odd-even}). 

  The need for attenuation of the D1S force within the framework of the 
CRHB+LN theory is not surprising since its pairing properties were adjusted 
by fitting only the odd-even mass differences of the Sn isotopes. Thus the 
quality of the description of pairing may deteriorate far from this mass region. 
Indeed, the moments of inertia of nuclei in mass regions closer to the Sn region, 
such as the rare-earth region \cite{J1Rare}, the superdeformed $A\sim 190$ mass 
region \cite{A190,CRHB} and neutron-deficient $A\sim 75$ region \cite{AF}, are 
described well by means of CRHB+LN calculations using the original D1S force.

 CRHB calculations (without LN) with original scaling factor $f=1.0$ 
provide a reasonable description of both moments of inertia before band 
crossing (see Fig.\ \ref{no254-j2j1-comp}b) and odd-even mass differences (see 
Table \ref{Table-odd-even}). This approach will be applied to the calculations 
of the quasiparticle spectra in odd nuclei (see Sects.\ \ref{Bk249} and 
\ref{Cf249}), for which the CRHB+LN calculations are numerically less 
stable. However, it is not justified for the calculations at large rotational 
frequencies because an unphysical pairing collapse takes place above the 
crossing between the ground and $S$-bands.

%%%%%%%%%%%%%%%%%%%%%%%%%%%%%%%%%%%%%%%%%%%%%%%%%%%%%%%%%%%%%%%%
\subsection{High-spin behavior}
%%%%%%%%%%%%%%%%%%%%%%%%%%%%%%%%%%%%%%%%%%%%%%%%%%%%%%%%%%%%%%%

 Alignment and backbending features of the rotational bands in the 
actinide region have been discussed in a number of publications; 
see Refs.\ \cite{CF.83,ER.84} and references quoted therein. In this mass 
region, two high-$j$ shells ($i_{13/2}$ for protons and $j_{15/2}$ 
for neutrons) come close to the Fermi surface and the angular momentum
of quasiparticles in either orbital can align with the axis of 
rotation. The CRHB+LN calculations  with the NL3 parametrization show 
that the alignment of the proton $i_{13/2}$ pair ($\pi [633]7/2$) and 
neutron $j_{15/2}$ 
pair ($\nu [734]9/2$) (see Fig.\ \ref{qp-routh-no254}) takes place 
simultaneously in $^{254}$No  at $\Omega_x 
\approx 0.32$ MeV (see Fig.\ \ref{no254-contr-toj1j2}). The total 
angular momentum gain at the band crossing is $\approx 17\hbar$, with proton and neutron 
contributions of $\approx 7\hbar$ and $\approx 10\hbar$, respectively. The 
alignment leads to a decrease of the mass quadrupole moment $Q_0$, to a
sign change of the mass hexadecapole moment $Q_{40}$ and to an appreciable 
increase of $\gamma$-deformation (see Fig.\ \ref{no254-nl3-full}a,b,c). 
A similar situation holds also in $^{252}$No, but the total angular
momentum gain at the band crossing is smaller ($\approx 11\hbar$). 

 The simultaneous alignment of proton and neutron pairs occurs also in 
calculations with NL1. The crossing frequency is shifted down by $\approx 0.01$ 
MeV (see caption of Fig.\ \ref{no5254-nl1nl3}), the angular momentum gain at 
the band crossing is slightly different, and the high-$j$ particles align 
in $^{252}$No more gradually. 

 Our results differ from those of the cranked HFB calculations based on the 
Gogny force \cite{ER.00}, which indicate in $^{254}$No upbending at $I\sim 30\hbar$ 
and backbending at $I\sim 38\hbar$. These calculations are performed without 
particle number projection, which result in a collapse of neutron pairing 
correlations at relatively low spin, $I\sim 20\hbar$. In our calculations, 
the pairing energies decrease with increasing rotational frequency due 
to the Coriolis anti-pairing effect, but there is no collapse 
of pairing (see Fig.\ \ref{no254-nl3-full}d). The experimental data do 
not extend up to predicted backbending and thus do not discriminate
between these calculations.

%%%%%%%%%%%%%%%%%%%%%%%%%%%%%%%%%%%%%%%%%%%%%%%%%%%%%%%%%%%
\subsection{$^{252}$No versus $^{254}$No}
%%%%%%%%%%%%%%%%%%%%%%%%%%%%%%%%%%%%%%%%%%%%%%%%%%%%%%%%%%%

  The kinematic and dynamic moments of inertia of the band in 
$^{252}$No show similar trends in rotational frequency as the 
experimental data (see Fig.\ \ref{no5254-nl1nl3}a,c). In 
experiment, the moments of inertia at low rotational frequencies 
are smaller for $N=150$ than for $N=152$, in contrast with the
calculations. One possible reason is the fact that the CRHB+LN 
calculations give deformed shell gaps at $N=148$ and/or 150 
(dependent on parametrization), rather than at $N=152$ as seen 
in experiment (see Sects.\ \ref{shell-structure} and \ref{qp-energies} 
for details). Indeed, in the calculations with the NL3 set the 
neutron contribution to the total moment of inertia (see Fig.\ 
\ref{no-contr-j1}) increases at the $N=148$ shell gap, most likely 
due to the weakening of the neutron pair correlations (see the 
pairing energies for the Fm isotopes in Fig.\ \ref{pairing-chains}, 
which are similar to those for the No isotopes). This suggests that 
if the calculations were to give a shell gap at $N=152$, the relative 
magnitudes of the moments of inertia in $^{252,254}$No would be 
reproduced.

%%%%%%%%%%%%%%%%%%%%%%%%%%%%%%%%%%%%%%%%%%%%%%%%%%%%%%%%
\subsection{Results for other nuclei and general trends of the  
moments of inertia as functions of the particle number}
\label{j1-general trends}
%%%%%%%%%%%%%%%%%%%%%%%%%%%%%%%%%%%%%%%%%%%%%%%%%%%%%%%%
 
 Figures \ref{j1-fm} and \ref{j1-cf-cm-no} demonstrate that the $N$-dependence
of the moments of inertia in $^{254,256}$Fm and $^{248,250,252}$Cf is rather 
well described in all parametrizations, whereas some problems exist for 
$^{242-250}$Cm (see Fig.\ \ref{j1-cf-cm-no}a). The absolute values are 
typically reproduced within a few \% in the Fm and Cf isotopes, but the 
discrepancy between experiment and calculations becomes somewhat larger 
for the Cm isotopes. The experimental values of the moments of inertia in 
the $N=152$ isotopes are reproduced rather well in all RMF parametrizations, 
with the exception of NLSH, which somewhat 
underestimates the moments of inertia (see Fig.\ \ref{j1-n152}).  One should 
note, however, that the maximum value of $J^{(1)}$ is at $Z=96$ in the calculations,
while available data show the maximum at $Z=98$. 

It is interesting to compare the present CRHB+LN results with those from other 
models. The calculations using the MM method in Ref.\ \cite{SMP.01} agree
reasonably well 
with experimental excitation energies of $E(2^+)$ states
and thus with the moments of inertia (see Figs.\ \ref{j1-fm}, \ref{j1-cf-cm-no}
and \ref{j1-n152})   
in this mass region.  
This is not surprising considering that these data have been used in 
the fit of the strengths  (monopole and isospin-dependent) of 
the proton and neutron pairing correlations. However, despite
the use of 4 adjustable parameters for pairing and better single-particle 
spectra obtained in the Woods-Saxon potential,
the  overall level of agreement is comparable with that obtained 
in the CRHB+LN calculations  (see Figs.\ \ref{j1-fm}, \ref{j1-cf-cm-no}
and \ref{j1-n152}).
For example,  a detailed examination indicates that the $N$-dependence
is not described correctly for $^{248,250}$Cf and $^{244,246}$Cm (see Table 1 
in Ref.\ \cite{SMP.01} and  Fig.\ \ref{j1-cf-cm-no} in the
present manuscript). A similar problem exists for the $Z$-dependence of 
the moments of inertia in $^{254}$No and $^{250}$Cf nuclei 
(see Fig.\ \ref{j1-n152}). This suggests that the description of the single-particle 
states within the Woods-Saxon potential, with the ``universal'' set of parameters 
\cite{CDNSW.87}, is still not completely correct in this mass region, despite the 
fact that the systematics of the experimental data on both spherical and deformed 
odd-mass nuclei were simultaneously taken into account in the fit of these 
parameters. 

 While there are several calculations of the moments of inertia by means of the 
MM method in the actinide and trans-actinide regions (see Ref.\ \cite{SMP.01} and 
references therein), little has been done in microscopic approaches so far. The 
rotational bands in $^{252,254}$No have been studied in the cranked HFB approaches 
based on the Skyrme  \cite{DBH.01,LSQP.01} and Gogny forces \cite{ER.00,Egido-private} 
(see discussion above). The relative magnitude of the moments of inertia of these 
two nuclei is not reproduced in either approach (see, for example, Fig.\ 6 in Ref.\ 
\cite{DBH.01}). In the calculations with Skyrme forces, this is most likely due to 
the fact that the deformed shell gap appears at $N=150$ \cite{BRBRMG.98}, instead of 
the experimentally observed value of $N=152$.

 The systematic calculations for the Fm isotopes and $N=152$ isotones permit 
the following general observations. Different RMF parametrizations give similar $N$- 
and  $Z$- dependencies of the moment of inertia (see Figs.\ \ref{j1-fm} and 
\ref{j1-n152}). In the Fm isotopes, the correlation between the calculated 
quadrupole deformations $\beta_2$ (see Fig.\ \ref{def-fm}) and  moments of inertia $J^{(1)}$ 
(see Fig.\ \ref{j1-fm}) is clearly seen with nearly constant values of $J^{(1)}$ 
and $\beta_{2}$ at $N=138-160$, followed by a pronounced drop of both for 
$N\geq 160$. 

 The situation is more complicated in the $N=152$ isotones, where the change 
of $J^{(1)}$ as a function of proton number does not correlate with  $\beta_2$. 
While $\beta_2$ is almost constant for $90 \leq Z \leq 110$ (see Fig.\ 
\ref{def-n152-allforces}), $J^{(1)}$ shows a maximum at $Z=96$ and a minimum at 
$Z=108$ (see Fig.\ \ref{j1-n152}). 

By comparing the different RMF parametrizations, one can see that the sets 
which produce a smaller quadrupole deformation also produce a smaller moment 
of inertia. NL1 and NL-Z give very similar  $J^{(1)}$. The same holds for NL3 
and NL-RA1, whereas NLSH provides smaller values of $J^{(1)}$. It can also be 
seen that differences in the underlying single-particle spectrum (see Sect.\ 
\ref{qp-energies}) do not lead to significant modifications of the moments of 
inertia. This suggests that many orbitals contribute to the angular momentum.

%%%%%%%%%%%%%%%%%%%%%%%%%%%%%%%%%%%%%%%%%%%%%%%%%%%%%%%%%%%%%%%%%%%%
\subsection{Summary for Section \ref{Rotation}}
%%%%%%%%%%%%%%%%%%%%%%%%%%%%%%%%%%%%%%%%%%%%%%%%%%%%%%%%%%%%%%%%%%%%

  In general, the moments of inertia for the heaviest nuclei are 
well described by the CRHB+LN theory.  However, it was necessary to
reduce the strength of the D1S Gogny force in the pairing channel 
by $\approx 12\%$, whereas in lighter nuclei with $A \approx 70-190$, 
the moments of inertia are well reproduced with a full strength D1S 
force. This points to a somewhat different $A$-dependence of the 
pairing strength than given by the Gogny force. The trends
around $A\sim 250$ with respect to neutron or proton numbers are 
reasonably well reproduced. The calculations with and without 
pairing indicate very weak dependence (within $\approx 5\%$) of 
the moments of inertia on the parametrization of the RMF Lagrangian. 
The moments of inertia in this mass region are highly collective. 
Since many single-particle orbitals contribute, they are insensitive 
to fine details of the single-particle states. On the other hand, 
deformed shell gaps affect moments of inertia to some extend leading 
to a larger values.

%%%%%%%%%%%%%%%%%%%%%%%%%%%%%%%%%%%%%%%%%%%%%%%%%%%%%%%%%%%%%%%%%%%%
\section{Deformations}
\label{Deformation}
%%%%%%%%%%%%%%%%%%%%%%%%%%%%%%%%%%%%%%%%%%%%%%%%%%%%%%%%%%%%%%%%%%%%

%%%%%%%%%%%%%%%%%%%%%%%%%%%%%%%%%%%%%%%%%
\subsection{Comparison with experiment}
%%%%%%%%%%%%%%%%%%%%%%%%%%%%%%%%%%%%%%%%%

  Direct experimental information on the deformations of nuclei from 
Coulomb excitation and lifetime measurements is quite limited 
\cite{Raman.87}. An alternative method is to derive a quadrupole 
moment from the $2^+ \rightarrow 0^+$ transition energy by 
employing the relation given by Grodzins \cite{G.62} or by later refinements
\cite{Raman.87,No252}. The prescription of Ref.\ \cite{No252} has an 
accuracy of about 10\%. In Figs.\  \ref{def-n152-allforces}, \ref{def-fm} 
and \ref{def-cf-cm-no}, the results of the CRHB+LN calculations are compared 
with deformations extracted by this method. From the calculated and 
experimental charge quadrupole moments $Q$, we derive the deformation 
parameters $\beta_2$ by the relation
\begin{eqnarray} 
Q=\sqrt{ \frac{16\pi}{5}} \frac{3}{4\pi} Z R_0^2 \beta_2,
\,\,\,\,\,\,\,{\rm where}\,\,\,\,\,\,\, R_0=1.2 A^{1/3}.
\end{eqnarray}
The simple linear expression is used to maintain consistency with 
earlier papers \cite{Raman.87}. It is sufficient for comparison
between calculations and experiment because the same relation is
used. Including higher powers of $\beta_2$, e.\ g. as in Ref.\ 
\cite{NZ-def}, yields values of $\beta_2$ that are $\approx 10\%$ 
lower.

 Figures \ref{def-n152-allforces}, \ref{def-fm} and \ref{def-cf-cm-no} 
demonstrate that the values of $\beta_2$ obtained from the 
$2^+ \rightarrow 0^+$ transition energies with the prescription of 
Ref.\ \cite{No252} are consistent with those from Coulomb excitation 
measurements. The CRHB+LN calculations with the NL3, NL-RA1, NL1, and 
NL-Z parametrizations agree rather well with these values. Considering the 
uncertainties on the extracted values of $\beta_2$ and the limited 
experimental data, it is difficult to give any preference for a
particular set. Only NLSH seems to systematically underestimate 
$\beta_2$.

%%%%%%%%%%%%%%%%%%%%%%%%%%%%%%%%%%%%%%%%%%%%%%%%%%%%%%%%%%%%%%%
\subsection{General trends}
%%%%%%%%%%%%%%%%%%%%%%%%%%%%%%%%%%%%%%%%%%%%%%%%%%%%%%%%%%%%%%%

Figures  \ref{def-n152-allforces} and \ref{def-fm} illustrate the 
general trends of deformation as functions of proton and neutron 
numbers for the $N=152$ isotones and for the Fm $(Z=100)$ isotopes. 
In the Fm chain, $\beta_2$ increases gradually from 
$N=138$ up to  $N\approx 150$ for all parametrizations except NLSH, 
which gives a slight decrease around $N=140$. For $N\geq 150$, there 
is a gradual decrease of the $\beta_2$ values, which becomes more
rapid above $N\approx 160$. These trends are more pronounced with 
NL3, NLSH and NL-RA1, which have been fitted to the data on neutron-rich 
nuclei as well. The variations are more gradual in the NL1 parametrization, 
which was obtained by fit to data from the beta-stability valley.
The mass 
hexadecapole moments $Q_{40}$ are similar for all parametrizations and 
decrease with increasing neutron number. While the changes of the slope 
of $\beta_2$ as a function of neutron number (see Fig.\ \ref{def-fm}a) 
correlate with the shell gaps at $N=148,150$ and $N=160,162$ (Sect.\ 
\ref{shell-structure}), no such correlations are seen for the $Q_{40}$ 
values (see Fig.\ \ref{def-fm}b).

 The $\beta_2$ and $Q_{40}$ values change more gradually with proton 
number $Z$ in the $N=152$ chain (see Fig.\ \ref{def-n152-allforces}). 
The $\beta_2$ values are almost constant as a function of proton number. 
The calculated mass hexadecapole moments $Q_{40}$ show a sinusoid-like 
curve as a function of proton number, with a maximum at $Z\approx 94$ 
and a minimum at $Z\approx 106$. 

 We expect that the trends are similar in the chains adjacent to
the Fm and $N=152$ chains, which is corroborated by the less systematic 
calculations for the Cm, Cf and No isotopes (see Fig.\ \ref{def-cf-cm-no}).
The equilibrium deformations are very similar for NL3 and NL-RA1 
on the one hand, and for NL1 and NL-Z  on the other hand. For this
reason, the results obtained with NL-RA1 and NL-Z are omitted in Figs.\ 
\ref{def-n152-allforces}, \ref{def-fm} and \ref{def-cf-cm-no}. The 
calculated $\beta_2$ increases as the parametrization changes from 
NL-SH to NL3 to NL1. This trend has previously been seen in the 
$A\sim 60, 150,$ and 190 regions of superdeformation 
\cite{CRHB,A60,ALR.98}.

%%%%%%%%%%%%%%%%%%%%%%%%%%%%%%%%%%%%%%%%%%%%%%%%%%%%%%%%%%%%%%%%
\subsection{Summary for Section \ref{Deformation}}
%%%%%%%%%%%%%%%%%%%%%%%%%%%%%%%%%%%%%%%%%%%%%%%%%%%%%%%%%%%%%%%%

 In summary, the CRHB+LN theory with the NL3, NL1, NL-RA1
and NL-Z parameter sets satisfactorily reproduces 
the magnitude of the $\beta_2$ deformation of the heaviest nuclei, 
where they have been measured, whereas NLSH systematically 
underestimates it.

%%%%%%%%%%%%%%%%%%%%%%%%%%%%%%%%%%%%%%%%%%%%%%%%%%%%%%%%%%%
\section{Shell structure}
\label{shell-structure}
%%%%%%%%%%%%%%%%%%%%%%%%%%%%%%%%%%%%%%%%%%%%%%%%%%%%%%%%%%%

The stability of the superheavy elements is due to a region of low 
level density in the single-particle spectrum. For deformed nuclei
all single-particle states are two-fold degenerate, and, thus, the
region of low level density generally correlates with the 'shell 
gap'. The situation is more complicated in spherical nuclei, where 
the shell correction energy depends not only on the size of the 
shell gap, but also on the degeneracy of single-particle states in 
the vicinity of the Fermi level \cite{KBNRVC.00,BNR.01}.
It is
a concern of this paper to study how well the different RMF 
parametrizations reproduce the shell gaps in the heavy deformed
elements. 
A simple intuitive measure for the level density at the Fermi surface
is the distance $E_{SP-GAP}$ between the last occupied and the first 
empty  levels. Another way is to consider the
two-neutron $S_{2n}(Z,N)$ and two-proton $S_{2p}(Z,N)$ 
separation energies 
\begin{eqnarray}
S_{2n}(Z,N)=B(Z,N)-B(Z,N-2) \nonumber \\
S_{2p}(Z,N)=B(Z,N)-B(Z-2,N),
\label{2p-sep-ener}
\end{eqnarray}
where $B(N,Z)$ is the binding energy. The separation energies  show 
a sudden drop at the shell gaps, if they are large. If the variations
of the level density are less pronounced, the quantity $\delta_{2n}(Z,N)$
related to the derivative of the separation energy is a more sensitive 
indicator of the localization of the shell gaps. For the neutrons (and 
similarly for the protons) it is defined as
\begin{eqnarray}
\delta_{2n}(Z,N)=S_{2n}(Z,N)-S_{2n}(Z,N+2)= \nonumber \\
=-B(Z,N-2)+2B(Z,N)-B(Z,N+2).
\label{2n-shell-gap}
\end{eqnarray}
We show in Appendix A that variations (but not their absolute
values) of $\delta_{2n}(Z,N)$ and $E_{SP-GAP}$ agree rather 
well. 

 In this section, we study the shell structure along both the Fm $(Z=100)$ 
and the $N=152$ chains.

%%%%%%%%%%%%%%%%%%%%%%%%%%%%%%%%%%%%%%%%%%%%%%%%%%%%
\subsection{Shell structure along the $Z=100$ line}
\label{sstruc100}
%%%%%%%%%%%%%%%%%%%%%%%%%%%%%%%%%%%%%%%%%%%%%%%%%%%%

 The results for two-neutron separation energies $S_{2n}(Z,N)$ in the Fm 
$(Z=100)$ chain with different RMF parametrizations are shown in Fig.\ 
\ref{sep-ener-fm}. There is a systematic difference between  the  
NL1, NL-Z sets and the NL3, NLSH, NL-RA1 sets. The former 
underestimate two-neutron separation energies, thus revealing 
their weakness in the description of isotopic trends. NL-Z is
somewhat better as compared with NL1. Considering that these two 
sets are fitted to the same  data, this result together with the 
one shown in Fig. 2 of Ref.\ \cite{Bend01}
possibly indicate the importance of a more microscopic treatment of 
the center-of-mass correction for the reproduction of isotopic 
trends.
On the other hand, 
NL3, NLSH and NL-RA1 better reproduce the  experimental $S_{2n}(Z,N)$ 
values (see Fig.\ \ref{sep-ener-fm}) indicating that the isovector 
component of the interaction has been treated more carefully in these
sets.

 The latter sets reasonably describe $S_{2n}(Z,N)$ for $144 \leq N \leq 148$ 
and $154 \leq N \leq 159$, but underestimate $S_{2n}(Z,N)$ for $N=150,152$. 
The shoulder in experimental $S_{2n}(Z,N)$ values at $N=152$ reveals a deformed 
shell gap \cite{CAFE.77}, which is better seen in the plot of $\delta_{2n}(Z,N)$  
(Fig.\ \ref{delta-fm}a). The size of this gap depends sensitively on proton number. 
As seen in Fig.\ \ref{delta-fm}, there are discrepancies with experiment: 
NL3 and NL-RA1 (NL1 and NL-Z) produce a gap at $N=148$ ($N=148,150$) instead of 
at $N=152$ and NLSH does not show a clear gap (see also Figs.\  \ref{neutron-no254-sp} 
and \ref{fm-sp-ener-neu}). The analysis of the neutron quasi-particle spectra in 
$^{249,251}$Cf (see Sect.\ \ref{Cf249}) also indicates that the calculated shell 
gaps do not correspond to the experimental ones.

 Earlier calculations predicted the presence of a deformed neutron shell gap at  
$N=162$ in superheavy nuclei (see, for example, Refs.\ \cite{CDHMN.96,BRBRMG.98} 
and references quoted therein). The presence of this gap in nuclei with $Z\approx 108$ 
was confirmed by recent experimental information \cite{HM.00,Og01}. The appearance 
of this gap and its size strongly depend on the parametrization of the specific theory 
and on the proton number. For example, in the Skyrme Hartree-Fock calculations 
\cite{BRBRMG.98}, this gap is pronounced in the SkI3 parametrization, where it is seen 
over the proton range of $Z=98-116$, but is absent in the SkP parametrization. This 
gap is clearly seen in RMF calculations with the NL3 and PL-40 parametrizations 
\cite{BRBRMG.98} but only at proton numbers around $Z=106$. The present CRHB+LN calculations 
in the Fm chain indicate a gap at $N=162$ for NL3 and at $N=160$ for NL1, NL-Z and 
NL-RA1 (see Fig.\ \ref{delta-fm}). However, the small value of $\delta_{2n}(Z,N)\approx 
0.8$ MeV suggests a small gap.

%%%%%%%%%%%%%%%%%%%%%%%%%%%%%%%%%%%%%%%%%%%%%%%%%%%%
\subsection{Shell structure along the $N=152$ line}
\label{sstrucn152}
%%%%%%%%%%%%%%%%%%%%%%%%%%%%%%%%%%%%%%%%%%%%%%%%%%%%

 To judge the reliability of predictions of superheavy nuclei, 
it is critical to see how different RMF parametrizations are able to 
describe the experimental shell gaps as a function of proton 
number. The calculations for the $N=152$ isotones are compared with 
experimental data in Figs.\ \ref{sep-ener-n152} and \ref{delta-n152}. 
As for the two-neutron separation energies, one can see that the two-proton 
separation energies $S_{2p}(Z,N)$ are best described by NL3, NLSH, 
NL-RA1. In contrast, the $S_{2p}(Z,N)$ values are overestimated 
by NL1 and NL-Z, which were obtained by fit to stable nuclei. 
The $S_{2p}(Z,N)$ plots do 
not show clearly where the proton deformed gaps are located, which
becomes visible in the $\delta_{2p}(Z,N)$ plots. The experimental data 
show a shell gap at $Z=100$. Only NLSH describes the position of this 
gap and the $\delta_{2p}(Z,N)$ values agree very well. However, the 
analysis of the quasi-particle spectra in Sect.\ \ref{qp-other-forces}
reveals that this gap lies between the wrong bunches of single-particle states. 
Calculations with NLSH also indicate  a gap at $Z=108$, 
which has not been observed so far. NL-RA1 does not show any deformed 
gap for $92 \leq Z \leq 108$ (see Fig.\ \ref{delta-n152}). NL3, NL1 and 
NL-Z give a shell gap at $Z=104$, in contradiction with experiment. The 
analysis of the proton quasi-particle spectra in $^{249}$Bk (see 
Sect.\ \ref{Bk249}) leads to the same conclusion.

  Many effective interactions not specifically fitted to the actinide
region encounter similar problems. For example, most of the Skyrme forces 
fail to reproduce the deformed $Z=100$ shell gap in the 
$N=152$ isotones (see Fig. 5 in Ref.\ \cite{BRBRMG.98}). SkI4 is the only 
force which shows this gap. The SkI3, SkI1 and Sly6 forces show a $Z=104$ 
shell gap, while the SkM$^{*}$ and SkP forces do not show any gap at 
$Z=100-104$. 
 
%%%%%%%%%%%%%%%%%%%%%%%%%%%%%%%%%%%%%%%%%%%%%%%%%%%%%%%%%%%%%
\subsection{Pairing along the $Z=100$ and $N=152$ lines}
\label{pairing-lines}
%%%%%%%%%%%%%%%%%%%%%%%%%%%%%%%%%%%%%%%%%%%%%%%%%%%%%%%%%%%%%

  Figure \ref{pairing-chains} shows the pairing  energies $E_{pairing}$ 
(see Eq.\ (\ref{Epair})) obtained with NL1 and NL3 as a function of neutron 
number along the $Z=100$ line and as a function of proton number along the 
$N=152$ line. The general trend as a function of nucleon number is the same
for both sets. 

  Let us first discuss the $Z=100$ line. In both RMF parametrizations, neutron 
pairing correlations are weakest at $N\approx 148$, reflecting the presence of 
a shell gap at $N=148$ in the NL3 parametrization and somewhat smaller gaps at 
$N=148$ and $N=150$ in the NL1 parametrization (see Fig.\ \ref{neutron-no254-sp}).  
Going away from these shell gaps, the neutron pairing energies increase
in absolute value, reflecting the increasing level density (Fig.\ 
\ref{fm-sp-ener-neu}). The neutron pairing is  weakened at $N\approx 160$ 
due to the presence of smaller shell gaps at $N=160$ (NL1, see Sect.\ 
\ref{sstruc100}) and at $N=162$ (NL3, see Fig.\ \ref{fm-sp-ener-neu}).
The weakening of the  neutron pairing at $N\approx 148$ and $N\approx 160$
is more pronounced in the NL3 parametrization as compared with NL1, reflecting 
larger shell gaps (see Fig.\ \ref{neutron-no254-sp}).

 The proton pairing shows the same trend as the neutron pairing, but with much 
smoother changes in neutron number. In both parametrizations, proton 
pairing is smaller and stays relatively constant at $N=138-150$, reflecting 
low and nearly constant level density below the Fermi level (see Figs.\ 
\ref{proton-no254-sp} and \ref{fm-sp-ener-prot}). For $N\geq 150$, the 
deformation modifications induced by changes in the neutron number 
increase the proton level density  near the Fermi level (see Fig.\ 
\ref{fm-sp-ener-prot}), enhancing the proton pairing.

  The pairing energies on the $N=152$ line exhibit the same features as 
discussed above.  Both in the neutron and proton subsystems they reflect 
the presence of the $Z=104$ and $Z=96$ shell gaps (see Figs.\ \ref{proton-no254-sp} 
and \ref{pairing-chains}).
 
%%%%%%%%%%%%%%%%%%%%%%%%%%%%%%%%%%%%%%%%%%%%%%%%%%%%%%%%%%%%%%%%%%%%%
\subsection{Summary for Section \ref{shell-structure}}
%%%%%%%%%%%%%%%%%%%%%%%%%%%%%%%%%%%%%%%%%%%%%%%%%%%%%%%%%%%%%%%%%%%%%

The experimental $\delta_{2n}(Z,N)$ quantity shows a distinct deformed shell gap at 
$N=152$, which is quite pronounced for No and Fm nuclei and 
less so for Cm and Cf (see Fig.\ \ref{delta2n-exp}). For the Fm isotopes, the 
CRHB+LN calculations predict a deformed shell gap(s) at 
$N=148$ (NL3 and NL-RA1) or at $N=148,150$ (NL1 and NL-Z); see Fig.\ 
\ref{delta-fm}. The NLSH parameterization does not give a clear gap.
The experimental data for the $N=152$ isotones show a shell gap at 
$Z=100$;  NL1, NL-Z and NL3 give a gap at $Z=104$ and NL-RA1
no gap at all. Only NLSH gives a gap at $Z=100$; however, it lies 
between the wrong bunches of single-particle states (see Sect.\ 
\ref{qp-energies}). This demonstrates
the fact that the usual analysis of shell structure, 
in terms of only $S_{2n}(Z,N)$ and/or $\delta_{2n}(Z,N)$, maybe insufficient
to judge the quality of the parametrization.

%%%%%%%%%%%%%%%%%%%%%%%%%%%%%%%%%%%%%%%%%%%%%%%
\section{Quasi-particle states}
\label{qp-energies}
%%%%%%%%%%%%%%%%%%%%%%%%%%%%%%%%%%%%%%%%%%%%%%%

 The investigation of the single-particle states in the $A\sim 250$ deformed 
mass region can shed additional light on the reliability of the RMF predictions 
of the energies of spherical subshells  responsible for 'magic' numbers in 
superheavy nuclei. This is because several deformed single-particle states 
experimentally observed in odd nuclei of this mass region (see Table 
\ref{qp-experiment}) originate from these subshells.  Considerable deviations 
between experiment and theory for a specific deformed state will indicate that 
the position of the spherical subshell from which this state originates is 
poorly described.

 In the past, the RMF studies of single-particle spectra have been 
mostly performed in spherical or near-spherical nuclei (see, for 
example, Ref.\ \cite{RBRMG.98} and references quoted therein),  
where a number of restrictions, such as the neglect of the currents 
or of the breaking of the Kramer's degeneracy, have been imposed in 
order to simplify the task. In addition, the BCS approximation
was used. Moreover, a direct comparison between experimental and 
theoretical single-particle states in spherical nuclei should 
include the particle-vibration coupling, which can modify the 
single-particle energies considerably \cite{MBBD.85}. The modification of the 
quasiparticle states by particle-vibration coupling is weaker in deformed 
nuclei \cite{Sol-book,Sol-book2,Malov-private}.

 Not much is known about the accuracy of the description of the 
quasiparticle states in deformed nuclei within the framework of 
the RMF theory. In most cases the analysis of odd nuclei was 
based on the  single-particle spectra calculated in 
neighboring even-even nuclei. To our best knowledge, a direct 
comparison between experimental and theoretical quasi-particle 
spectra has not been published.

 The most detailed attempt to analyze the single-particle properties 
at finite deformation and their connections with those at spherical 
shape has been done within the cranked RMF theory in Ref.\ \cite{ALR.98} 
for superdeformed (SD) bands in the $A\sim 150$ mass region. Although 
a direct comparison between experimental and calculated single-particle 
energies was not possible, some general conclusions were 
drawn  based on a systematic analysis of experimental data and the 
expected response of specific single-particle states to a change of 
deformation. It was found that the RMF theory provides a reasonable 
description of the single-particle states in the vicinity of the SD shell 
gaps. However, some deviations between experiment and theory were
detected, which could reach $\approx 1$ MeV for some states. For 
example, the relative positions of the $\nu[651]1/2$ and $\nu[642]5/2$ 
states from the $\nu 2g_{9/2}$ and $\nu 1i_{13/2}$ spherical subshells 
are not reproduced. This problem exists in the NL1, NL3 and NLSH 
parametrizations of the RMF Lagrangian.

%%%%%%%%%%%%%%%%%%%%%%%%%%%%%%%%%%%%%%%%%%%%%%%%%%
\subsection{Computational details}
%%%%%%%%%%%%%%%%%%%%%%%%%%%%%%%%%%%%%%%%%%%%%%%%%

 In the present manuscript, we address for the first time the question 
of a fully self-consistent description of quasiparticle states in the 
framework of the RMF theory. A proper description of odd nuclei implies 
the loss of the time-reversal symmetry of the mean-field, which is broken by 
the unpaired nucleon.  The BCS approximation has to be replaced by the 
Hartree-(Fock-)Bogoliubov method, with time-odd mean fields taken into account. 
The breaking of time-reversal symmetry leads to the loss of the double 
degeneracy (Kramer's degeneracy) of the quasiparticle states. This requires 
the use of the signature or simplex basis in numerical calculations, thus 
doubling the computing task. Furthermore, the breaking of the time-reversal 
symmetry leads to nucleonic currents, which cause the {\it nuclear magnetism}
discussed in Sect.\ \ref{CRHB-eq}.

 The CRHB(+LN) theory takes all these effects into account. Thus, the 
effects of blocking due to odd particle are included in a fully 
self-consistent way. The CRHB computer code is set up in a signature 
basis and in three-dimensional Cartesian coordinates. The latter allows 
one to  study also the $\gamma$-deformation. In 
order to specify the detailed structure of blocked orbitals, the 
existing CRHB code \cite{CRHB} has been extended to describe 
odd and odd-odd nuclei. The blocking procedure is implemented according 
to Refs.\ \cite{RS.80,RBM.70,EMR.80}. The blocked orbital can be 
specified either by its dominant main oscillator quantum number $N$
or by the dominant $\Omega$ quantum number ($\Omega$ is the projection
of the total angular momentum on the symmetry axis) of the wave function, 
or by combination of both. In addition, it can be specified by 
the particle or hole nature of the blocked orbital. Note that $\Omega$ is 
not a conserved quantum number in the CRHB code. As a consequence, convergence 
problems, emerging from the interaction of the blocked orbital with others, 
appear somewhat more frequently than in a computer code restricted to axial 
symmetry. Convergence problems appear more frequently when approximate particle 
number projection by means of the Lipkin-Nogami method is imposed, which
is most likely due to additional non-linearities.

 As illustrated in Fig.\ \ref{qp-cf249}, the quasi-particle spectra
calculated within the CRHB+LN (with LN) framework and using the scaled D1S Gogny 
force given in Table \ref{Table-scaling} are very similar to those 
obtained by means of the CRHB (without LN) with the original D1S force. The difference 
in the energies of the quasi-particle states is typically less than 150 keV 
and the level ordering is the same. Thus in order to avoid the convergence 
problems in the calculations with LN, all other calculations of quasi-particle 
states were performed in the CRHB framework with the original D1S force. 

%%%%%%%%%%%%%%%%%%%%%%%%%%%%%%%%%%%%%%%%%%%%%%%%%%%%%%%%%%%%%%%%%%%%%%%%%%
\subsection{Particle-vibration coupling and other effects}
\label{PVP}
%%%%%%%%%%%%%%%%%%%%%%%%%%%%%%%%%%%%%%%%%%%%%%%%%%%%%%%%%%%%%%%%%%%%%%%%%%

  Figures \ref{qp-cf249}, \ref{qp-cf251} and \ref{qp-bk249} show that the 
calculated quasiparticle spectra are less dense than in experiment.
The average level density of the single-particle states is related to 
the effective mass (Lorentz mass in the notation of Ref.\ \cite{JM.89} 
for the case of RMF theory) of the nucleons at the Fermi surface 
$m^*(k_F)/m$.  The RMF theory gives a low effective mass $m^*(k_F)/m \approx 0.66$ 
 \cite{BRRMG.99}. The experimental density of the quasiparticle levels 
corresponds to an effective mass $m^*(k_F)/m$ close to one. This 
discrepancy appears also for non-relativistic mean-field models 
\cite{R.96}. It has been demonstrated for spherical nuclei that the 
particle-vibration coupling brings the average level density in 
closer agreement with experiment \cite{MBBD.85}, which means $m^*(k_F)/m$ 
closer to one. In a similar way, the particle-vibration coupling leads 
to a compression of the quasi-particle spectra in deformed nuclei 
\cite{Malov-private}. The surface vibrations are less collective in 
deformed nuclei than in spherical ones because they are more 
fragmented \cite{Sol-book,BM}. As a consequence, the corrections to the 
energies of quasiparticle states in odd nuclei due to particle-vibration 
coupling are less state-dependent in deformed nuclei. Hence the comparison 
between experimental and mean field single-particle states is less ambiguous 
in deformed nuclei as compared with spherical ones \cite{MBBD.85,BM}, at 
least at low excitation energies, where vibrational admixtures in the wave 
functions are small. Calculations within the quasiparticle-phonon model 
\cite{GIMS.71,IKMS.73} indicate that in the $A\sim 250$ mass region the 
lowest states have mainly quasi-particle nature and the corrections to 
their energies due to  particle-vibration coupling are typically at the 
level of 150 keV or less. The states above $\approx 700$ keV contain  
very large vibrational admixtures \cite{Malov-private} and thus experimental 
states above this energy should not be compared with the pure quasiparticle 
states obtained in the CRHB calculations.

Since particle-vibration coupling is not included, it is important to 
estimate how large is  the discrepancy between calculated and experimental 
quasiparticle energies due to the low effective mass $m^*(k_F)/m$ of the 
RMF theory. Assuming for an estimate that the effective mass just 
stretches the energy scale, the difference between the energies of 
quasiparticle states obtained in the calculations with $\frac{m^*(k_F)}{m}$ 
and $\frac{m^*(k_F)}{m}=1$ is
\begin{eqnarray}
\Delta E_{qp} = E_{qp}(\frac{m^*(k_F)}{m})(1-\frac{m^*(k_F)}{m}),
\end{eqnarray}
which remains below $\sim 200$ keV as long as  the calculated state is 
located in the energy window of 700 keV with respect to the Fermi surface, 
whereas it grows for higher excitation energies.

We are mostly interested in how well the positions of spherical subshells 
are described in the RMF calculations. One may reduce the error by comparing 
the experimental and calculated averages of two (or more) deformed 
single-particle states emerging from the same spherical subshell. The 
average of deformed states has the same energy when extrapolated to
spherical shape as each of these states. The advantage  is that the average 
of the states has smaller excitation energy than at least one of these 
states. As a result, the energy of the spherical subshell can be estimated 
more precisely (within 200 keV  if $|(E_{qp,1}+ E_{qp,2})/2| \leq 700$ keV). 
Such an approach is especially useful when one of the states has particle 
($E_{qp}^p > 0$) and the other  hole ($E_{qp}^h < 0$) nature, since their 
average energy can be well within the 700 keV energy window even if the 
excitation energy of each is far outside this window.

An additional source of uncertainty is the Coriolis interaction between 
the quasi-particle states, which  is neglected. It is relatively modest, 
affecting the energies of quasi-particle states by at most $100-200$ keV 
\cite{NR.book,S.75}. As a whole, the uncertainty of our estimate for
the spherical subshell energies in Sects.\ \ref{Bk249} and \ref{Cf249} is 
around 300 keV.

%%%%%%%%%%%%%%%%%%%%%%%%%%%%%%%%%%%%%%%%
\subsection{General observations}
%%%%%%%%%%%%%%%%%%%%%%%%%%%%%%%%%%%%%%%%

\begin{figure}[t]
\epsfxsize 16.0cm
\epsfbox{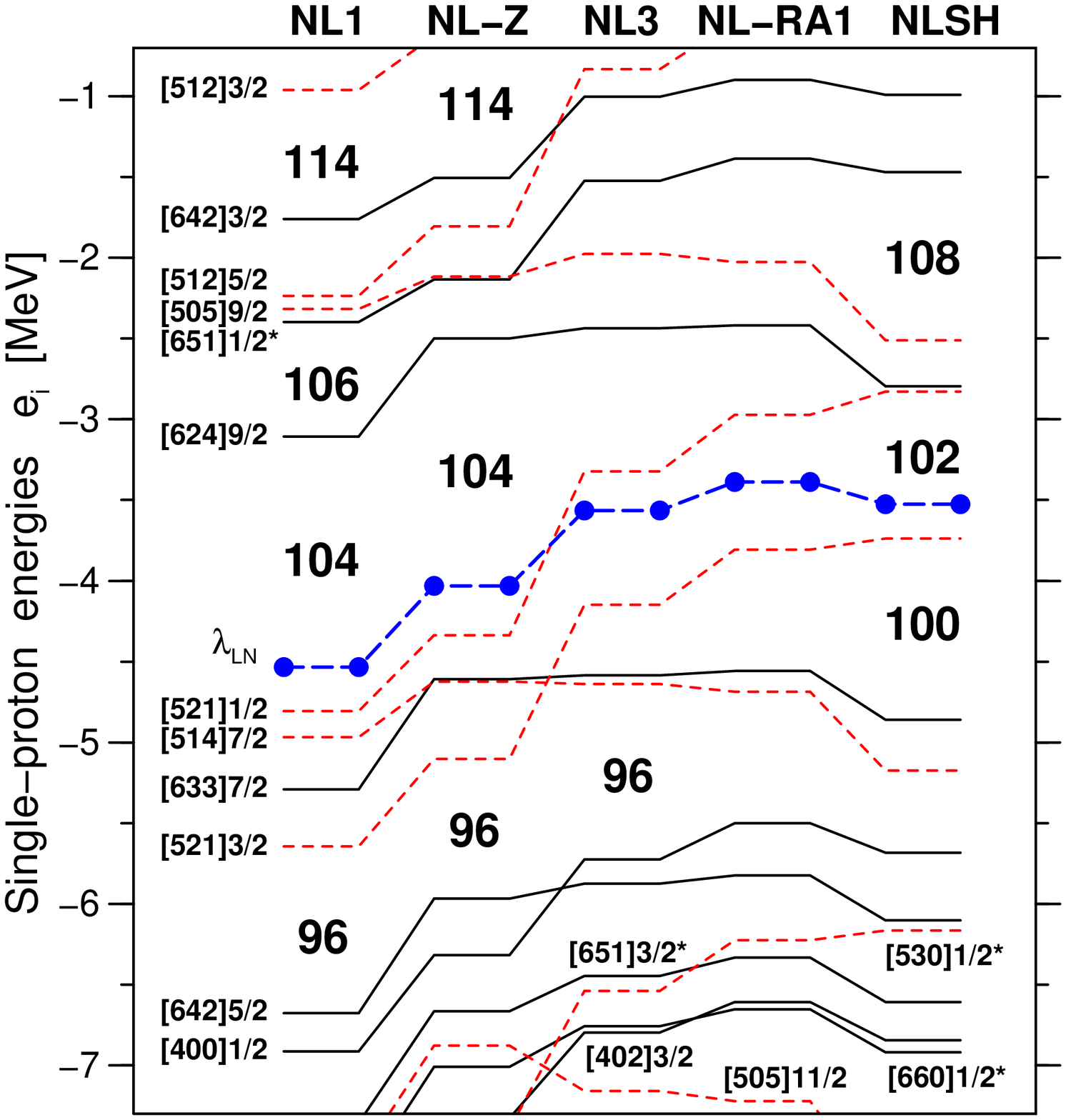}
\vspace{0.5cm}
\newpage
\caption{Single-proton energies in $^{254}$No obtained at the 
equilibrium deformation in the CRHB+LN calculations with different
RMF parametrizations. Solid and dashed lines are used for positive 
and negative parity orbitals, respectively. The $\lambda_{LN}$ 
values are shown by the long-dashed line with solid 
circles. The single-particle orbitals are labeled by means of the 
asymptotic quantum numbers $[N n_z \Lambda]\Omega$ (Nilsson quantum 
numbers) of the dominant component of the wave function. The asterisk
(*) at the Nilsson label indicates that the wave function is fragmented
and the weight of dominant component is below 50\%. In this case,
the Nilsson label does not characterize the wave function but is an indicator
of the position of the orbital within the $[N]\Omega$ group (see
Ref.\ \protect\cite{NR.book}), where $N$ is the main oscillator quantum 
number of the dominant $N$-shell and $\Omega$ the projection of total 
angular momentum onto the symmetry axis. $N$ can be considered as a good 
approximate quantum number in almost all cases, since typically the 
weight of a specific $N$-shell in the structure of the wave function
is at least 85\% or larger. However, the $\pi [402]3/2$ and $\pi [651]3/2^*$ 
orbitals are strongly mixed by $\Delta N= \pm 2$ interaction. 
}
\label{proton-no254-sp}
\end{figure}

\begin{figure}[t]
\epsfxsize 16.0cm
\epsfbox{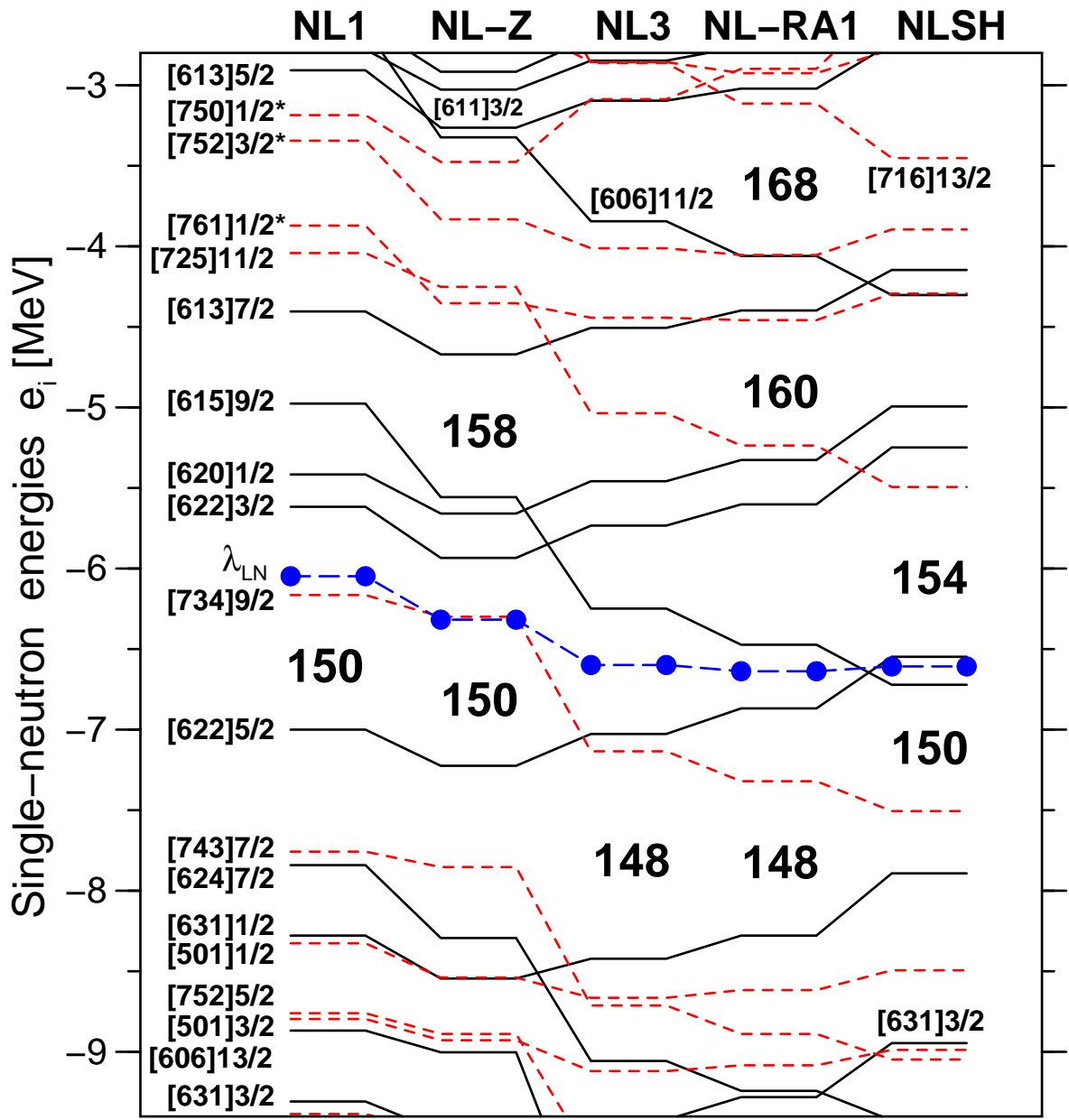}
\vspace{0.5cm}
\caption{The same as Fig.\ \protect\ref{proton-no254-sp} but
for the single-neutron energies.}
\label{neutron-no254-sp}
\end{figure}

\begin{figure}
\epsfxsize 15.0cm
\epsfbox{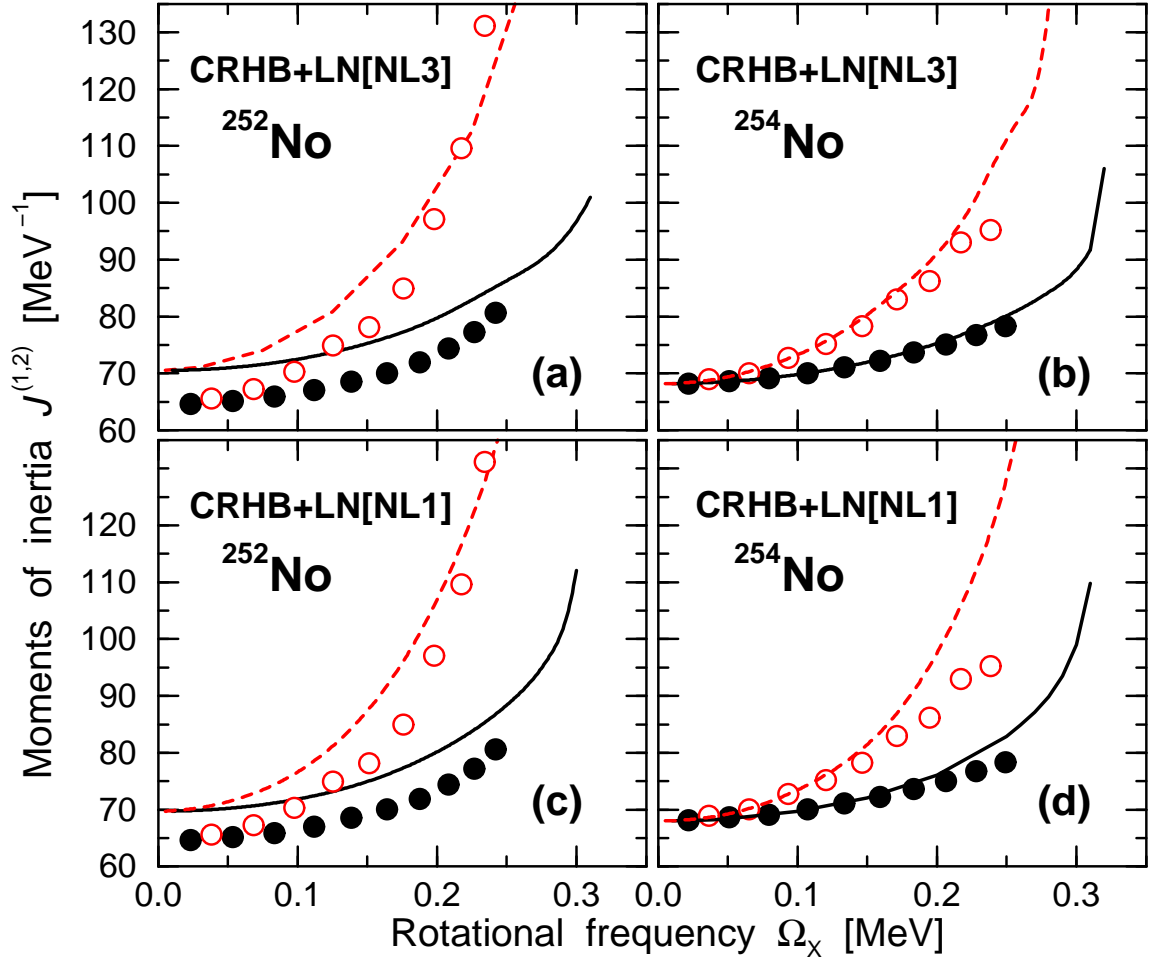}
\vspace{0.5cm}
\caption{Experimental and calculated dynamic and kinematic moments 
of inertia of the normal-deformed bands in $^{252,254}$No. The
experimental $J^{(1)}$ and $J^{(2)}$ values (from Refs.\
\protect\cite{No254-exp1,No254-exp2,No254-exp3,Julin2001})
are shown by solid and open circles, respectively. Solid and 
dashed lines are used for the calculated $J^{(1)}$ and $J^{(2)}$ values, 
respectively. In the calculations, the D1S Gogny force is attenuated 
by the scaling factor $f$ given in Table \protect\ref{Table-scaling}.
The results of the calculations with NL3 and NL1 parametrizations are 
displayed in the upper and the bottom panels, respectively. They are 
shown only up to a rotational frequency where a sharp band crossing 
takes place. } 
\label{no5254-nl1nl3}
\end{figure}

\begin{figure}
\epsfxsize 13.0cm
\epsfbox{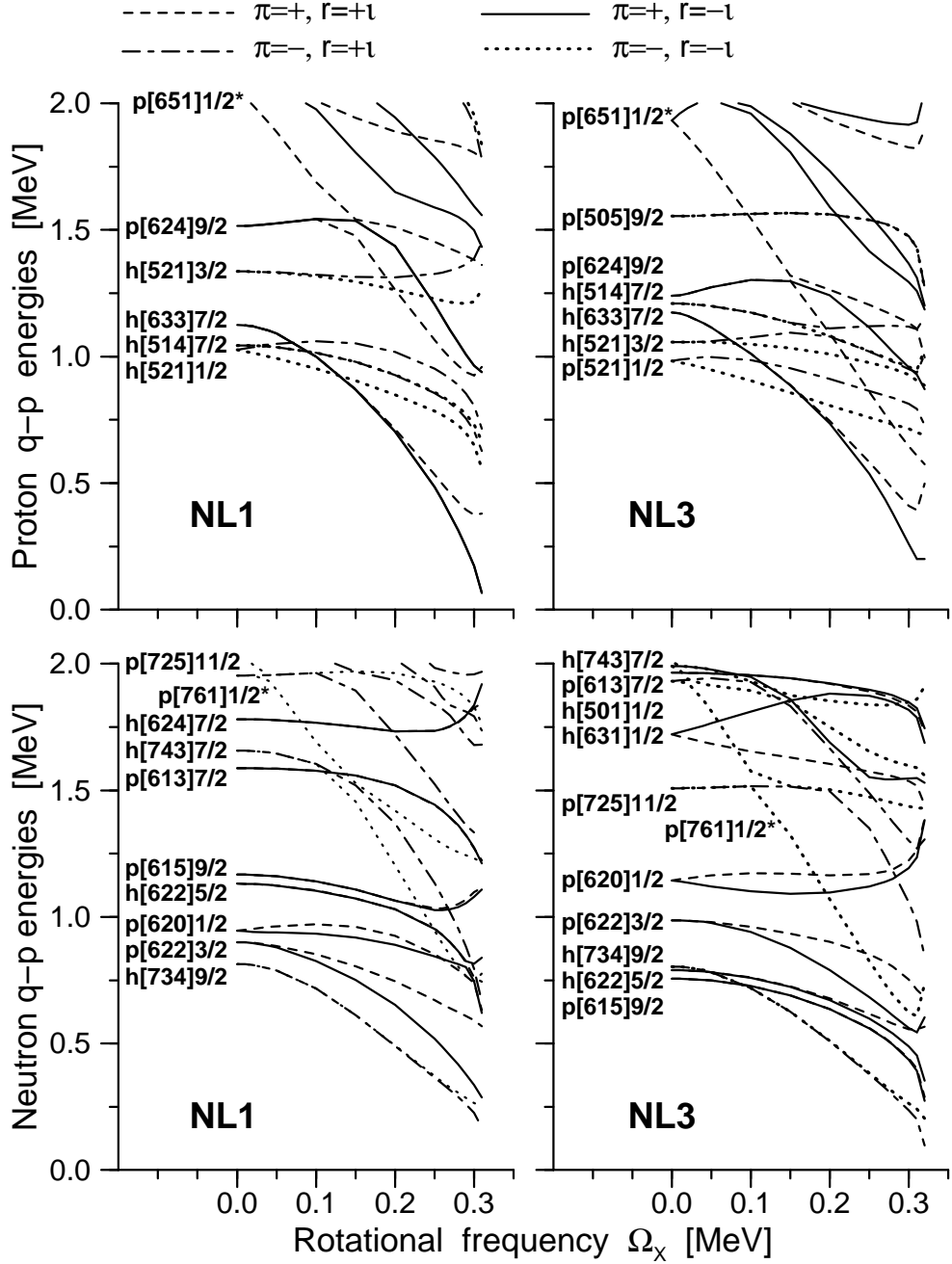}
\vspace{0.5cm}
\caption{Proton (top panels) and neutron (bottom panels)
quasiparticle energies corresponding to the lowest configuration
in $^{254}$No. The CRHB+LN calculations have been performed 
with the NL1 (left panels) and NL3 (right panels) parametrizations. 
The letters 'p' and 'h' before the Nilsson labels are used to 
indicate whether a given Routhian is of particle or hole type.} 
\label{qp-routh-no254}
\end{figure}

\begin{figure}[t]
\epsfysize 13.5cm
\epsfbox{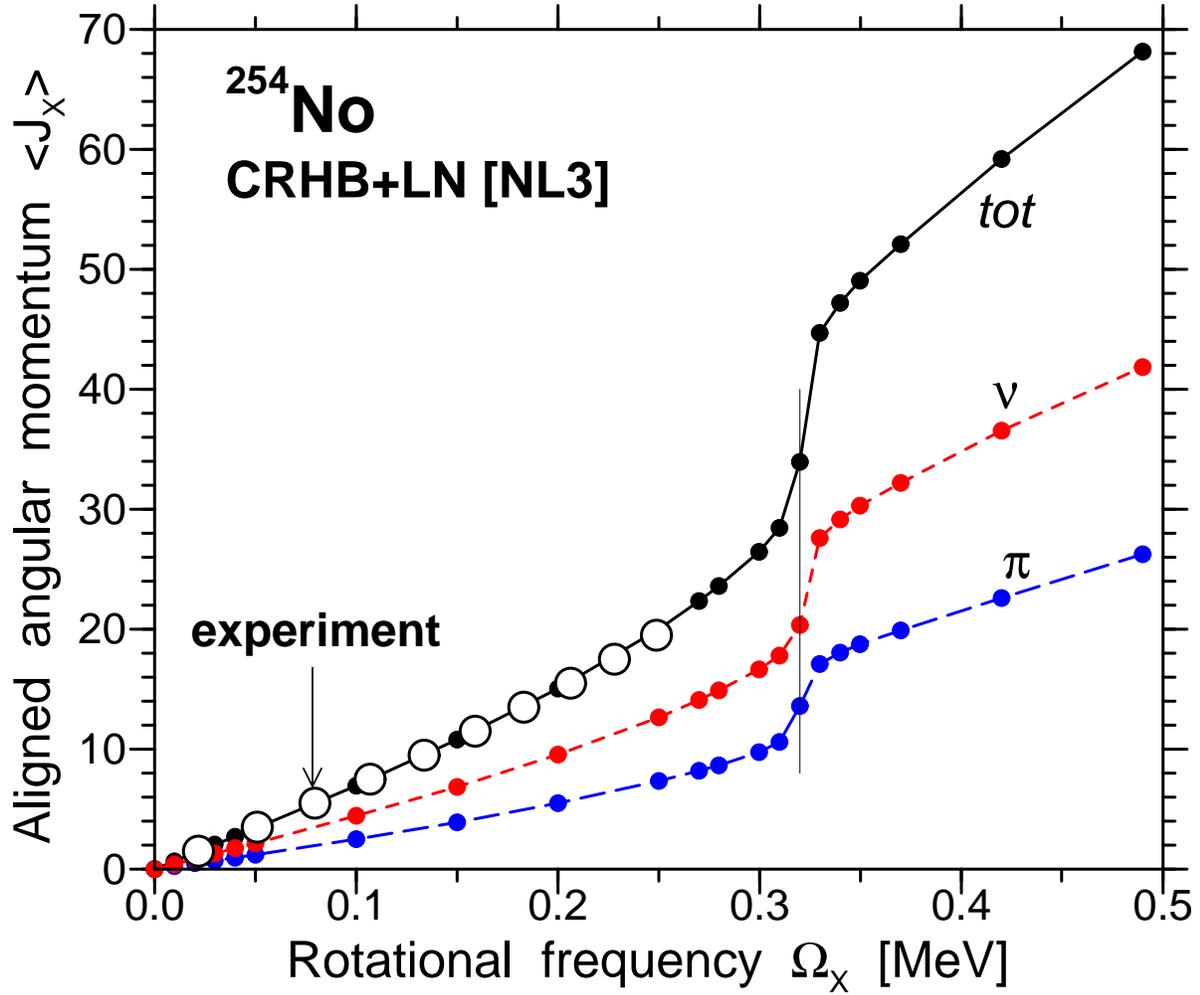}
\vspace{0.5cm}
\caption{Aligned angular momenta (in units $\hbar$) versus 
rotational frequency $\Omega_x$ in $^{254}$No. Proton, neutron 
and total $\left < J_x \right >$ are shown. Small solid circles 
on lines show the frequencies at which the CRHB+LN calculations 
have been performed. The experimental quantities, defined as 
$<J_x>=\protect\sqrt{I(I+1)} \approx I+1/2$ (see Ref.\ 
\protect\cite{CF.83}), are shown by open large circles. The crossing 
frequency $\Omega_x \approx 0.32$ MeV is indicated by a thin 
vertical line.}
\label{no254-contr-toj1j2}
\end{figure}

\begin{figure}
\epsfxsize 15.0cm
\epsfbox{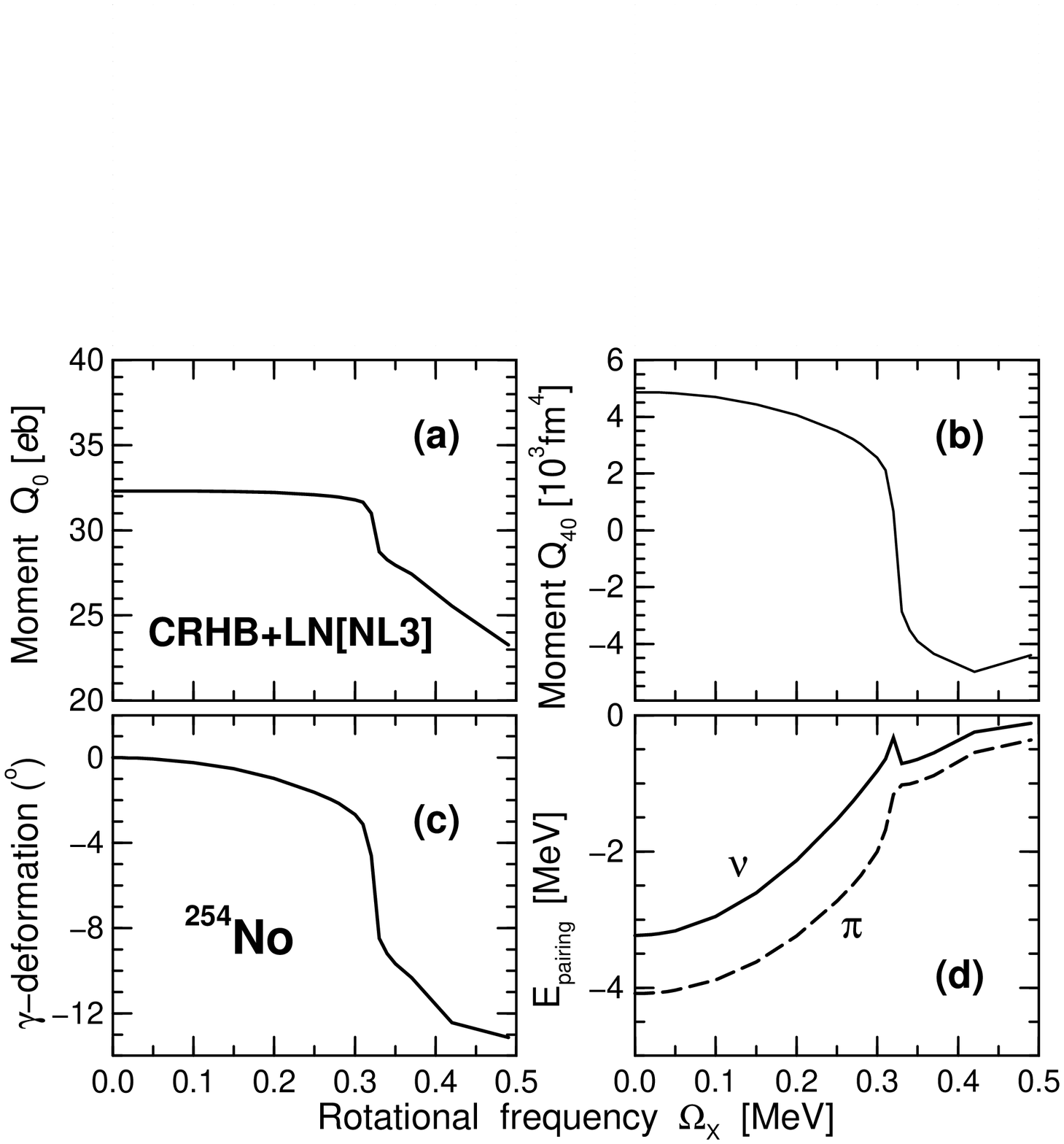}  
\vspace{0.5cm}
\caption{The results of CRHB+LN calculations with the NL3 
parametrization for $^{254}$No. Mass quadrupole moments $Q_0$, 
mass hexadecapole moments $Q_{40}$, and $\gamma$-deformation as 
a function of rotational frequency $\Omega_x$ are given on 
panels (a),(b) and (c), respectively. Proton and neutron pairing 
energies are shown in panel (d).}
\label{no254-nl3-full}
\end{figure}

\begin{figure}[t]
\epsfysize 14.0cm
\epsfbox{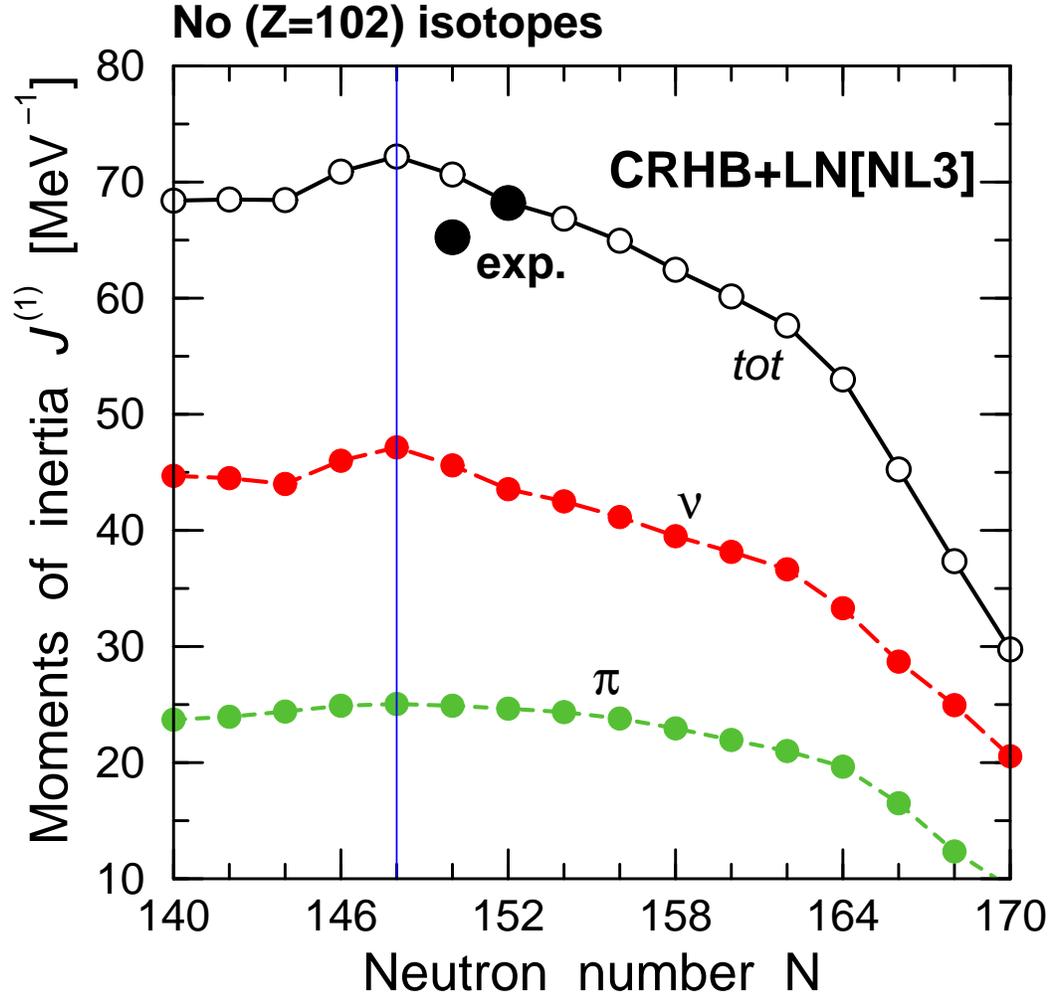}
\vspace{0.5cm}
\caption{Kinematic moments of inertia $J^{(1)}$ (total, neutron 
and proton contributions) in the No isotopes as a function of the 
neutron number $N$. The CRHB+LN calculations are carried out with 
the NL3 parametrization at $\Omega_x=0.02$ MeV. The vertical line 
shows the position of the $N=148$ shell gap.}
\label{no-contr-j1}
\end{figure}

\begin{figure}[t]
\epsfxsize 16.0cm
\epsfbox{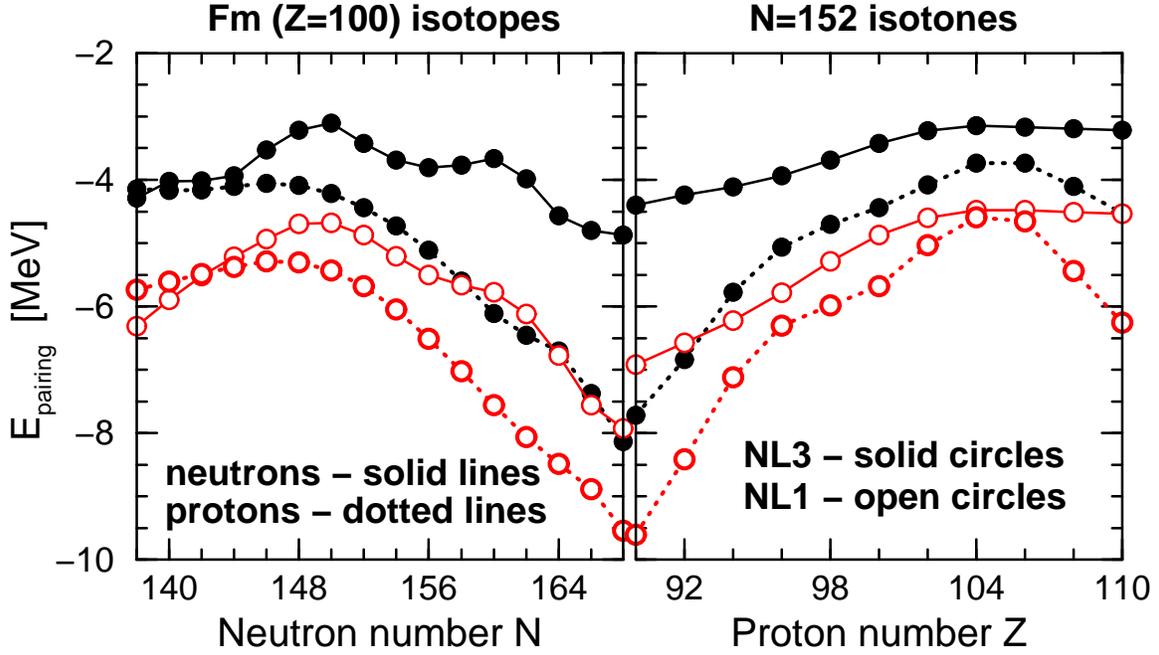}
\vspace{0.5cm}
\caption{Proton and neutron pairing energies 
$E_{pairing}=-\frac{1}{2}Tr (\Delta\kappa)$ obtained in the CRHB+LN 
calculations with the NL1 and NL3 parametrizations 
for the Fm ($Z=100$) isotopes and the $N=152$ isotones.} 
\label{pairing-chains}
\end{figure}

\begin{figure}[t]
\epsfxsize 16.0cm
\epsfbox{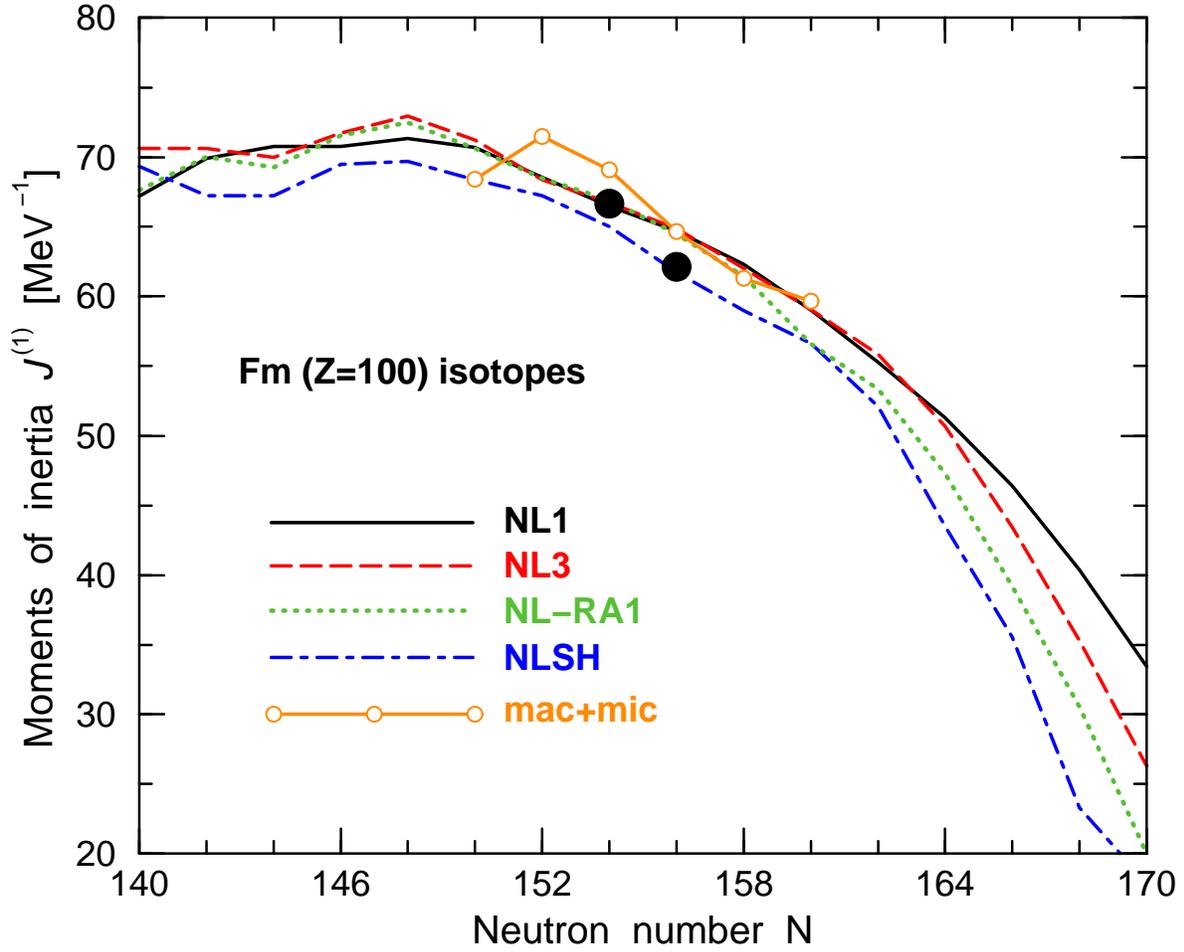}
\vspace{0.5cm}
\caption{Kinematic moments of inertia $J^{(1)}$ in the Fm isotopes 
as a function of the neutron number $N$. The results of the CRHB+LN 
calculations  at $\Omega_x=0.02$ MeV with different RMF parametrizations 
are  presented by lines.  The results with the NL-Z parametrization
follow those with NL1, but are systematically lower by $\approx 1$ MeV$^{-1}$.
The results of the calculations of Ref.\ \protect\cite{SMP.01} within 
the macroscopic+microscopic (mac+mic) method  are shown by solid lines 
with open circles. Experimental data are shown by solid circles.
}
\label{j1-fm}
\end{figure}

\begin{figure}[t]
\epsfxsize 16.0cm
\epsfbox{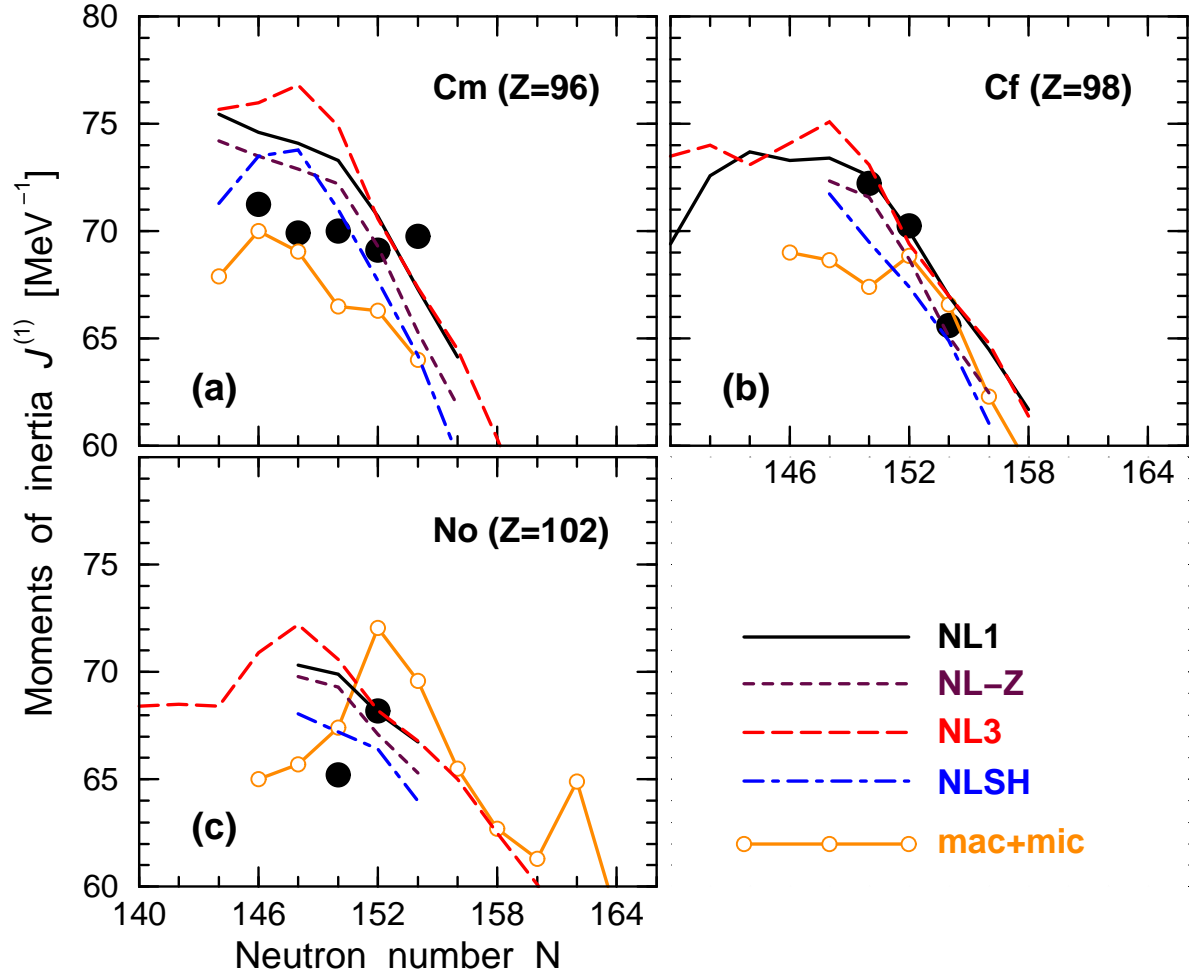}
\vspace{0.5cm}
\caption{The same as in Fig.\ \protect\ref{j1-fm}, but for
Cm, Cf and No isotopes. The results with NL-RA1 almost 
coincide with those for NL3 that are displayed.}
\label{j1-cf-cm-no}
\end{figure}

\begin{figure}[t]
\epsfxsize 16.0cm
\epsfbox{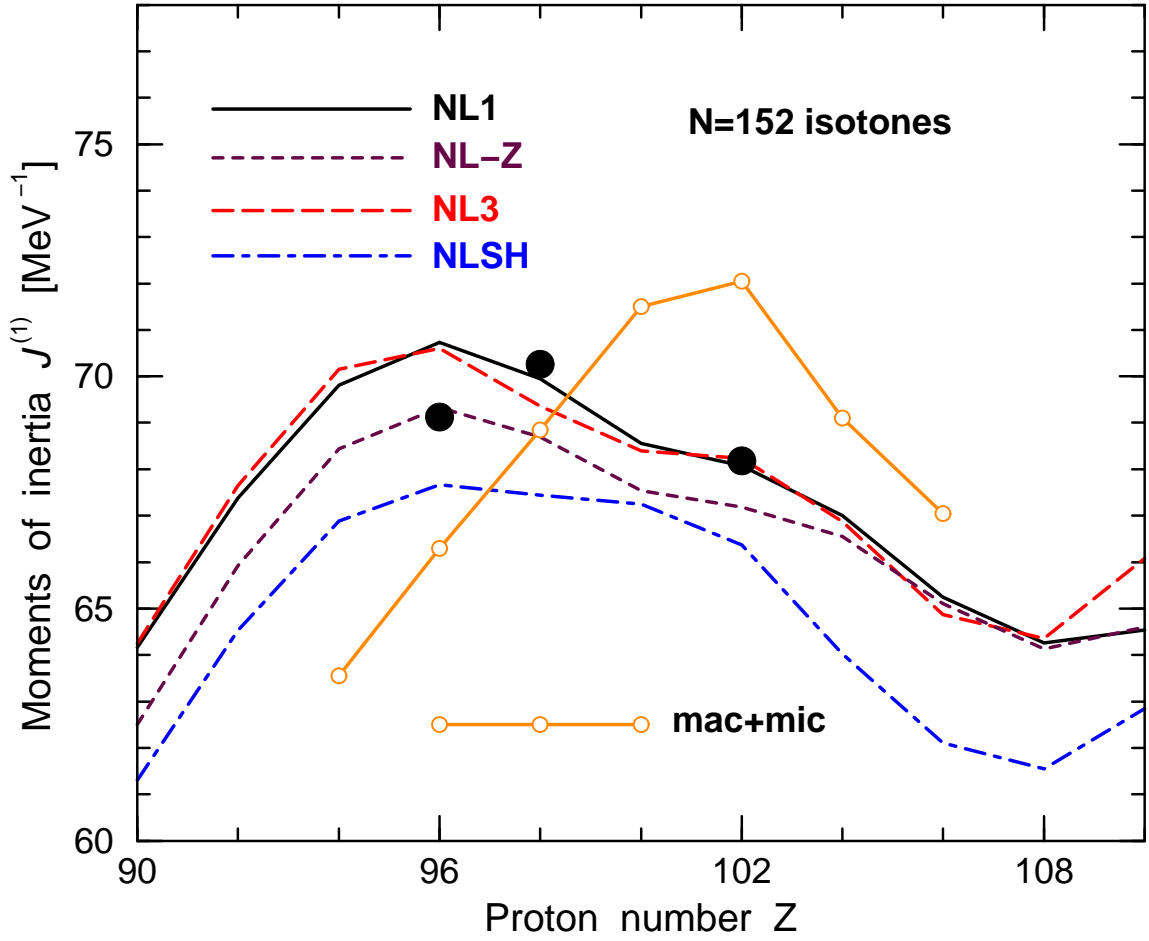}
\vspace{0.5cm}
\caption{The same as Fig.\ \protect\ref{j1-fm}, but for the 
$N=152$ isotones as a function of proton number $Z$. The results 
with NL-RA1 are very close to those with NL3.}
\label{j1-n152}
\end{figure}

\begin{figure}[t]
\epsfxsize 12.0cm
\epsfbox{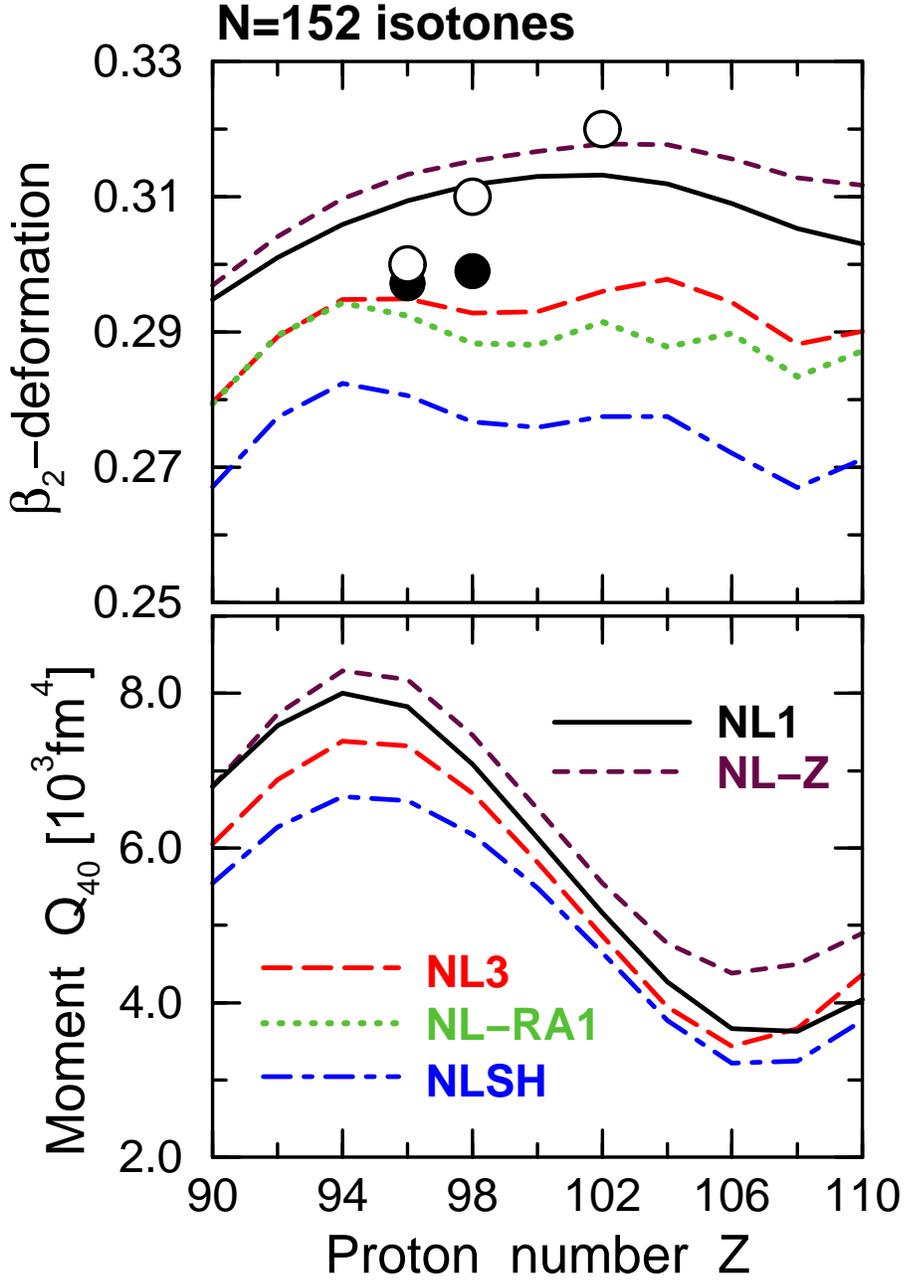}
\vspace{0.5cm}
\caption{The calculated (lines) and experimental (circles) 
deformation parameters $\beta_2$ (top panel) and calculated mass 
hexadecapole moments $Q_{40}$ (bottom panel) in the chain of $N=152$ 
isotones.  The experimental values of $\beta_2$ obtained in the 
direct measurements \protect\cite{Raman.87} are shown by solid 
circles, while those deduced from the $2^+ \rightarrow 0^+$ transition 
energies, with the prescription of Ref.\ \protect\cite{No252}, are 
given by open circles. Since the results with NL-RA1 for 
$Q_{40}$ coincide with those with NL3, they are not shown in the 
bottom panel.} 
\label{def-n152-allforces}
\end{figure}

 Figures \ref{proton-no254-sp} and \ref{neutron-no254-sp}, 
where the single-particle states in $^{254}$No are shown
for different parametrizations, reveal important trends. The 
energies of some single-particle states depend strongly on the 
parametrization. For  example, for $\pi [521]3/2$ and 
$\pi [521]1/2$ \footnote{ These states are of special interest since 
they originate from the spin-orbit partner spherical subshells $\pi 2f_{7/2}$ 
and $\pi 2f_{5/2}$, which define the size and the position of the magic 
spherical proton shell gap in superheavy nuclei \cite{BRRMG.99}. Their 
splitting, defined primarily by the spin-orbit splitting and their response 
to deformation, almost does not depend on the RMF parametrization. In part, 
this is due to the fact that their interaction with other orbitals 
is rather weak (see, for example, the Nilsson diagram (Fig.\ 4) in 
Ref.\ \cite{CAFE.77}).}, the  single-particle energies calculated with 
NL1 and NLSH differ by $\approx 2$ MeV. The small differences in the 
self-consistent deformations cannot explain the differences in the 
single-particle energies. 

 Another observation is that the relative energies of the different $\Omega$ 
states, which emerge from the same spherical subshell, almost do not depend 
on the parametrization. For protons, these are, for example, the $\pi [642]5/2$, 
$\pi [633]7/2$, $\pi [624]9/2$ states from the $\pi 1i_{13/2}$ spherical subshell 
and the $\pi [514]7/2$, $\pi [505]9/2$ states from the $\pi 1h_{9/2}$ subshell. 
This is expected because the splitting of different $\Omega$ states from the 
same spherical subshell is a consequence of the deformation of the mean field,
which is not very sensitive to the parametrization (see Sect.\ \ref{Deformation}). 
On the other hand, the single-particle energies of the $\pi 2f_{7/2}$ 
and $\pi 2f_{5/2}$ spherical subshells (but not their splitting), as well 
as the deformed states emerging from them, change considerably when going from the NL1 
to the NLSH parametrization (see Fig.\ 20 in Ref.\ \cite{ALR.98} and Fig.\ 
\ref{proton-no254-sp} in the present manuscript for their deformed counterparts). 
This leads  to a $Z=104$ deformed shell gap in the NL1 and NL-Z parametrizations 
and a $Z=100$ gap in the NLSH parametrization. 

 The situation is analogous for the neutron states (see Fig.\ \ref{neutron-no254-sp}). 
For example, the relative energies of the $\nu [624]7/2$, $\nu [615]9/2$ 
and $\nu [606]11/2$ states, emerging from the $\nu 1i_{11/2}$ spherical subshell, 
almost do not depend on the RMF parametrization. The same is true for relative energies
of the states $\nu [743]7/2$, $\nu [734]9/2$ and $\nu [725]11/2$, emerging from the 
$\nu 1j_{15/2}$ subshell, and the states $\nu [761]1/2^*$ and $\nu [752]3/2^*$ 
from the $\nu 1j_{13/2}$ subshell. It is interesting that the relative energies 
of the states $\nu [631]1/2$ ($\nu 2g_{7/2}$), $\nu [622]5/2$ ($\nu 2g_{9/2}$), 
$\nu [622]3/2$ $(\nu 2 g_{7/2})$, $\nu [620]1/2$ ($\nu 3d_{5/2})$, $\nu [613]7/2$ 
($\nu 2g_{9/2}$) and $\nu [613]5/2$ $(\nu 2g_{7/2})$, originating from the different 
spherical subshells shown in parenthesis after the Nilsson labels, are almost 
independent on the RMF parametrization. This indicates that the relative energies 
of the $\nu 2g_{9/2}$, $\nu 2g_{7/2}$ and $\nu 3d_{5/2}$ spherical subshells only 
marginally depend on the RMF parametrization, which is clearly seen in the 
single-particle spectra of spherical nuclei (see Fig.\ 20 in Ref.\ \cite{ALR.98} 
and Fig.\ 1 in Ref.\ \cite{BRRMG.99} for spectra of $^{208}$Pb and Figs.\ 
4, 9 and 15 in Ref.\ \cite{BRRMG.99} for spectra of superheavy nuclei). 
On the other hand, the increase of the separation between the $\nu [761]1/2^*$ 
and $\nu [750]1/2^*$ states when going on from the NL1 to the NLSH parametrizations 
shows that the separation between the $\nu 1j_{13/2}$ and $\nu 2h_{11/2}$ spherical 
subshells increases.

  In a previous study in the $A\sim 150$ mass region
of superdeformation \cite{ALR.98}, we also concluded that the dependence 
of the single-particle energies of deformed states on the RMF 
parametrizations reflect their energy displacement at spherical shape. 
Since different RMF sets give similar spin-orbit splittings (see Fig.\ 2 
in Ref.\ \cite{BRRMG.99} and Fig.\ \ref{proton-no254-sp} for the splitting 
and position of the $\pi [521]3/2$ and $\pi[521]1/2$ states), the 
dominant modification is a shift of the position of the $l$-shells.

 There are similarities between the single-particle spectra (see 
Figs.\ \ref{proton-no254-sp} and \ref{neutron-no254-sp}) obtained 
with the NL1 and NL-Z parametrizations, on the one hand, and NL3 
and NL-RA1, on the other hand. Thus detailed study of quasi-particle 
spectra will be performed with only the NL1 and NL3 parametrizations. 
They can be considered as representative examples of the two groups of 
RMF parametrizations discussed in Sect.\ \ref{RMFforces}.

%%%%%%%%%%%%%%%%%%%%%%%%%%%%%%%%%%%%%%%%%%%%%%%%%%%%%%%%%%%%
\subsection{Odd-neutron nuclei $^{249,251}$Cf}
\label{Cf249}
%%%%%%%%%%%%%%%%%%%%%%%%%%%%%%%%%%%%%%%%%%%%%%%%%%%%%%%%%%%%

 The quasiparticle spectra of these two nuclei with neutron 
numbers $N=151,153$ are presented in Figs.\  \ref{qp-cf249} 
and \ref{qp-cf251}. The ground state configuration of $^{249}$Cf 
is correctly reproduced in both parametrizations. Only NL1 gives 
the correct ground state $\nu [620]1/2$  in $^{251}$Cf, 
whereas NL3 gives the $\nu [615]9/2$.

 The $\nu [622]5/2$ and $\nu [613]7/2$ (and $\nu [604]9/2$ 
in $^{251}$Cf) states emerge from the $\nu 2g_{9/2}$ spherical 
subshell. The $\nu [622]5/2$ energy is reproduced within 300 keV 
in both nuclei by both parametrizations. However, in both nuclei 
the excitation energy of the $\nu [613]7/2$ state is overestimated 
by $\approx 0.55$ MeV and by $\approx 1.0$ MeV in the NL1 and NL3 
parametrizations, respectively. The comparison of the average 
energies of the experimental and calculated $\nu [613]7/2$ and 
$\nu [622]5/2$  states suggests that the energy of the $\nu 2g_{9/2}$ 
subshell has to be decreased by $\approx 0.15$ MeV and by
$\approx 0.6$ MeV in the NL1 and NL3 
parametrizations, respectively.

 The relative position of the $\nu [613]7/2$ and $\nu [615]9/2$ states, 
emerging from the $\nu 2 g_{9/2}$ and $\nu 1i_{11/2}$ spherical subshells, 
is not reproduced in both parametrizations. In addition, the excitation 
energy of the $\nu [624]7/2$ state from the $\nu 1i_{11/2}$ subshell is 
overestimated by $\geq 1$ MeV in $^{249}$Cf (see Fig.\ \ref{qp-cf249}).
An analysis similar to the one given above suggests that the energy 
of the $\nu 1i_{11/2}$ spherical subshell has to be increased by 
$\approx 0.3$ MeV and by $\approx 1$ MeV in the NL1 and NL3 parametrizations,
respectively, in order to bring the calculations in agreement with 
experiment.

  In $^{251}$Cf, the relative positions of the $\nu [622]3/2$ 
$(\nu 2g_{7/2})$ and $\nu [620]1/2$ ($\nu 3d_{5/2}$) states (see 
Fig.\ \ref{qp-cf251}) differ in experiment and calculations
(with NL3). In addition, NL3 fails to reproduce the $\nu [620]1/2$ 
energy. The states $\nu [622]3/2$ and $\nu [620]1/2$ are almost 
degenerate in energy in the NL1 parametrization, while in experiment
the excitation energy difference between the $\nu [622]3/2$ and 
$\nu [620]1/2$ states is 178 keV. An increase (decrease) of the energy 
of the $\nu 2g_{7/2}$ spherical subshell by $\approx 0.15$ MeV in NL1 
(NL3) and a decrease of the energy of the $\nu 3d_{5/2}$ spherical 
shell by $\approx 0.5$ MeV in the NL3 parametrization would remove 
these discrepancies.

  In addition, the $\nu [761]1/2^*$ \footnote{The use of asterisk at 
the Nilsson labels is explained in caption of Fig.\ \ref{proton-no254-sp}.} 
state has been observed in $^{251}$Cf. Its quasiparticle energy is well 
described (within 150-200 keV) in both parametrizations, suggesting that 
the energy of the $\nu 1j_{13/2}$ spherical subshell is correctly 
accounted for in both parametrizations. 

The comparison of the average energies of the experimental 
and calculated $\nu [734]9/2$ and $\nu [725]11/2$ states in $^{251}$Cf 
(see Fig.\ \ref{qp-cf251}), from the $\nu 1j_{15/2}$  subshell, 
suggests that the energy of this subshell has to be decreased by 
$\approx 0.7$ MeV and by $\approx 0.45$ MeV in the NL1 and NL3 
parametrizations, respectively. However, the $\nu [734]9/2$ state
would still remain the ground state of $^{249}$Cf after all the 
modifications discussed above. 
 
%%%%%%%%%%%%%%%%%%%%%%%%%%%%%%%%%%%%%%%%%%%%%%%%%%%%%%%%%%%%
\subsection{Odd-proton nucleus $^{249}$Bk}
\label{Bk249}
%%%%%%%%%%%%%%%%%%%%%%%%%%%%%%%%%%%%%%%%%%%%%%%%%%%%%%%%%%%%

 The quasiparticle spectrum of $^{249}$Bk ($Z=98,N=152$) is presented 
in Fig.\ \ref{qp-bk249}. Three states $\pi [642]5/2$, 
$\pi [633]7/2$ and $\pi [624]9/2$, from the $\pi 1i_{13/2}$ 
subshell, have been observed in experiment. We select this subshell 
as a reference, with respect to which the positions of other spherical 
subshells will be compared, because in the NL1 and NL3 parametrizations the 
$\pi [633]7/2$ state is close to experiment, and the $\pi [642]5/2$ 
and $\pi [624]9/2$ states are located below and above the Fermi level, 
respectively. The $\pi [642]5/2$ and $\pi [633]7/2$ 
states are reasonably well reproduced. The $\pi [624]9/2$ state is
excluded from the direct comparison since vibrational admixtures are
expected to be large due to high experimental excitation energy of 
this state.

 The energy difference between the $\pi[521]3/2$ and $\pi[521]1/2$ 
states is well reproduced, suggesting that the spin-orbit splitting 
between the $\pi f_{5/2}$ and $\pi f_{7/2}$ spherical subshells is 
correctly described. However, their positions with respect to the 
Fermi level depend on the parametrization: for only NL1 is the 
$\pi[521]3/2$ state lowest in energy, in agreement with experiment. 
The energy difference between the  [633]7/2 and [521]3/2 states is 
small, around 250 keV  (see Fig.\ \ref{qp-bk249}). A decrease 
(increase) of the energy of the $2f_{7/2}$ spherical subshells by 
$\approx 0.25$ MeV in the NL3 (NL1) parametrizations would bring 
the relative positions of these states in agreement with experiment. 
This also implies the same shift of its spin-orbit partner 
$\pi 2f_{5/2}$.

 The $\pi [514]7/2$ state, from the $\pi 1h_{9/2}$ subshell, 
is too low in energy in both parametrizations with respect to 
the $\pi [633]7/2$ state. An increase of the energy of the $1h_{9/2}$ 
spherical subshell by $\approx 0.85$ MeV for NL3 and by $\approx 0.6$ 
MeV for NL1 would bring calculations in agreement with experiment. The 
analyses of odd-proton nuclei around $^{208}$Pb (see Fig.\ 7 in Ref.\ 
\cite{RBRMG.98}) and of shape coexistence in the Pt-Hg-Pb isotopes 
\cite{NVRL.02} also point to this deficiency in the description of 
the $1h_{9/2}$ spherical subshell energy.

  The calculations underestimate the position of the $\pi [530]1/2$
state, from the $\pi 2f_{7/2}$ subshell,  by $\approx 1.5$ MeV. 
If the energy of the $1h_{9/2}$ subshell were increased as discussed above, 
this would push the $\pi [530]1/2$ state closer to the Fermi level due 
to the interaction with the $\pi [541]1/2$ state, from the 
$\pi 1h_{9/2}$ subshell (see Fig.\ 4 in Ref.\ \cite{CAFE.77}). This 
would lead to a better agreement between calculations and experiment.

 The $\pi [400]1/2$ state, from the $\pi 3s_{1/2}$ subshell, is also 
reasonably well reproduced (somewhat better with NL3 than with NL1). 
However, the relative positions of the $\pi [400]1/2$ and $\pi [642]5/2$
states (the latter from the $\pi 1i_{13/2}$ subshell) suggest that the 
energy of the $\pi 3s_{1/2}$ subshell has to be increased by $\approx 0.3$ 
MeV in the NL1 parametrization. 

%%%%%%%%%%%%%%%%%%%%%%%%%%%%%%%%%%%%%%%%%%%%%%%%%%%%%%%%%%%%%%%%%%
\subsection{Consequences for deformed shell gaps in the $A\sim 250$ 
mass region.}
%%%%%%%%%%%%%%%%%%%%%%%%%%%%%%%%%%%%%%%%%%%%%%%%%%%%%%%%%%%%%%%%%%

  Figure \ref{def-sp-cor} shows how the proton and neutron spectra 
in $^{254}$No are modified if the spherical subshells are 
shifted as discussed in Sects.\ \protect\ref{Bk249} and 
\protect\ref{Cf249}. Similar corrected spectra are obtained
with NL3 and NL1, indicating that the shifts are correctly defined. 
These shifts would lead to the deformed shell gaps at $N=152$ and 
$Z=100$, as seen in experiment (Sects.\ \ref{sstruc100} and 
\ref{sstrucn152}). Neutron gaps at $N=148,150$ (NL1) ($N=148$ in NL3) 
seen in uncorrected spectra disappear, while the proton gap at $Z=104$ 
becomes smaller. 

 In addition, the ordering of the neutron and proton single-particle 
states below and above these shell gaps would be more similar to the 
Woods-Saxon potential (see Fig.\ 8 in Ref.\ \cite{Md255}), whose 
parameters were defined by an overall fit to the single-particle 
spectra in heavy actinide nuclei \cite{CA.97}.

 These examples illustrate that in the region of high level density 
and small shells gaps,  a shift of the energies of one or two 
single-particle states by a modest energy of 0.5 MeV can modify 
the nucleon number of the shell gap by two or four units. A new 
parametrization of the RMF Lagrangian, which implements the shifts 
discussed in Sects.\ \protect\ref{Bk249} and \protect\ref{Cf249} 
naturally, is called for.

%%%%%%%%%%%%%%%%%%%%%%%%%%%%%%%%%%%%%%%%%%%%%%%%%%%%%%%%%%%%%%%%%%
\subsection{Estimates for other parametrizations}
\label{qp-other-forces}
%%%%%%%%%%%%%%%%%%%%%%%%%%%%%%%%%%%%%%%%%%%%%%%%%%%%%%%%%%%%%%%%%%

 The calculations of odd-$A$ nuclei performed with the NL1 and NL3 sets 
indicate that, in general, the results are quite similar but somewhat
better agreement is obtained in the NL1 parametrization. Moreover, some 
conclusions about the accuracy of the description of the quasi-particle 
states in other RMF parametrizations can be drawn with the aid of the spectra 
presented in Figs.\ \ref{proton-no254-sp} and \ref{neutron-no254-sp}.
The NL-Z parametrization gives single-particle spectra 
in between those for NL1 and NL3 and, thus, a similar accuracy of the 
description of quasi-particle states is expected. 

 The agreement with experiment is worse for the NLSH and NL-RA1 parametrizations. 
Let us illustrate this by a few examples for NLSH, which 
deviate most from experiment. The energy splitting between $\nu [613]7/2$ 
and $\nu [615]9/2$ increases from $\approx 0.5$ MeV up to $\approx 2.5$ MeV 
when going from NL1 to NLSH (see Fig.\ \ref{neutron-no254-sp}). Thus in 
order to reproduce the relative positions of these states in $^{249,251}$Cf 
(see Figs.\ \ref{qp-cf249} and \ref{qp-cf251}), the relative distance between 
the spherical $\nu 2 g_{9/2}$ and $\nu 1 i_{11/2}$ subshells should be corrected 
by roughly 2 MeV. 

 The Fermi level for the odd-proton (Z=97) $^{249}$Bk nucleus will be 
located somewhere in the vicinity of the $\pi [633]7/2$ and 
$\pi [514]7/2$ states (see Fig.\ \ref{neutron-no254-sp}). Thus, the 
$\pi [521]3/2$ and $\pi [521]1/2$ states (and the corresponding 
$\pi 2 f_{7/2}$ and $\pi 2 f_{5/2}$ spherical subshells) should be 
lowered by roughly 1 MeV with respect of the $\pi [633]7/2$ state 
(the $\pi 1i_{13/2}$ subshell) in order to reproduce the experimental 
spectra (see Fig.\ \ref{qp-bk249}). In addition, the position of the 
$\pi [514]7/2$ state ($\pi 1h_{9/2}$ spherical subshell) should be 
raised by $\approx 700$ keV with respect of the $\pi [633]7/2$ state 
(the $\pi 1i_{13/2}$ subshell). 

 Thus the empirical shifts required to reproduce experimental quasiparticle 
energies are much larger for NLSH  than the ones needed for NL1 and NL3 
(see Sects.\ \ref{Cf249} and \ref{Bk249}). Only after these shifts will the 
$\pi [633]7/2$ and $\pi [521]3/2$ states be located in the vicinity 
of the proton Fermi level and there will be a gap at $Z=100$  between these 
states and $\pi [521]1/2$ and $\pi [514]7/2$ in agreement with an analysis 
based on the Woods-Saxon potential \cite{Md255}. Although the NLSH 
parametrization is the only one parametrization which reproduces the $Z=100$ 
gap (see Fig.\ \ref{delta-n152}), this gap is created between the wrong 
bunches of states.

%%%%%%%%%%%%%%%%%%%%%%%%%%%%%%%%%%%%%%%%%%%%%%%%%%%%%%%%%%%%%%%%%%
\subsection{Consequences of nuclear magnetism for quasiparticle 
states}
\label{nucl-magn}
%%%%%%%%%%%%%%%%%%%%%%%%%%%%%%%%%%%%%%%%%%%%%%%%%%%%%%%%%%%%%%%%%%
 
 The influence of nuclear magnetism on the binding energies of 
one-quasiparticle states in $^{249}$Cf and $^{249}$Bk is shown in 
Table \ref{Table-ENM}. In all cases, it provides small additional binding. 
It is state dependent and lies between $-16$ and $-69$ keV, depending
weakly on the strength of the pair correlations and on particle number 
projection. This indicates that these heavy nuclei are rather 
robust against polarization effects induced by nuclear magnetism. 
Thus, if nuclear magnetism was neglected, the quasiparticle spectra 
in this mass region would only be marginally modified. The influence 
of nuclear magnetism on quasiparticle energies is larger in lighter 
systems \cite{AF} (see also Ref.\ \cite{DBHM.02} for a study
of the effects of the time-odd mean fields on the quasiparticle 
energies within the Skyrme Hartree-Fock approach) and in 
two-particle configurations \cite{Lund2000}.

  While the neglect of nuclear magnetism seems to be a reasonable 
approximation for the one-quasiparticle energies,  it has to be taken 
into account when the strength of the pairing correlations is fitted 
to experimental odd-even mass differences because it modifies 
$\Delta^{(3)}$ by $\approx 10\%$ (see Table \ref{Table-odd-even}).

%%%%%%%%%%%%%%%%%%%%%%%%%%%%%%%%%%%%%%%%%%%%%%%%%%%%%%%%%%%%%%
\subsection{Implications for the study of superheavy nuclei}
\label{implication}
%%%%%%%%%%%%%%%%%%%%%%%%%%%%%%%%%%%%%%%%%%%%%%%%%%%%%%%%%%%%%%

  In the NL1 and NL3 parametrizations, the energies of the spherical 
subshells, from which the deformed states in the vicinity of the Fermi 
level of the $A\sim 250$ nuclei emerge, are described with an accuracy 
better than 0.5 MeV for most of the subshells (see Figs.\ \ref{z120-proton} 
and \ref{z120-neutron} where required corrections for single-particle 
energies are indicated). The discrepancy reaches 0.6-1.0 MeV 
for the  $\pi 1h_{9/2}$ (NL3, NL1), $\nu 1i_{11/2}$ (NL3),  
$\nu 1j_{15/2}$ (NL1) and $\nu 2 g_{9/2}$ (NL3) spherical subshells.
Considering that the RMF parametrizations were fitted only to bulk properties 
of spherical nuclei, this level of agreement is good. However, the accuracy 
of the description of single-particle states is unsatisfactory in the 
NLSH and NL-RA1 parametrizations (see Sect.\ \ref{qp-other-forces}).

 The single-particle levels of spherical magic superheavy nuclei are not modified 
much with the empirical shifts of Sects.\ \ref{Bk249} and \ref{Cf249}
(see Figs.\ \ref{z120-proton} and \ref{z120-neutron} for the calculated 
and corrected single-particle spectra of a $^{292}_{172}$120 nucleus).
This conclusion relies on the assumption that the shifts should 
be similar in the deformed $A\sim 250$ mass region and in superheavy 
nuclei. The corrected spectra from the NL1 and NL3 calculations 
are very similar, with minor differences coming from the limited 
amount of information on quasiparticle states used in the analysis. 
More systematic study of quasiparticle states in deformed nuclei 
are required to determine these corrections more precisely.

   Let us consider the calculations for the nucleus with $Z=120$, $N=172$. 
The corrected single-particle levels still suggest that $N=172$ and 
$N=184$ are candidates for magic neutron numbers in superheavy nuclei. 
The position of the $\nu 4s_{1/2}$ spherical subshell and the spin-orbit 
splitting of the 
$3d_{5/2}$ and $3d_{3/2}$ subshells will decide which of these numbers (or 
both of them) is (are) actually magic. The  corrected proton levels indicate that 
the $Z=120$ gap is large whereas the $Z=114$ gap is small. Hence, on the 
basis of the present investigation we predict that $Z=120$ is the magic proton 
number. This conclusion is based on the assumption that the NL1 and NL3 sets
predict the position of the $\pi 1i_{11/2}$ and $\pi 3p_{1/2,3/2}$ subshells 
within 1 MeV. The positions of $\pi 1 j_{15/2}$ and $\pi 2g_{9/2}$ 
seems less critical, because they are located well above this group of states 
both in Skyrme and RMF calculations \cite{BRRMG.99}. It seems possible to obtain 
information about the location of the $\pi 1i_{11/2}$ subshell, which may have 
been observed through its deformed state $(\pi [651]1/2^*)$ in superdeformed 
rotational bands of Bi-isotopes \cite{Bi-SD1,Bi-SD2}. An CRHB analysis may 
provide this critical information. 

 In this context it is important to mention that the RMF parametrizations,
NL-SH and NL-RA1,  which are the only ones to predict $Z=114$ as the magic 
proton number \cite{LSRG.96,NL-RA1}, provide poor descriptions  of the 
single-particle states (see Sect.\ \ref{qp-other-forces}).
                                                        
  The Nilsson diagrams given, for example, in Figs.\ 3 and 4 of Ref.\ 
\cite{CAFE.77}, suggest that spectroscopic studies of deformed odd nuclei 
with proton and neutron numbers up to $Z\approx 108$ and $N\approx 164$ 
may lead to  observation of the deformed states with $\Omega=1/2$, 
emerging from the $\pi 1i_{11/2}$ and $\pi 1j_{15/2}$ spherical subshells 
(located above the $Z=120$ shell gap) and from $\nu 1k_{17/2}$ and either 
$\nu 2 h_{11/2}$ or $\nu 1j_{13/2}$ subshells (located above the $N=184$ 
shell gap). This will further constrain microscopic models and effective 
interactions. 

 No information on low-$j$ states, such as $\pi 3p_{3/2}$, $\pi 3p_{1/2}$, 
$\nu 3d_{3/2}$ and $\nu 4s_{1/2}$, which decide whether $Z=120$ or $Z=126$ 
and $N=172$ or $N=184$ are magic numbers in microscopic theories (see Refs.\ 
\cite{BRRMG.99,RBM.01} and references quoted therein), will come from the
study of deformed nuclei (see Table \ref{qp-experiment}).

  The measured and calculated energies of the single-particle states at 
normal deformation provide constraints on the spherical shell gaps of 
superheavy nuclei. In particular, the small splitting between the 
$\pi [521]1/2$ and $\pi [521]3/2$ deformed states, emerging from the 
$\pi 2f_{5/2}$ and $\pi 2f_{7/2}$ spherical subshells that straddle proton 
number 114, suggests that the $Z=114$ shell gap is not large.

%%%%%%%%%%%%%%%%%%%%%%%%%%%%%%%%%%%%%%%%%%%%%%%%%%%%%%%%%%%%%%%%%%%%%
\subsection{Concluding remarks to Section \ref{qp-energies}}
%%%%%%%%%%%%%%%%%%%%%%%%%%%%%%%%%%%%%%%%%%%%%%%%%%%%%%%%%%%%%%%%%%%%%

In order to judge the reliability of the energies of 
single-particle states predicted for superheavy nuclei by self-consistent 
mean-field theories, it is necessary to check the theoretical energies 
against the experimental ones in the heaviest nuclei where data exist.
The energies of quasiparticle states have been calculated in a fully
self-consistent manner with the CRHB method for $^{249}$Bk and $^{249,251}$Cf
and compared with experiment. The calculated single-particle 
energies depend on the Lagrangian parameterization; NL1, NL3 and NL-Z provide 
good
descriptions of the measured energies, whereas NLSH and NL-RA1 do not.  For
the former set, the quasiparticle energies are generally reproduced for most
orbitals within $\approx 500$ keV.  However, for some orbitals originating
from a few
specific spherical subshells, the discrepancy between theory and experiment can
reach 1 MeV.  Empirical shifts of the energies of these orbitals can be
introduced to fit the experimental data. Including these shifts,
the next spherical shell gaps beyond $^{208}$Pb are predicted at $Z=120$ and 
$N=172, 184$; no gap is seen at $Z=114$. 
%Since the level densities for the heaviest nuclei are high, it is difficult to
%accurately predict the locations of shell gaps, especially the small gaps
%corresponding to deformed shapes. 
The occurrence of some sizeable discrepancies in single-particle energies 
calls for an improved Lagrangian parametrization, which can better 
describe single-particle energies and, thereby, give more reliable predictions 
about the properties of superheavy nuclei.

%%%%%%%%%%%%%%%%%%%%%%%%%%%%%%%%%%%%%%%%%%%%%%%%%%%%%%%%%%%%%%%%%  
\section{Conclusions}
\label{Conclusion}
%%%%%%%%%%%%%%%%%%%%%%%%%%%%%%%%%%%%%%%%%%%%%%%%%%%%%%%%%%%%%%%%%

 The cranked relativistic Hartree+Bogoliubov theory has been applied for 
a systematic study of the nuclei around $^{254}$No, the heaviest elements 
for which detailed spectroscopic data are available. The deformations, 
rotational response, pair correlations, quasiparticle spectra, shell 
structure and the two-nucleon separation energies have been studied. The 
goal was to see how well the theory describes the experimental data and 
how this description depends on the RMF parametrization. Although the 
relativistic mean field theory has been used extensively for 
the predictions of the properties of superheavy nuclei, it has not yet 
been demonstrated how well it describes spectroscopic data in the heaviest 
nuclei, which are the gateway to superheavy nuclei. The present investigation 
provides a basis for better judging the reliability of extrapolations to 
superheavy nuclei.

 The calculations with the NL3, NL1, NL-RA1 and NL-Z parameter sets 
reproduce well the experimental quadrupole deformations of the Cm, Cf, 
Fm and No nuclei, whereas the NLSH set underestimate them.

 In order to reproduce the moments of inertia in the $A\sim 250$ mass 
region, the strength of the D1S Gogny force in the particle-particle 
channel has to be attenuated by $\approx 12\%$. In contrast, the moments 
of inertia of lighter nuclei can be well described with the full strength 
D1S force.

 With the attenuated D1S force, the rotational response is well described.
In $^{252,254}$No nuclei, the alignment of the proton $i_{13/2}$ 
($\pi [633]7/2$) and neutron $j_{15/2}$ ($\nu [734]9/2$) pairs takes place 
simultaneously at $\Omega_x \approx 0.31$ MeV. While the crossing frequency 
depends only weakly on the RMF parametrization, the gain of aligned angular 
momentum at the band crossing and the sharpness of the band crossing  are 
more sensitive to it. The moments of inertia at low spin in the Cm, Cf, Fm 
and No isotopes and their variations with nucleon number are reproduced. 

 The two-particle separation energies are best described by the NL3, NLSH and
NL-RA1 parametrizations, which were derived by fitting experimental information on 
neutron-rich nuclei. The calculated deformed shell gaps occur at nucleon numbers 
which may deviate by as much as 4 from those observed in experiment.

 The quasiparticle-states calculated for odd-$A$ nuclei are the same as 
those identified in 
experiment. For many states, the difference between experimental and 
theoretical energies calculated with the NL1 and NL3 sets is less than 0.5 MeV, 
but may reach 1 MeV in some cases. The spectrum is less compressed in the 
calculations as compared with experiment, which reflects the low effective 
mass of the RMF theory. Inclusion of particle-vibration coupling  may correct 
that. The agreement between experiment and theory can be considered quite good, 
considering that the 6 or 7 parameters of the RMF theory have been adjusted 
to the ground-state properties of spherical nuclei, without taking into 
account the experimental information on the single-particle states. 

 Concerning the predictions for superheavy nuclei we conclude the following.

 (i)  Among the investigated RMF sets,  NL1, NL3 and NL-Z provide best 
description of single-particle states so they seem to be most promising  
for the study of superheavy nuclei. The corresponding self-consistent calculations 
predict as likely candidates for magic numbers $N=172$ and $N=184$ for neutrons
and $Z=120$ for protons. No significant shell gap is found for $Z=114$. These 
conclusions take into account the possible shifts of spherical subshells that 
are suggested by the discrepancies between calculations and experiment found 
in our analysis of deformed odd-mass actinide nuclei.

 (ii) NL-SH and NL-RA1, which are the only RMF sets predicting $Z=114$ as a 
magic proton number, provide poor descriptions of single-particle states and 
thus are not considered as reliable for study of superheavy nuclei. 

 (iii) Experimental studies of deformed odd nuclei with proton and neutron numbers 
up to $Z\approx 108$ and $N\approx 164$ may lead to  observation of the deformed 
states emerging from the high-$j$ spherical subshells located above $Z=120$ and 
$N=184$. Their observation will provide a crucial constraint on the magic 
numbers.

 (iv) The study of deformed states will not provide access to a number of 
low-$j$ subshells, which largely define whether $Z=120$ or $Z=126$ and 
$N=172$ or $N=184$ are magic numbers.

 (v) More systematic studies of the splitting between the $\pi [521]1/2$ and 
$\pi [521]3/2$ deformed states, which originate from the $\pi 2f_{5/2}$ and 
$\pi 2f_{7/2}$ spherical subshells, may provide more stringent information 
on whether a shell gap exists at $Z=114$.

 The present results demonstrate the limitations of adjusting the RMF parameters 
only to the bulk properties of spherical nuclei and may point to missing 
components in the effective Lagrangian. A new fit to both the bulk and 
single-particle properties should lead to a more accurate theory.

 Like many Hartree(-Fock) calculations based on effective interactions, the RMF 
theory underestimates  the single-particle level density. This indicates that 
some important mechanism is missing, which may be the particle-vibration 
coupling.

%%%%%%%%%%%%%%%%%%%%%%%%%%%%%%%%%%%%%%%%%%%%%%%%%%%%%%%%%%%%%%%%%%%%%%%%%%
\section{Acknowledgments}
%%%%%%%%%%%%%%%%%%%%%%%%%%%%%%%%%%%%%%%%%%%%%%%%%%%%%%%%%%%%%%%%%%%%%%%%%%

 The authors would like to thank P.\ Ring, R.\ R.\ Chasman and A.\ Malov 
for valuable discussions. This work was supported in part by the U. S. 
Department of Energy, Grants No. W-31-109-ENG38 and DE-FG02-95ER40934.
The numerical calculations were performed in part on the Cray PVP Cluster 
at the National Energy Research Scientific Computing Center.

%%%%%%%%%%%%%%%%%%%%%%%%%%%%%%%%%%%%%%%%%%%%%%%%%%%%%%%%%%%%%%%%%%%%%%%%%%
\section{Appendix A: The quantity $\delta_{2n}(Z,N)$. }
%%%%%%%%%%%%%%%%%%%%%%%%%%%%%%%%%%%%%%%%%%%%%%%%%%%%%%%%%%%%%%%%%%%%%%%%%%

 In order to understand the quantity $\delta_{2n}(Z,N)$  related to
the second derivative of the binding energy as a function of nucleon
number better, we first discuss the case when pairing is neglected. Then 
we perform a detailed analysis with pairing included in the CRHB+LN 
framework. 

 Figure \ref{delta-no-pairing} compares $\delta_{2n}(Z,N)$ obtained in the 
RMF calculations without pairing with $2E_{SP-GAP}$, where $E_{SP-GAP}$ 
is the energy gap between the last occupied and the first unoccupied 
single-particle level in the $(Z,N)$ system. One can see that 
$\delta_{2n}(Z,N)$ is shifted down by  $0.56^{+0.12}_{-0.10}$ MeV with 
respect to $2E_{SP-GAP}$. This shift can be understood within the 
shell-correction method \cite{NR.book,S.67,BDJPSW.72}, in which the total energy of 
the system $E_{tot}$ in the absence of pairing is given by
\begin{eqnarray}
E_{tot}=E_{LD} + E_{sh}^{\pi} + E_{sh}^{\nu}
\end{eqnarray}
where $E_{LD}$ is a liquid drop energy and $E_{sh}$ is a shell
energy (superscript $\nu$ stands for neutrons, $\pi$ for protons). 
The latter is given by
\begin{eqnarray}
E_{sh} = 2 \sum_{i-occ} e_i - 2 \int^{\tilde{\lambda}} e \tilde {g}(e) de =
2  \sum_{i-occ} e_i - \tilde{E}
\end{eqnarray}
where $e$ is a single-particle energy, $\tilde {g}(e)$ the
smeared level density and $\tilde{E}$ the Strutinsky smoothed
sum. In this equation $\tilde{\lambda}$ is the Fermi energy
corresponding to $\tilde {g}(e)$ and is determined from the
condition of number conservation
\begin{eqnarray}
N = 2 \int^{\tilde{\lambda}}  \tilde {g}(e) de 
\end{eqnarray}
Neglecting the variations in $E_{sh}^{\pi}$
when the neutron number changes, one can write
\begin{eqnarray}
\delta_{2n}(Z,N) = 2 E_{SP-GAP} - \delta_{2n}^{\tilde{E}}(Z,N) + \delta_{2n}^{E_{LD}}(Z,N).   
\end{eqnarray}
$E_{LD}(Z,N)$ and $\tilde{E}(Z,N)$ are smooth functions, which weakly depend on 
the particle number \cite{NR.book}, and the $\delta_{2n}^{E_{LD}}(Z,N)$ 
and $\delta_{2n}^{\tilde{E}}(Z,N)$ quantities related to their second derivatives 
as a function of nucleon number are nearly constant within the considered interval.
Hence, $\delta_{2n}(Z,N)$ differs from $2E_{SP-GAP}$ by  nearly constant.
Although the shell-correction method is an approximation to the fully variational 
many-body approach such as the RMF theory, it elucidates the main physics in a simple 
way. This example clearly shows that $\delta_{2n}(Z,N)$ is not a direct measure of 
$2E_{SP-GAP}$.

 The pairing smoothes the variations of $\delta_{2n}(Z,N)$ because there is
gradual change of occupation numbers from 1 to 0. Comparing Figs.\ 
\ref{delta-no-pairing} and \ref{delta-contr-fm} one sees that the pairing 
reduces the height of the maximum of $\delta_{2n}(Z,N)$ at the $N=148$ shell
gap by approximately a factor of two and increases $\delta_{2n}(Z,N)$ at 
the neutron numbers away from it. This illustrates that the values of 
$\delta_{2n}(Z,N)$ cannot be taken as direct measure of $2 E_{SP-GAP}$  
since they are strongly dependent on pairing; see also Ref.\ \cite{DNW.95}
and Fig.\ \ref{delta-fm}b for a comparison of the results of CRHB and CRHB+LN 
calculations.

 Let us now consider the chain of the Fm isotopes within the CRHB+LN
theory as an illustrative example how the $\delta_{2n}(Z,N)$ is built
from different contributions. If we neglect the spurious center-of-mass 
correction, which in harmonic oscillator approximation does not contribute 
to $\delta_{2n}(Z,N)$, then the total energy in the laboratory frame is
given by (see Eqs.\ (21-24,43) in Ref.\ \cite{CRHB} for details)
\begin{eqnarray}
\begin{array}{lll}
E= & -~\frac{1}{2}g_\sigma\,\int d {\bff r} \,\sigma (\bff r)
\rho_{\sl s} (\bff r) 
-~\frac {1}{2}\, g_\omega \int d {\bff r}\,\omega_0(\bff r)
\rho_{\sl v}^{\it i\sl s} (\bff r) \quad & \Big\} = E_{S+V} \\
&-~\frac{1}{2}g_\sigma\,\int d {\bff r} \,
\left [ \frac{1}{3}g_2\sigma^3(\bff r)
+\frac{1}{2}g_3\sigma^4(\bff r) \right ] 
&\Big\} = E_{\sigma NL} \\
&-~\frac{1}{2}\, g_\rho \int d {\bff r}\,\rho_0(\bff r)
\rho_{\sl v}^{\it i\sl v} (\bff r) \,&\Big\}=E_{\rho} \\
&-~\frac{1}{2}\, e \int d {\bff r} A_0(\bff r) \rho_{\sl v}^p (\bff r)
&\Big\}=E_{Coul} \\
&+\, Tr(h_D \rho) 
&\Big\}=E_{part} \\
& -\,\frac{1}{2}Tr(\Delta \kappa) &\Big\}=E_{pairing}  \\ 
& -\,\lambda_2 \langle (\Delta \hat{N})^2 \rangle  &\Big\}=E_{LN}
\end{array}
\label{Etot}
\end{eqnarray}
where first four terms represent the contributions from bosonic degrees 
of freedom, while last three terms from fermionic degrees of freedom.
The $E_{S+V}$ is the sum of the energies of the fields associated with 
the linear part of the $\sigma$-meson and the $\omega$-meson. This sum 
represents the main part of the nucleonic potential \cite{R.96,Reinh89}. 
The $E_{\sigma NL}$ term is the energy of the non-linear part of the 
$\sigma$-meson, while $E_{\rho}$ is the energy of the $\rho$-meson
field and the $E_{Coul}$ is the energy of the Coulomb field.
Finally, the $E_{part}$, $E_{pairing}$, $E_{LN}$ terms are the particle 
and the  pairing energies as well as the  energy 
correction entering into particle-number projection by means of 
the Lipkin-Nogami method, respectively.

\begin{figure}[t]
\epsfxsize 12.0cm
\epsfbox{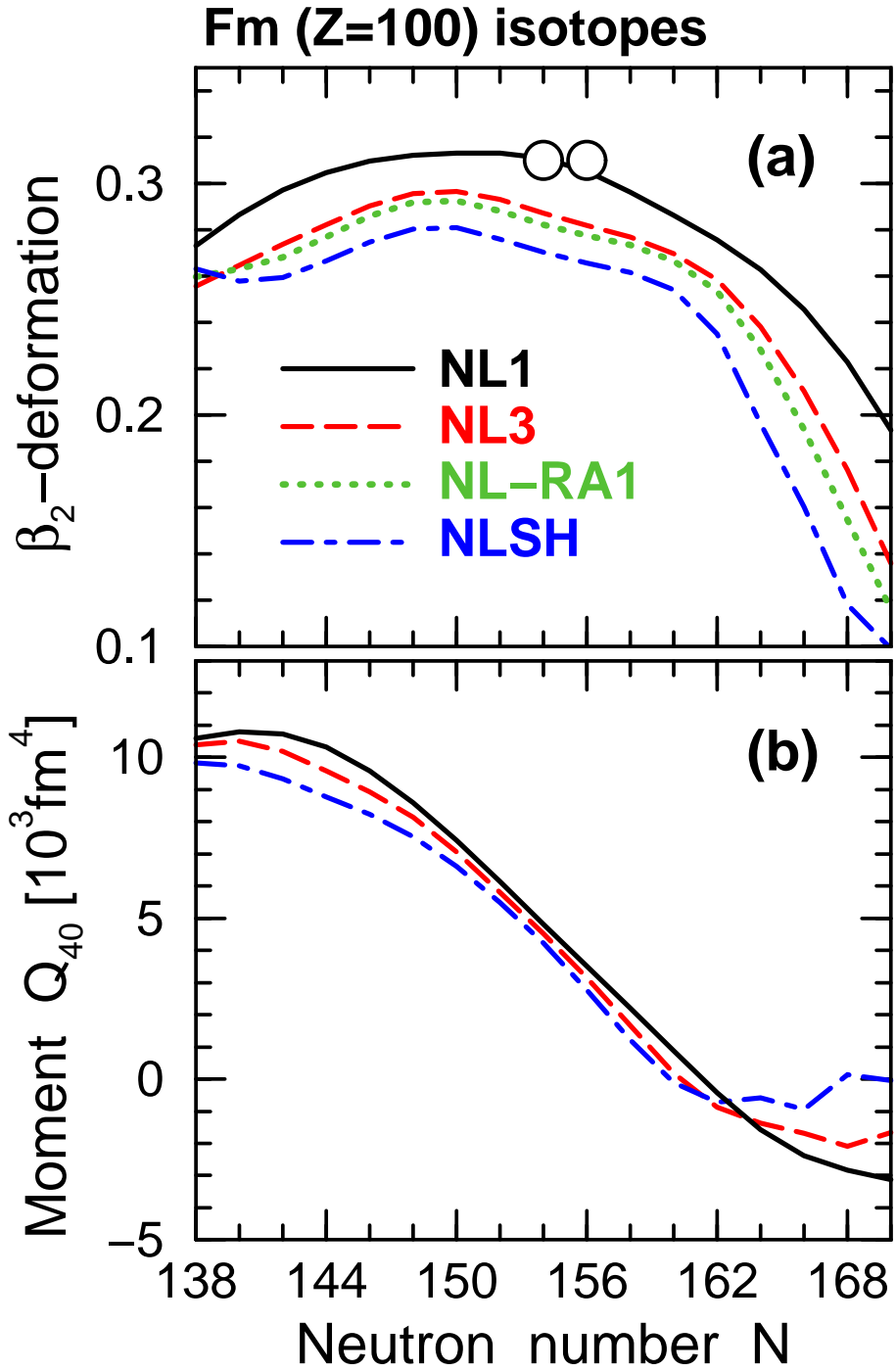}
\vspace{0.5cm}
\caption{The same as in Fig.\ \protect\ref{def-n152-allforces}, but for
the chain of the Fm isotopes. The values of $\beta_2$ calculated 
with NL1 and NL-Z are very close to each other, thus the values
obtained with NL-Z are omitted. The results for $Q_{40}$ obtained 
with NL-RA1 and NL-Z are very close to those with NL3 and NL1.}
\label{def-fm}
\end{figure}

\begin{figure}[t]
\epsfxsize 16.0cm
\epsfbox{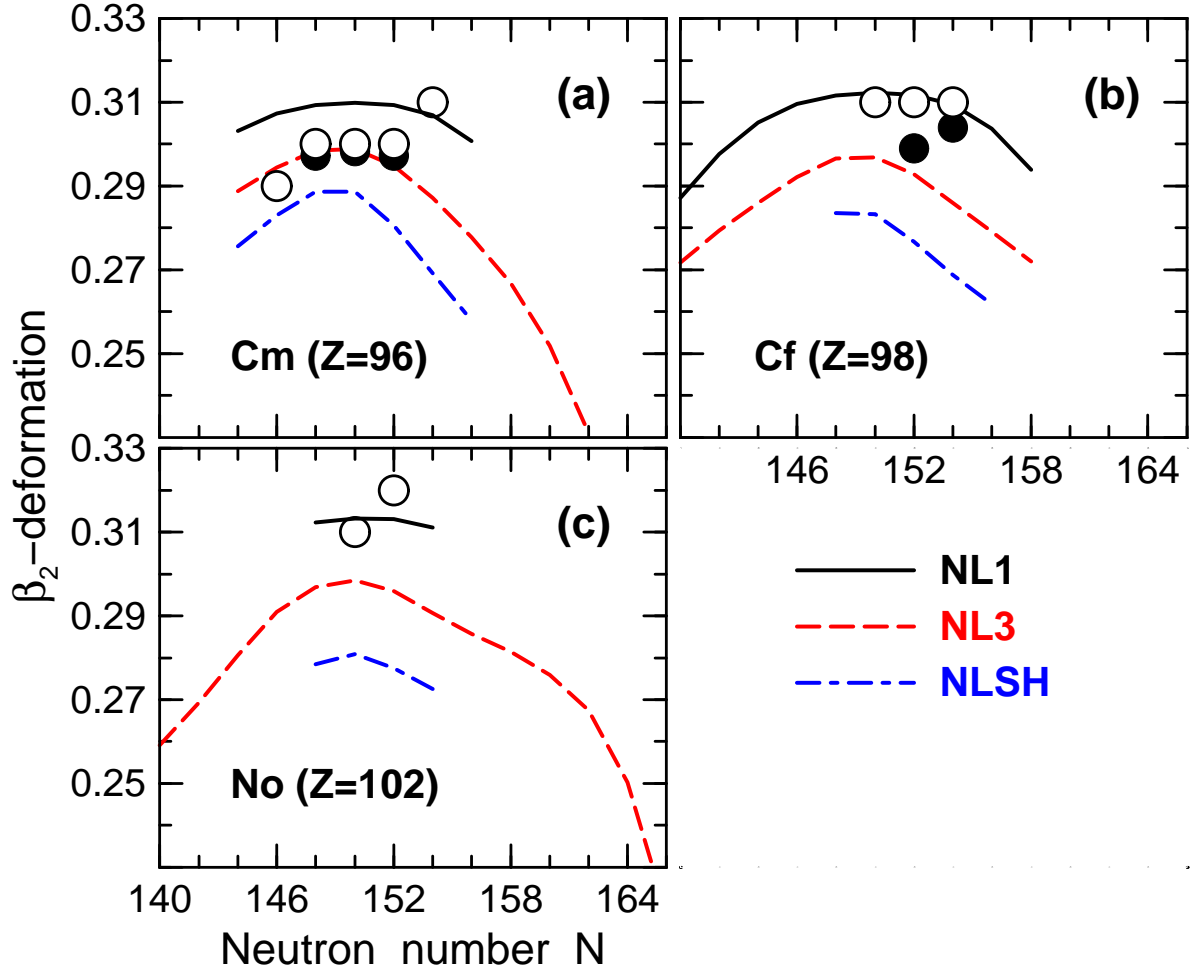}
\vspace{0.5cm}
\caption{The same as in Fig.\ \protect\ref{def-n152-allforces}, 
but for the chains of the Cm, Cf and No isotopes. }
\label{def-cf-cm-no}
\end{figure}

\begin{figure}
\epsfxsize 16.0cm
\epsfbox{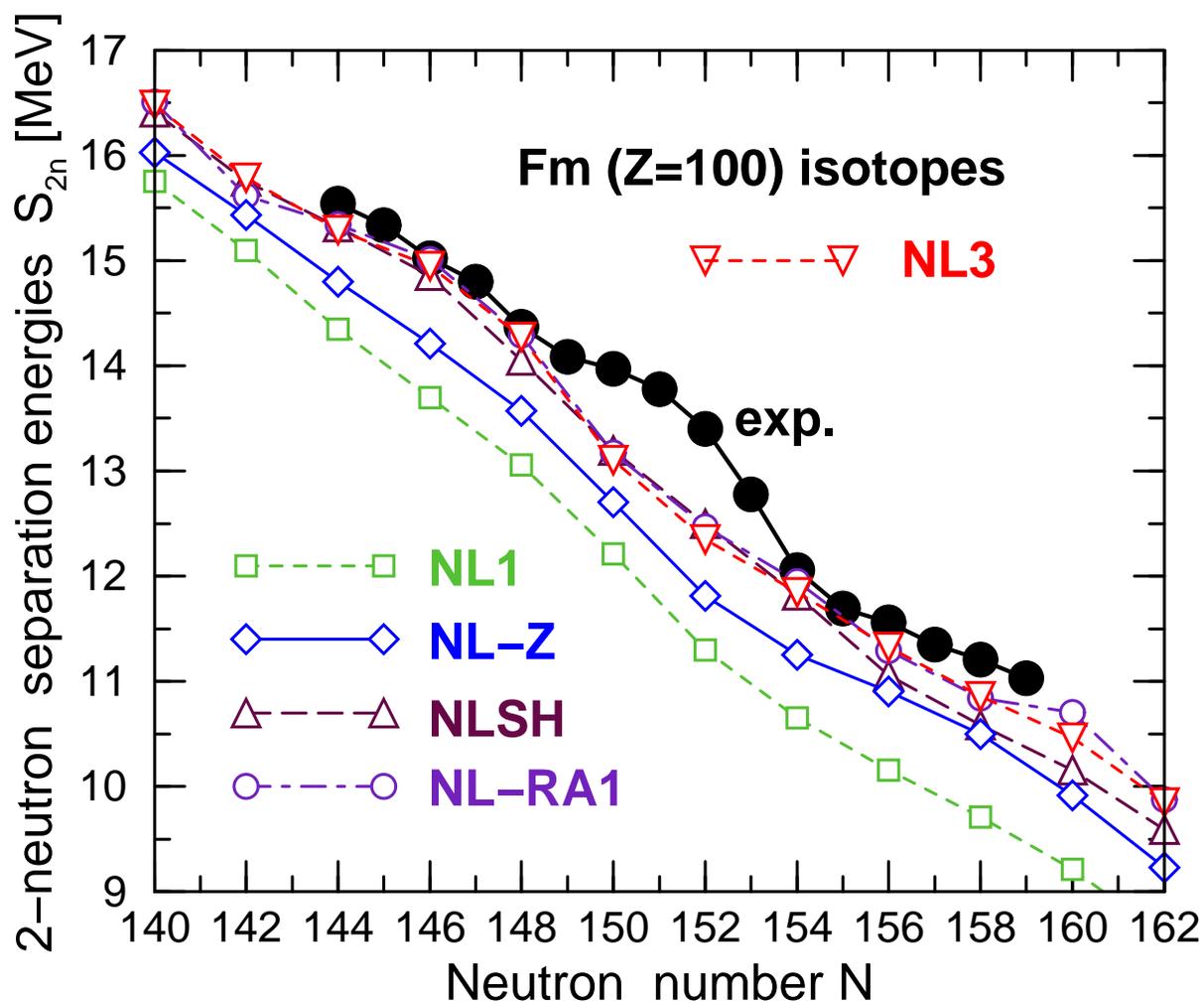}
\vspace{0.5cm}
\caption{The two-neutron separation energies $S_{2n}(Z,N)$ obtained 
in the CRHB+LN calculations for Fm $(Z=100)$ isotopes with different 
RMF parametrizations. Solid circles are used for experimental data, while 
open symbols for the theoretical results.} 
\label{sep-ener-fm}
\end{figure}

\begin{figure}
\epsfxsize 14.0cm
\epsfbox{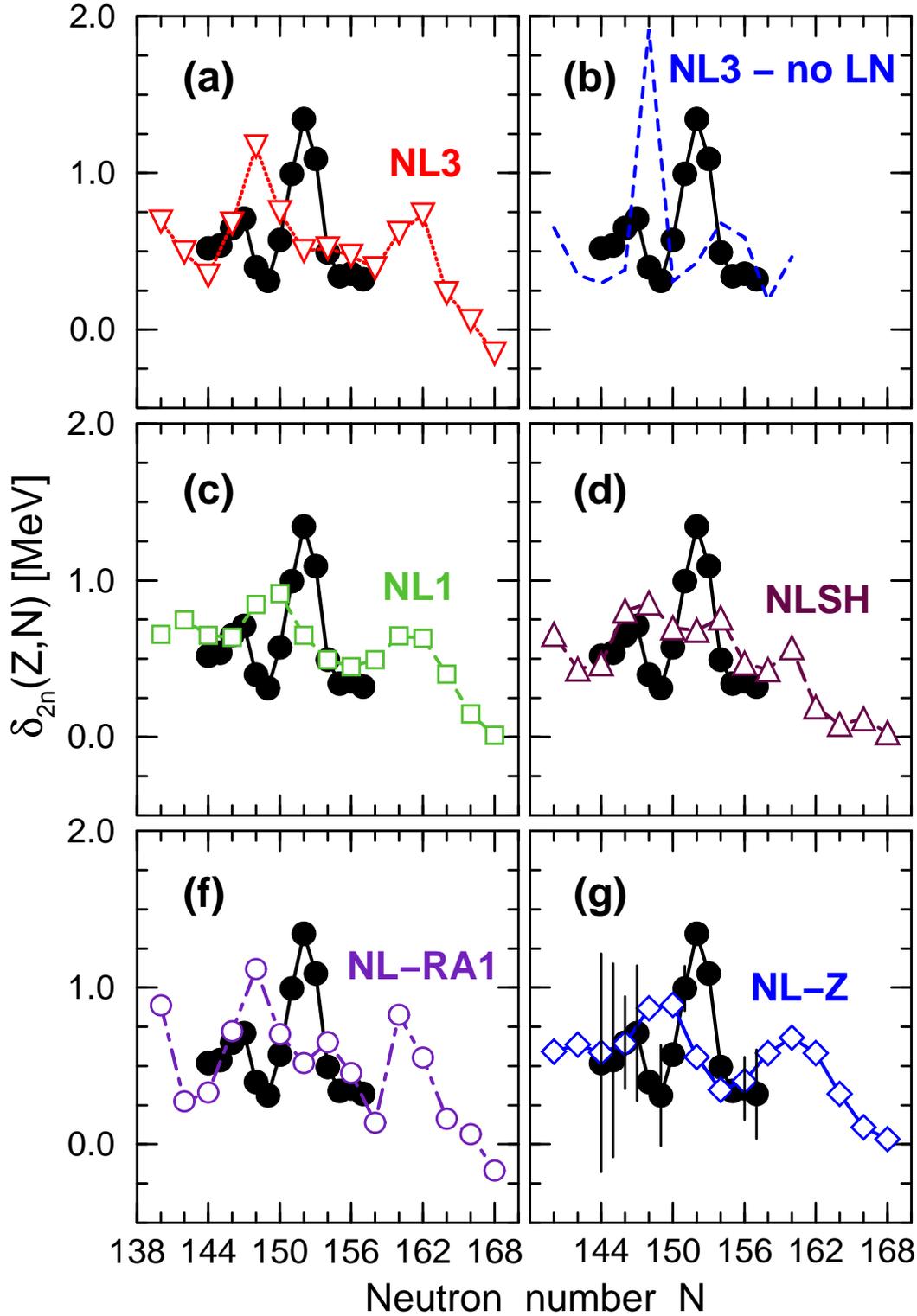}
\vspace{0.5cm}
%\epsfbox{new-delta-fm.eps}
\caption{The quantity $\delta_{2n}(Z,N)$ for Fm nuclei.
The experimental data (solid circles) are 
compared with the results (open symbols) obtained in the CRHB+LN 
calculations with the indicated RMF parametrizations in panels (a,c-g). 
The results of the calculations without LN are shown 
in panel (b) by the dashed line. The experimental error bars are 
shown in panel (g). } 
\label{delta-fm}
\end{figure}

\begin{figure}
\epsfxsize 14.0cm
\epsfbox{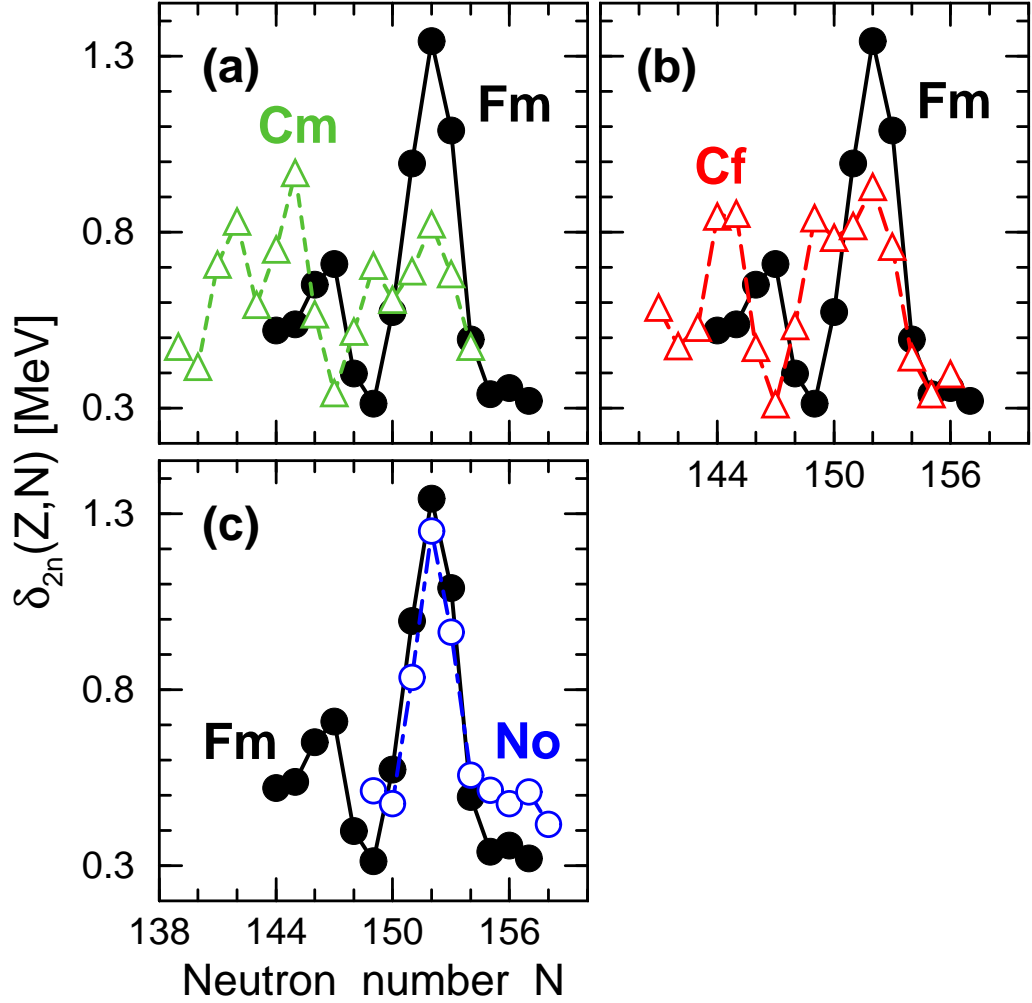}
\vspace{0.5cm}
%\epsfbox{new-delta-exp.eps}
\caption{ Experimental $\delta_{2n}(Z,N)$ values, 
shown by open symbols for Cm (Z=96) (panel (a)) , Cf (Z=98) 
(panel(b)), and No (Z=102) (panel(c)) nuclei. In all panels the 
experimental values for Fm (Z=100) nuclei are shown by solid circles 
in order to indicate the variations of $\delta_{2n}(Z,N)$ with change 
of proton number $Z$.
} 
\label{delta2n-exp}
\end{figure}

\begin{figure}[t]
\epsfysize 18.0cm
\epsfbox{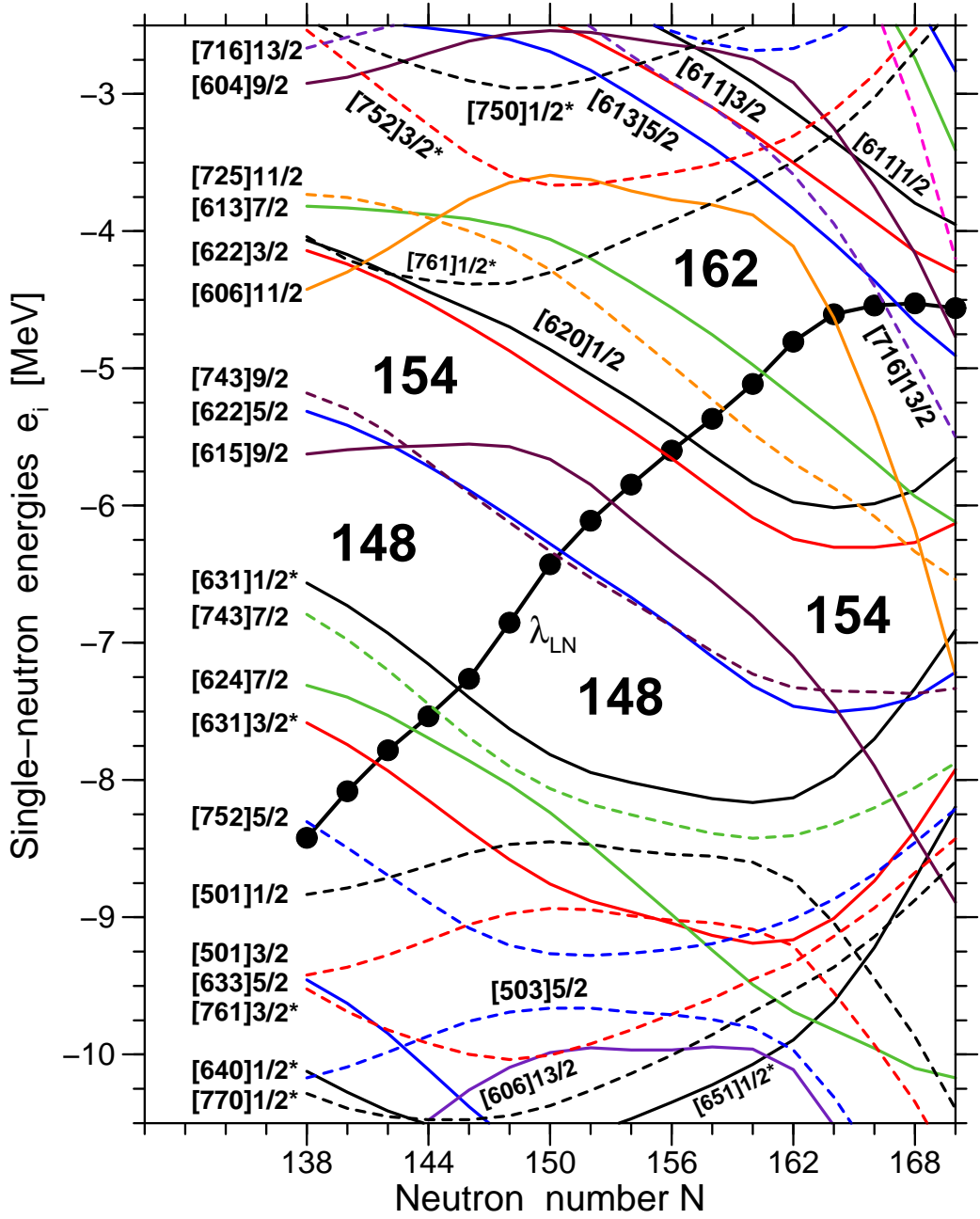}
\vspace{0.5cm}
\caption{Single-neutron energies in the Fm isotope chain as a 
function of neutron number $N$ obtained at the equilibrium deformation 
in the CRHB+LN calculations with the NL3 parametrization. Solid and 
dashed lines are used for positive and negative parity orbitals, 
respectively. The $\lambda_{LN}$ values are shown by solid line 
with solid circles. For other details, see caption of Fig.\ 
\protect\ref{proton-no254-sp}.}
\label{fm-sp-ener-neu}
\end{figure}

\begin{figure}
\epsfxsize 16.0cm
\epsfbox{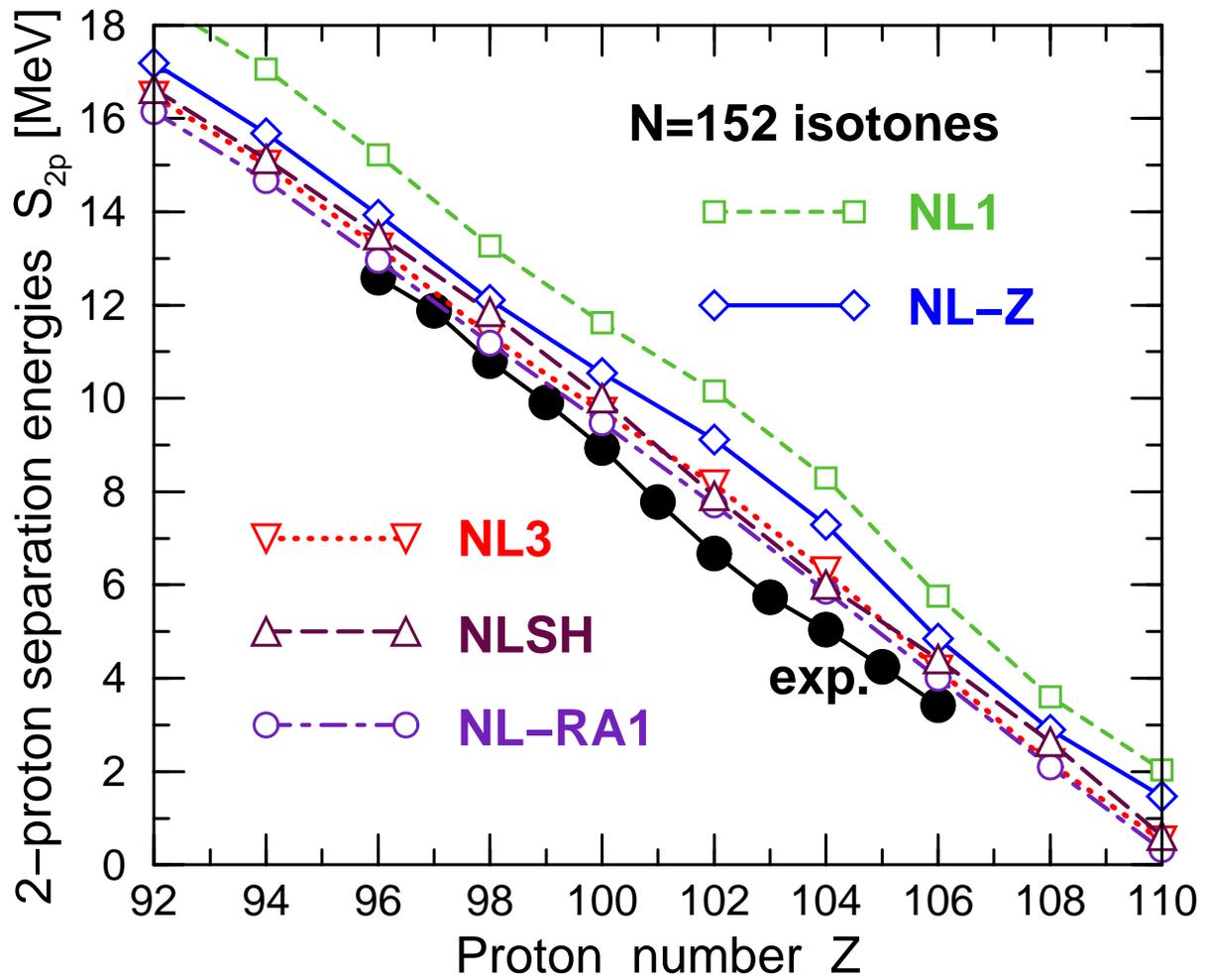}
\vspace{0.5cm}
\caption{The same as in Fig.\ \protect\ref{sep-ener-fm}, but
for  the two-proton separation energies $S_{2p}(Z,N)$ obtained for the 
$N=152$ isotones.} 
\label{sep-ener-n152}
\end{figure}

\begin{figure}
\epsfxsize 16.0cm
\epsfbox{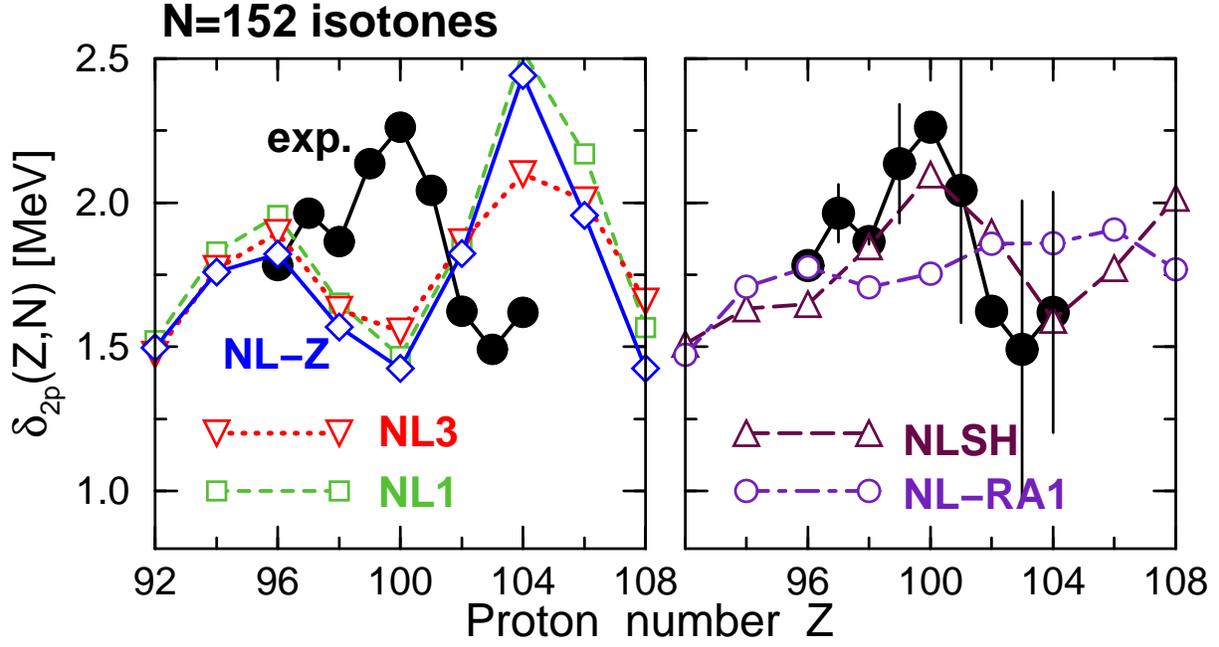}
\vspace{0.5cm}
\caption{The quantity $\delta_{2p}(Z,N)$ for the chain of $N=152$ 
isotones obtained in the CRHB+LN calculations with indicated RMF 
parametrizations. Solid circles are used for experimental data, 
while open symbols for theoretical results. The experimental error 
bars are shown in panel (b).} 
\label{delta-n152}
\end{figure}

\begin{figure}[t]
\epsfysize 18.0cm
\epsfbox{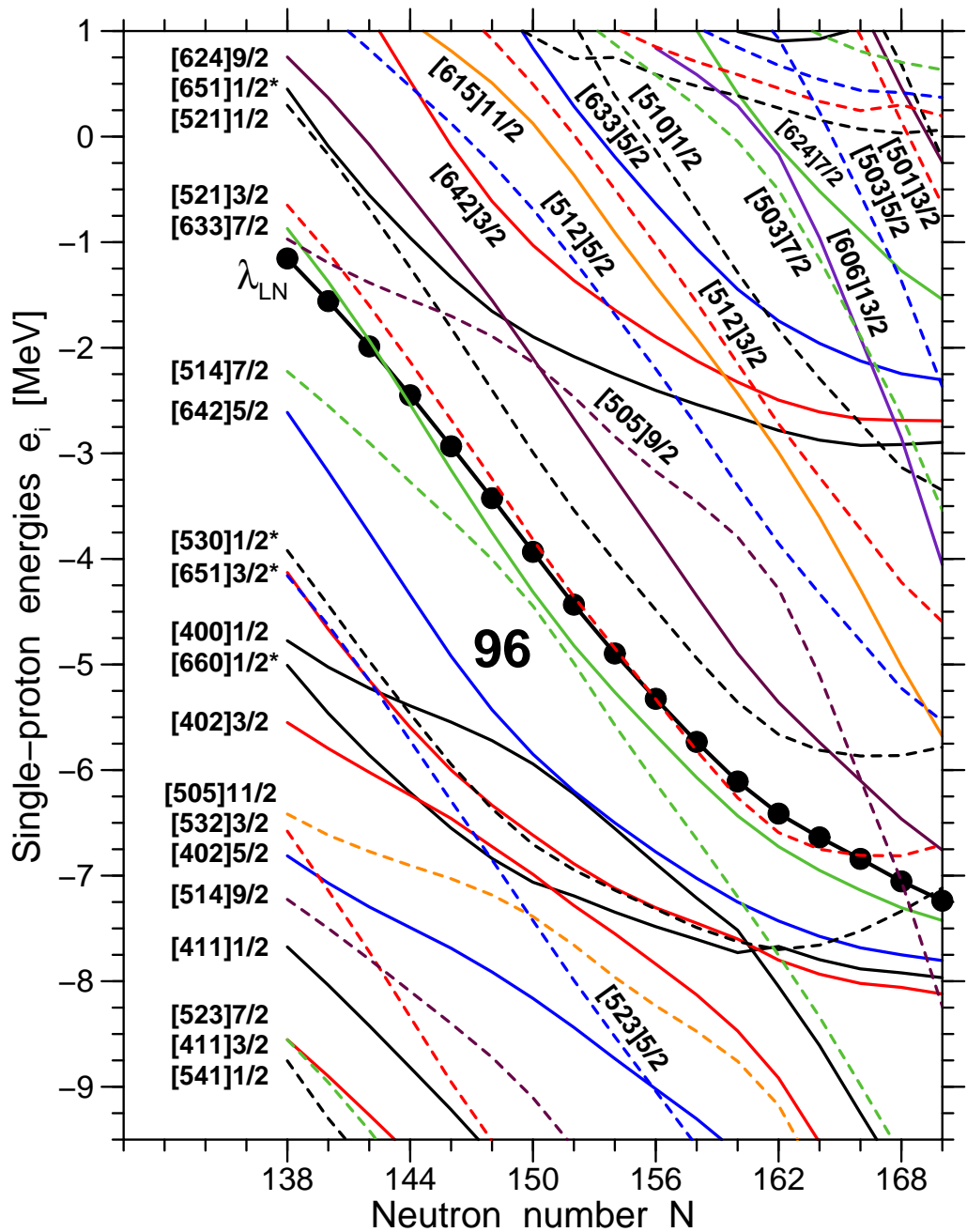}
\vspace{0.5cm}
\caption{The same as Fig.\ \protect\ref{fm-sp-ener-neu}, but 
for the single-proton energies.}
\label{fm-sp-ener-prot}
\end{figure}

\begin{figure}[t]
\epsfxsize 14.0cm
\epsfbox{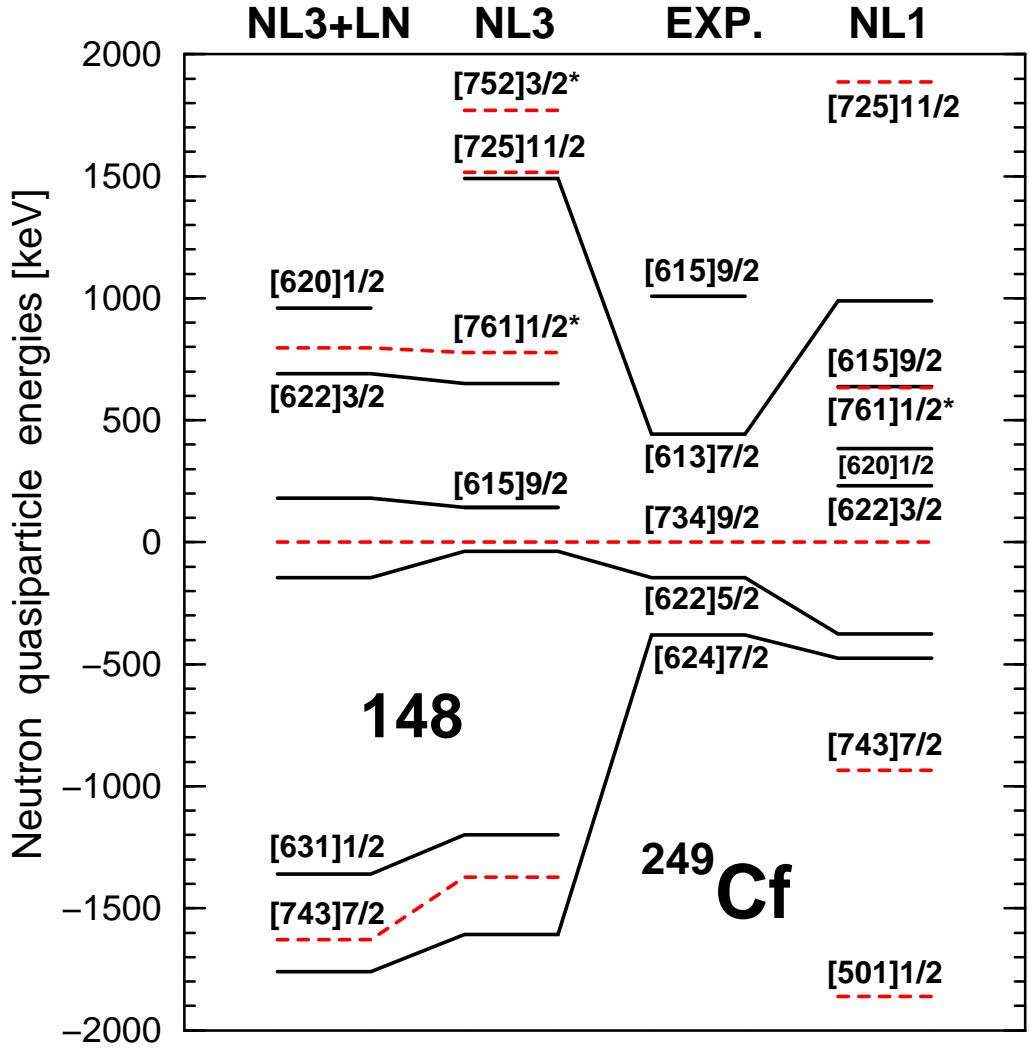}
\vspace{0.5cm}
\caption{Experimental and theoretical quasiparticle energies of 
neutron states in $^{249}$Cf. Positive and negative energies are
used for particle and hole states, respectively. The experimental 
data are taken from Ref.\ \protect\cite{249Cf}. Solid and dashed 
lines are used for positive and negative parity states, respectively. 
The symbols 'NL3' and 'NL1' indicate the RMF parametrization. 
The CRHB results shown below them were obtained with original D1S 
Gogny force $(f=1.0)$ used in pairing channel and 
without particle number projection. 'NL3+LN' 
indicates results with the LN method, the NL3 parametrization and 
scaling $f$ of the strength of D1S force (given in Table 
\protect\ref{Table-scaling}). In each calculational scheme, attempts 
were made to obtain solutions for every state shown in figure. 
The absence of a state indicates that convergence was not reached.}
\label{qp-cf249}
\end{figure}

\begin{figure}[t]
\epsfxsize 16.0cm
\epsfbox{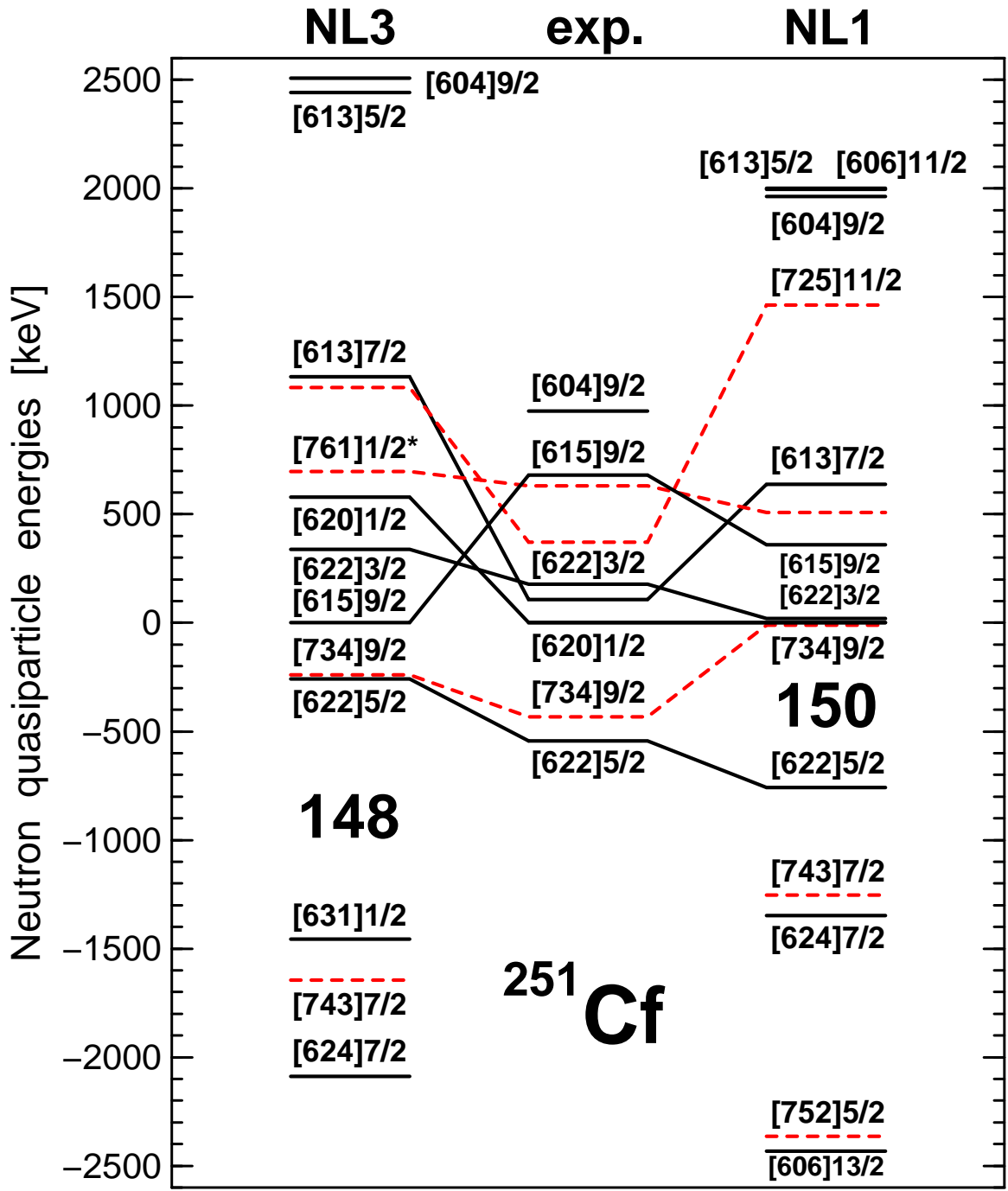}
\vspace{0.5cm}
\caption{The same as in Fig.\ \protect\ref{qp-cf249} but for 
neutron states in $^{251}$Cf. The experimental data are  
from Refs.\ \protect\cite{251Cf,251Cf-old,251Cf-a}.}
\label{qp-cf251}
\end{figure}

\begin{figure}[t]
\epsfxsize 16.0cm
\epsfbox{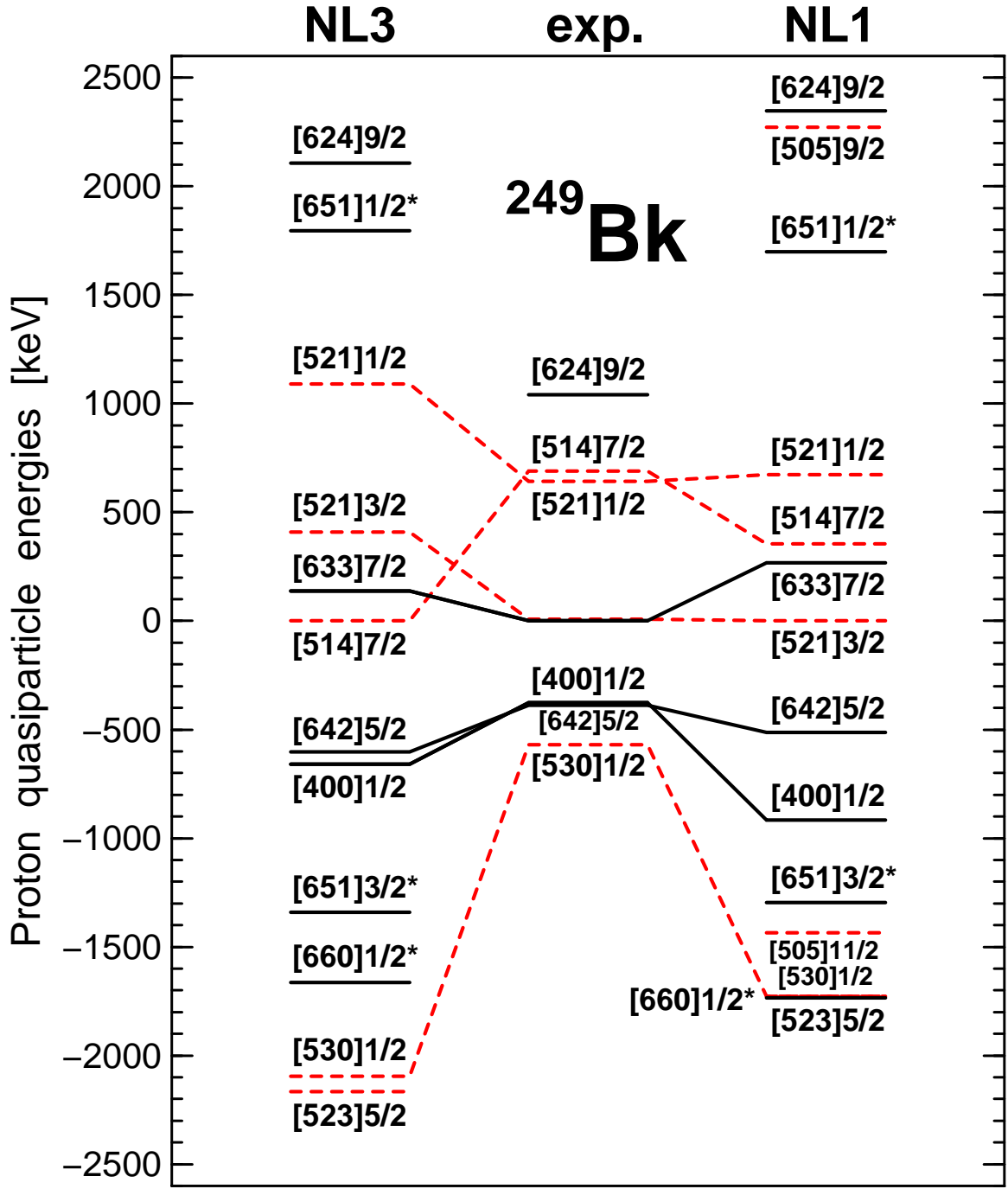}
\vspace{0.5cm}
\caption{The same as in Fig.\ \protect\ref{qp-cf249}
but for proton states in $^{249}$Bk. The experimental data 
are from Ref.\ \protect\cite{249Bk}.}
\label{qp-bk249}
\end{figure}

\begin{figure}[t]
\epsfxsize 14.0cm
\epsfbox{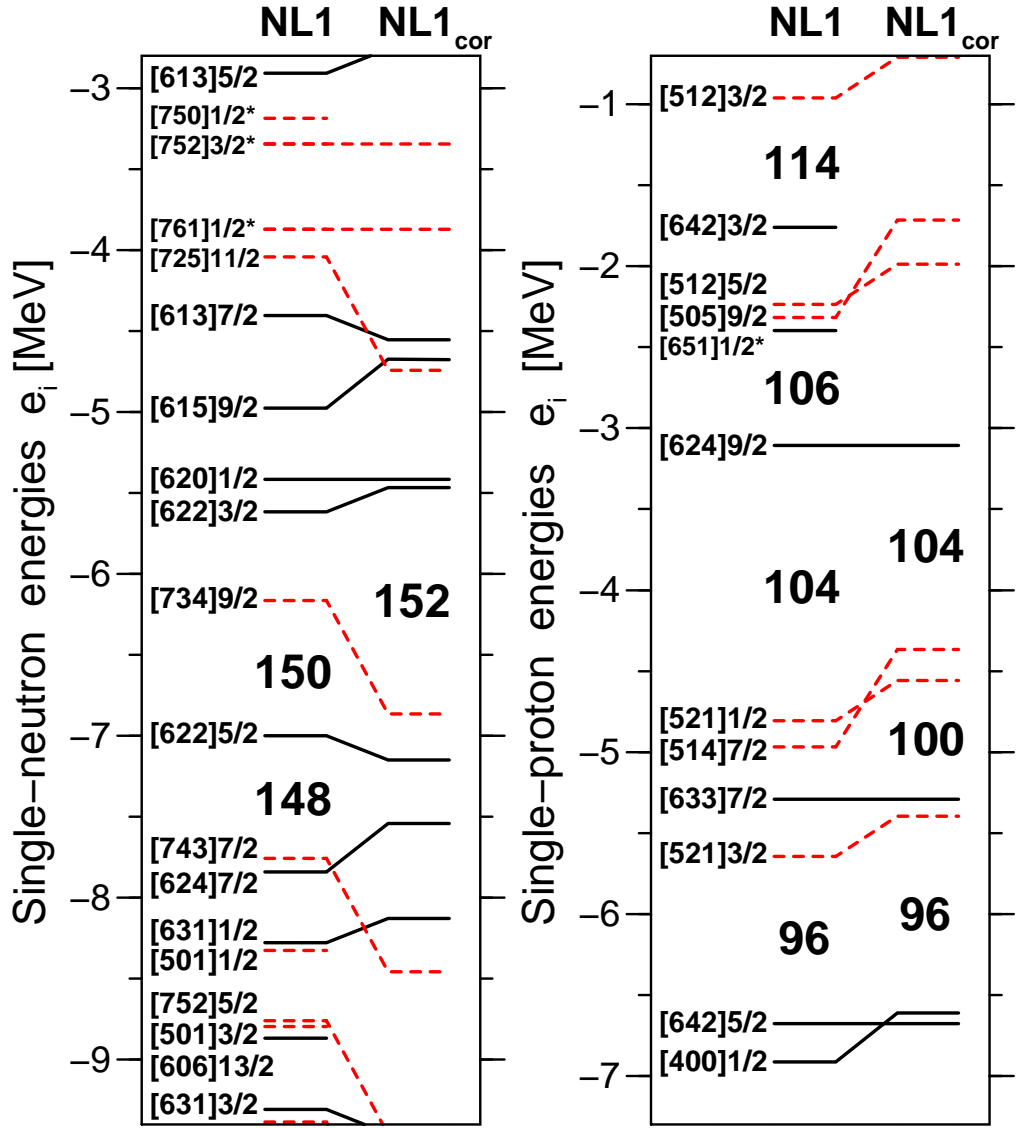}
\vspace{0.5cm}
\caption{Neutron and proton single-particle energies in 
$^{254}$No. The columns marked by 'NL1' show the original 
spectra obtained with the NL1 parametrization (see Figs.\ 
\protect\ref{proton-no254-sp} and 
\protect\ref{neutron-no254-sp}). The columns 'NL1$_{\rm cor}$' show how 
the spectra are modified if the energies were shifted as discussed in 
Sect.\ \protect\ref{Bk249} and \protect\ref{Cf249}. Solid and dashed 
lines are used for positive and negative parity states. Deformed 
gaps are indicated.}
\label{def-sp-cor}
\end{figure}

\begin{figure}[t]
\epsfxsize 14.0cm
\epsfbox{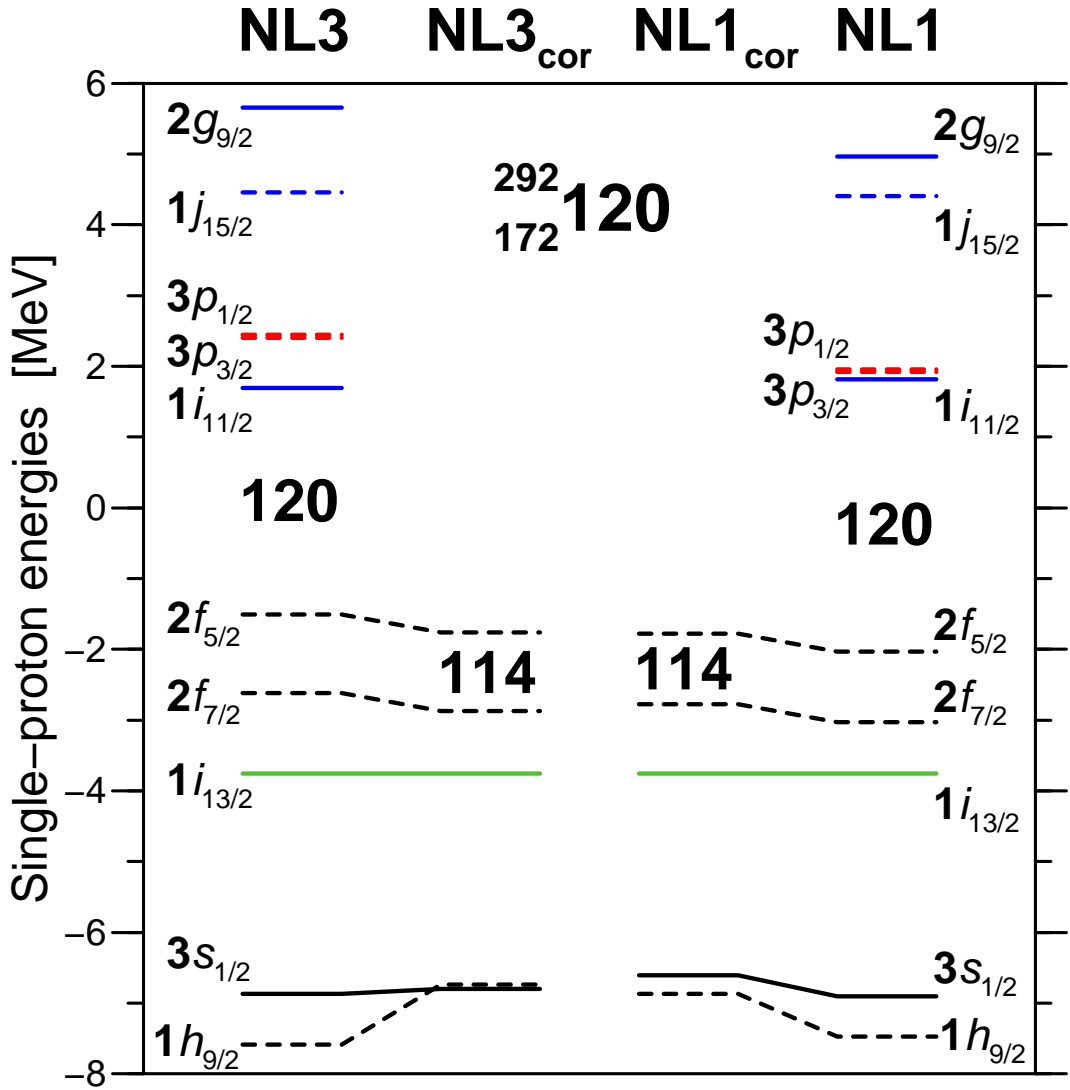}
\vspace{0.5cm}
\caption{Proton single-particle states in a $^{292}_{172}120$ 
nucleus. Columns 'NL3' and 'NL1' show the states obtained in 
the RMF calculations at spherical shape with the indicated 
parametrizations. The energy of the $1i_{13/2}$ state in the 
NL1 parametrization is set to be equal to that in NL3, 
which means that the energies of all states in NL1 (last 
column) are increased by 0.78 MeV. The columns 'NL3$_{\rm cor}$' 
and 'NL1$_{\rm cor}$' show how the spectra are modified if 
empirical shifts were introduced based on discrepancies between 
calculations and experiment for quasiparticle spectra in 
deformed $^{249}$Bk (see Sect.\ \protect\ref{Bk249} for 
numerical values). Solid and dashed lines are used for positive 
and negative parity states. Spherical gaps at $Z=114$ and 
$Z=120$ are indicated.}
\label{z120-proton}
\end{figure}

\begin{figure}[t]
\epsfxsize 14.0cm
\epsfbox{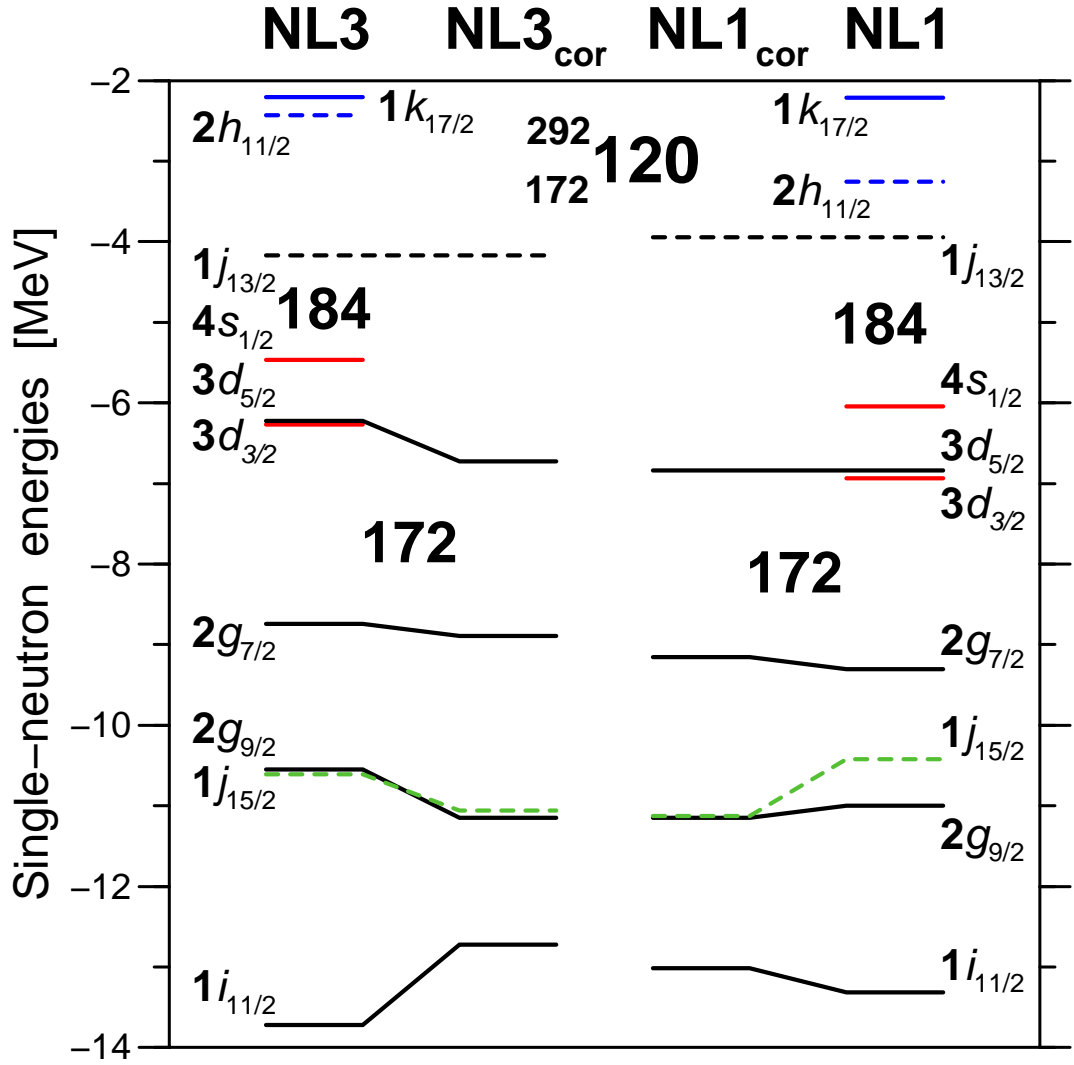}
\vspace{0.5cm}
\caption{The same as Fig.\ \protect\ref{z120-proton}, but for 
neutron single-particle states. The energies of all states 
obtained with the NL1 parametrization (last column) are increased 
by 0.76 MeV in order to have the same energies of the $2g_{9/2}$ 
states in the second and third columns. Spherical gaps at $N=172$ 
and $N=184$ are indicated.}
\label{z120-neutron}
\end{figure}

\begin{figure}[t]
\epsfxsize 16.0cm
\epsfbox{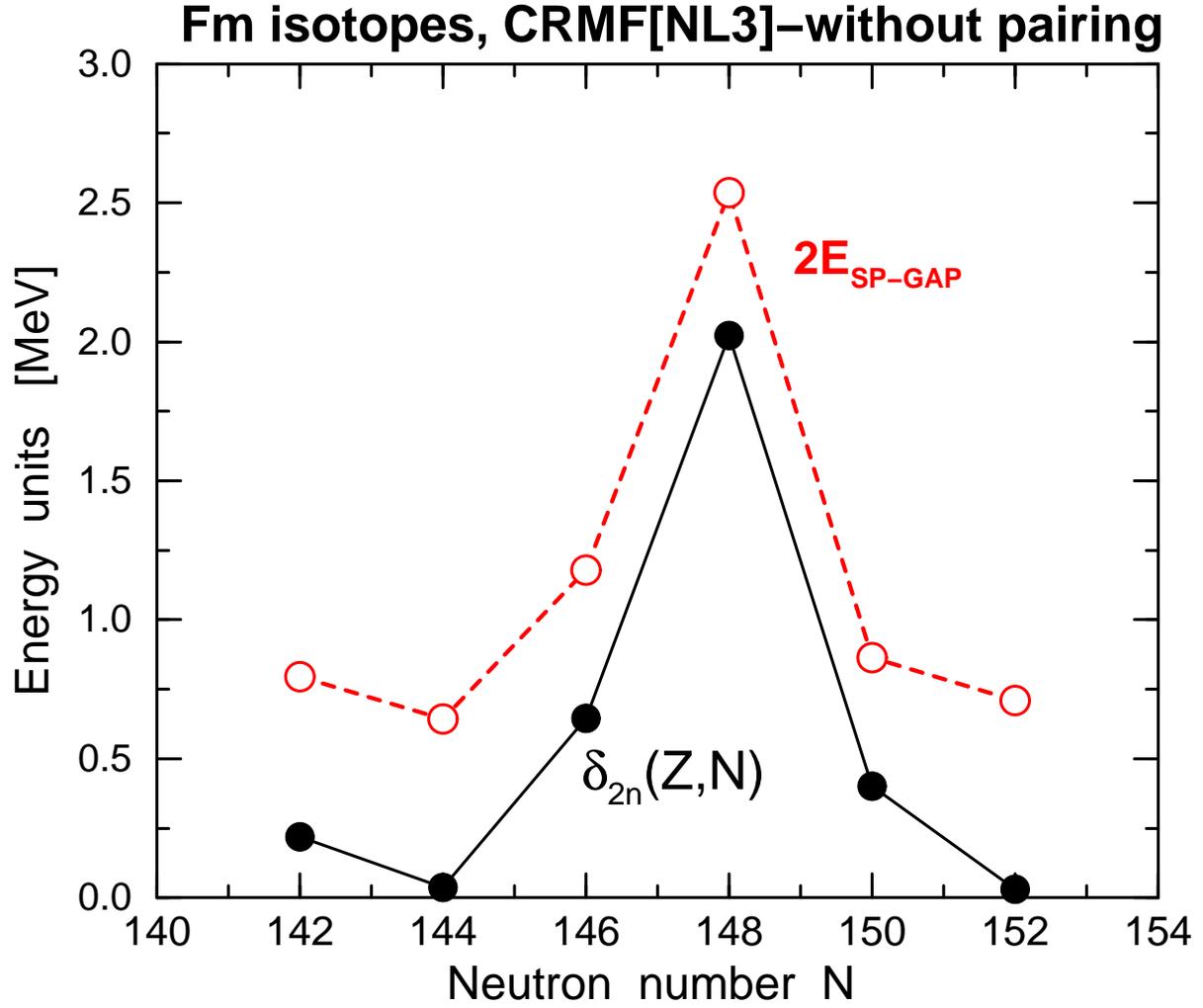}
\vspace{0.5cm}
\caption{Comparison of the $\delta_{2n}(Z,N)$ quantity with the 
twice the size of the shell gap $2E_{SP-GAP}$, which is the distance
between the last occupied and first unoccupied orbitals. Both
quantities are calculated with the NL3 parametrization and without 
pairing.}
\label{delta-no-pairing}
\end{figure}

\begin{figure}[t]
\epsfxsize 14.0cm
\epsfbox{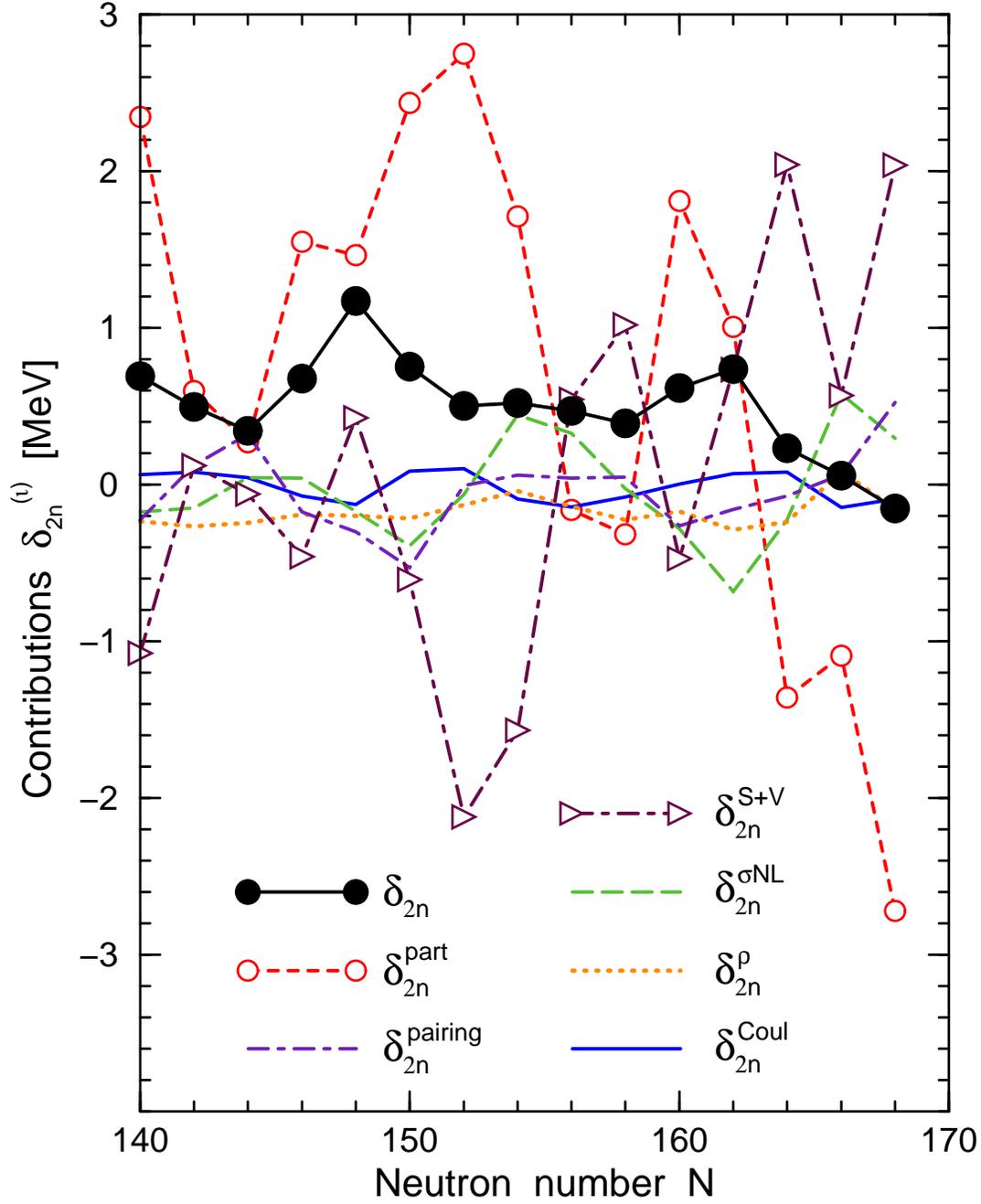}
\vspace{0.5cm}
\caption{The contributions $\delta_{2n}^{(i)}(Z,N)$ of the 
different terms of the CRHB+LN theory to the
quantity $\delta_{2n}(Z,N)$ as a function of the neutron 
number $N$ for the chain of the Fm isotopes. The contribution 
$\delta_{2n}^{LN}$ is not shown since it typically lies in the 
range from $-30$ keV up to $+30$ keV. The only exceptions are 
$N=148$ and $N=162$, where $\delta_{2n}^{LN}= 83$ and 
62 keV, respectively.}
\label{delta-contr-fm}
\end{figure}

 Based on Eq.\ (\ref{Etot}), one can express $\delta_{2n}(Z,N)$ as a sum 
of the contributions of different terms of the RMF Lagrangian
\begin{eqnarray}
\delta_{2n}(Z,N) &=& \delta_{2n}^{S+V}(Z,N) + 
\delta_{2n}^{\sigma NL}(Z,N) + \delta_{2n}^{\rho}(Z,N)
+ \delta_{2n}^{Coul}(Z,N) \nonumber \\
&+& \delta_{2n}^{part}(Z,N) + \delta_{2n}^{pairing}(Z,N)
+ \delta_{2n}^{LN}(Z,N).
\label{delta-2n}
\end{eqnarray}
An analysis of these contributions to $\delta_{2n}(Z,N)$ is presented 
for the chain of the Fm isotopes in Fig.\ \ref{delta-contr-fm}. 
The largest contributions to $\delta_{2n}(Z,N)$ come from the 
particle energies ($\delta_{2n}^{part}(Z,N)$) and from the main part 
of the nucleonic potential $(\delta_{2n}^{S+V}(Z,N))$. 
They are generally in opposite phase as a function of neutron 
number and thus they cancel each other to a large extent. It is interesting
to see that the maximum (in absolute value) of these contributions is 
located at neutron number $N=152$, while the large shell gap is seen in 
the single-neutron spectra at $N=148$ (see Fig.\ \ref{fm-sp-ener-neu}). It 
is difficult to understand why the maximum of $\delta_{2n}^{part}(Z,N)$ and 
$\delta_{2n}^{S+V}(Z,N)$ does not correlate with the $N=148$ shell gap, 
but a plausible reason is related to the trend of deformation changes. The 
calculated $\beta_2$-deformation increases at neutron number $N=138-148$, 
and then decreases at $N\geq 150$, see Fig.\ \ref{def-fm}a. 

 The contributions to the $\delta_{2n}(Z,N)$ coming from the 
$\rho$-meson $(\delta_{2n}^{\rho}(Z,N))$, the non-linear self-coupling 
of the $\sigma$-meson $(\delta_{2n}^{\sigma NL}(Z,N))$, the Coulomb 
potential $(\delta_{2n}^{Coul}(Z,N))$ and the pairing interaction 
$(\delta_{2n}^{pairing}(Z,N))$ are non-negligible and at some particle 
numbers some of them are comparable with the size of the total 
$\delta_{2n}(Z,N)$. 

 It is interesting that total $\delta_{2n}(Z,N)$ can become negative, 
as seen in the Fm isotopes at $N=168$ in the CRHB+LN calculations with 
the NL3 and NL-RA1 parametrizations; see Figs.\ \ref{delta-contr-fm}b and f.
This is a region of neutron numbers where the deformation changes are 
considerable; see Fig.\ \ref{def-fm}. This result reflects their 
importance in the definition of $\delta_{2n}(Z,N)$ and again underlines 
the fact that many factors beyond the size of the shell gap contribute 
to $\delta_{2n}(Z,N)$. Since by definition the shell gap has to have 
positive value and because of the reasons mentioned above, it is clear 
that $\delta_{2n}(Z,N)$ cannot be a direct measure of the shell gap. 

 However, this quantity, being related to the second derivative of the 
binding energy as a function of nucleon number, is more sensitive to 
the local decrease in the single-particle density associated with a shell 
gap than the two-nucleon separation energy $S_{2n}(Z,N)$.

\newpage
%------------------------------------------------------------
%-------------------------------------------------------------

\newpage
%------------------------------------------------------------
%-------------------------------------------------------------

\newpage
%------------------------------------------------------------
%-------------------------------------------------------------

\newpage
%----------------------------------------------------------------
%-----------------------------------------------------------------

\newpage
%------------------------------------------------------------
%------------------------------------------------------------

\newpage
%------------------------------------------------------------
%-------------------------------------------------------------

\newpage
%------------------------------------------------------------
%-------------------------------------------------------------

\newpage
%------------------------------------------------------------
%-------------------------------------------------------------

\newpage
%------------------------------------------------------------
%-------------------------------------------------------------

\newpage
%------------------------------------------------------------
%-------------------------------------------------------------

\newpage
%------------------------------------------------------------
%-------------------------------------------------------------

\newpage
%------------------------------------------------------------
%-------------------------------------------------------------

\newpage
%------------------------------------------------------------
%-------------------------------------------------------------

\newpage
%------------------------------------------------------------
%-------------------------------------------------------------

\newpage
%------------------------------------------------------------
%-------------------------------------------------------------

\newpage
%------------------------------------------------------------
%------------------------------------------------------------

\newpage
%------------------------------------------------------------
%------------------------------------------------------------

\newpage
%------------------------------------------------------------
%------------------------------------------------------------

\newpage
%------------------------------------------------------------
%-------------------------------------------------------------

\newpage
%------------------------------------------------------------
%------------------------------------------------------------

\newpage
%------------------------------------------------------------
%------------------------------------------------------------

\newpage
%------------------------------------------------------------
%-------------------------------------------------------------

%------------------------------------------------------------
%-------------------------------------------------------------

%------------------------------------------------------------
%-------------------------------------------------------------

%------------------------------------------------------------
%-------------------------------------------------------------

%------------------------------------------------------------
%-------------------------------------------------------------

%------------------------------------------------------------
%-------------------------------------------------------------

%------------------------------------------------------------
%-------------------------------------------------------------

\newpage
%------------------------------------------------------------
%-------------------------------------------------------------

\newpage
%------------------------------------------------------------
%-------------------------------------------------------------

\newpage
\begin{table}
\caption{The scaling factors $f$ of the Gogny D1S force (see Eq.\ (\protect\ref{Vpp})) 
used for different parametrizations of the RMF Lagrangian in the CRHB+LN 
calculations.}
\begin{center}
\begin{tabular}{|c|c|c|c|c|c|} 
Parametrization  & NL1   & NL-Z  &  NL3  & NL-RA1  & NLSH  \\ \hline
$f$    & 0.893 & 0.880 & 0.864 &  0.861  & 0.876 \\ 
\end{tabular}
\end{center}
\label{Table-scaling}
\end{table}

\begin{table}
\caption{Three-point indicators of the odd-even staggering of binding energies 
$\Delta^{(3)} (N) = \frac{(-1)^N}{2} \left[ B(N-1,Z) + B(N+1,Z) -2 B(N,Z) \right]$,
where $B(N,Z)$ is the (negative) binding energy of a system with $N$ neutrons
and $Z$ protons. An analogous proton indicator $\Delta^{(3)} (Z)$ is obtained by 
fixing the neutron number $N$ and replacing $N$ by $Z$. Column 1 indicates the type 
of indicator (proton $\Delta^{(3)} (Z)$ or neutron $\Delta^{(3)} (N)$) and 
nucleus with $Z$ (proton indicator) and $N$ (neutron indicator). Columns 3 and 
4 give the results obtained in CRHB calculations with $f=1.0$ for the D1S 
force and with the NL3 and NL1 parametrizations, while columns 5 and 6
give those obtained in CRHB+LN calculations with $f$ values given in Table 
\protect\ref{Table-scaling}.}
\begin{center}
\begin{tabular}{|c|c|c|c|c|c|} 
  $\Delta^{(3)} (...)$               &  exp   & NL3   &  NL1  & NL3+LN   & NL1+LN \\ \hline
    1                              &   2    &  3    &  4    &   5      &   6    \\ \hline
$\Delta^{(3)} (Z)$ [$^{249}$Bk]    & 0.399  & 0.516 & 0.515 &          &        \\ 
$\Delta^{(3)} (N)$ [$^{249}$Cf]    & 0.519  & 0.481 & 0.559 &  0.458   &  0.515 \\
$\Delta^{(3)} (N)$ [$^{251}$Cf]    & 0.531  & 0.491 & 0.605 &          &        \\ 
\end{tabular}
\end{center}
\label{Table-odd-even}
\end{table}

\begin{table}
\caption{Spherical subshells active in superheavy nuclei and their deformed 
counterparts active in the $A\sim 250$ mass region. The left column shows the 
spherical subshells active in the vicinity of the ``magic'' spherical gaps 
$(Z=120, N=172)$. Their ordering is given according to the RMF calculations 
with the NL3 parametrizations in the $^{292}_{172}120$ system (see Figs.\ 
\protect\ref{z120-proton} and \protect\ref{z120-neutron}). Although the gaps 
depend on the specific RMF parametrization, the same set of spherical subshells 
is active with other parametrizations (see, for example, Fig.\ 4 in Ref.\ 
\protect\cite{BRRMG.99}). The right column shows the deformed quasiparticle 
states observed in $^{249}_{\,\,\,97}$Bk$_{152}$ \protect\cite{249Bk} and 
$^{249,251}_{\,\,\,98}$Cf$_{151,153}$ \protect\cite{249Cf,251Cf,251Cf-old}. 
The bold style is used for the states which may be observed in nuclides with
$N\approx 162$ and/or $Z\approx 108$. The symbols 'N/A' (not accessible) are 
for the deformed states which typically increase their energy with  
deformation and thus are not likely to be seen experimentally.}  
\begin{center}
\begin{tabular}{|c|c|}
Spherical subshell      & Deformed state \\ \hline
Proton states           &               \\    
$\pi 1j_{15/2}$         &  {\bf $\pi$ [770]1/2} \\
$\pi 3p_{1/2} $         &    N/A        \\
$\pi 3p_{3/2} $         &    N/A        \\
$\pi 1i_{11/2}$         &  {\bf $\pi$ [651]1/2} \\
$Z=120$                 &               \\
$\pi 2f_{5/2} $         &  $\pi [521]1/2$     \\
$\pi 2f_{7/2} $         &  $\pi [521]3/2$,\,\,$\pi [530]1/2$ \\
$\pi 1i_{13/2}$         &  $\pi [642]5/2$,\,\,$\pi [633]7/2$,\,\,$\pi [624]9/2$ \\ 
$\pi 3s_{1/2} $         &  $\pi [400]1/2$ \\
$\pi 1h_{9/2} $         &  $\pi [514]7/2$ \\ \hline
%                       &                                     \\
Neutron states          &                                      \\
$\nu 1k_{17/2}$         &  {\bf $\nu$ [880]1/2}              \\
$\nu 2h_{11/2}$         &  {\bf $\nu$ [750]1/2}              \\
$\nu 1j_{13/2}$         &  $\nu [761]1/2$                    \\
$N=184        $         &                                  \\
$\nu 4s_{1/2} $         &  N/A                             \\
$\nu 3d_{5/2} $         &  $\nu [620]1/2$                  \\
$\nu 3d_{3/2} $         &  N/A                             \\
$N=172        $         &                                  \\
$\nu 2g_{7/2} $         &  $\nu [622]3/2$                  \\ 
$\nu 2g_{9/2} $         &   $\nu [622]5/2$, $\nu [613]7/2$, $\nu [604]9/2$ \\
$\nu 1j_{15/2}$         &   $\nu [734]9/2$, $\nu [725]11/2$ \\
$\nu 1 i_{11/2}$        &   $\nu [615]9/2$, $\nu [624]7/2$ \\
\end{tabular}
\end{center}
\label{qp-experiment}
\end{table}

\begin{table}
\caption{ Additional binding ($\Delta E_{NM}$) induced by nuclear 
magnetism for different neutron ($^{249}$Cf) and proton ($^{249}$Bk)
quasiparticle states obtained in the CRHB calculations with the NL3 
parametrization and full strength (scaling factor $f=1.0$) of the 
Gogny force. The quantity $\Delta E_{NM}$ is defined as the difference 
of binding energies obtained in the calculations with and without 
nuclear magnetism. As tested for a number of states, the change of 
the scaling factor $f$ 
to the one given in Table \protect\ref{Table-scaling} and/or the use 
of the LN method modifies $\Delta E_{NM}$ only marginally.}
\begin{center}
\begin{tabular}{|c|c|c|c|} 
neutron state    &  $\Delta E_{NM}$ (keV) & proton state     &  $\Delta E_{NM}$ (keV) \\ \hline
$ \nu[734]9/2 $  &     -36                & $\pi [521]1/2$   & -16  \\ 
$ \nu[615]9/2 $  &     -55                & $\pi [514]7/2$   & -35  \\
$ \nu[624]7/2 $  &     -56                & $\pi [633]7/2$   & -22  \\
$ \nu[622]3/2 $  &     -27                & $\pi [624]9/2$   & -23  \\
$ \nu[622]5/2 $  &     -33                & $\pi [521]3/2$   & -27  \\ 
$ \nu[734]7/2 $  &     -37                & $\pi [523]5/2$   & -33  \\
$ \nu[613]7/2 $  &     -29                & $\pi [642]5/2$   & -23  \\
$ \nu[725]11/2$  &     -34                &                  &  \\
$ \nu[761]1/2^*$ &     -69                &                  &  \\
$ \nu[752]3/2^*$ &     -53                &                  &  \\ 
\end{tabular}
\end{center}
\label{Table-ENM}
\end{table}

\end{document}